\documentclass[letter,11pt]{article}
 
\usepackage{jheppub}

\usepackage{graphicx}
\usepackage{dcolumn}
\usepackage{bm}
\usepackage{float}
\usepackage{hyperref}
\usepackage{subfig}
\usepackage{dsfont}
\usepackage{slashed}
\usepackage{color}
\usepackage{amsmath}
\usepackage[section]{placeins}
\usepackage{braket}
\usepackage{upgreek}
\usepackage[bottom]{footmisc}
\usepackage[normalem]{ulem}

\newcommand{\tr}{\text{Tr}}

\newcommand{\pphase}{\frac{d^4p}{(2\pi)^4}\delta(p^2)}
\newcommand{\qphase}{\frac{d^4q}{(2\pi)^4}\delta(q^2)}
\newcommand{\sinc}{\text{sinc}}

\newcommand{\bea}{\begin{eqnarray}}
\newcommand{\eea}{\end{eqnarray}}

\title{Factorization for energy-energy correlator in heavy ion collision}
\author{Balbeer Singh and}
\author{Varun Vaidya}
\affiliation{Department of Physics, \\ 
University of South Dakota}

\emailAdd{balbeer.singh@usd.edu}
\emailAdd{Varun.Vaidya@usd.edu}

\date{\today}
\preprint{}

\abstract{We present a factorization formula for the energy-energy correlator in the collinear limit for the case of heavy ion collisions. Employing Soft Collinear Effective Theory, we provide a complete framework for jet production and evolution by separating the jet dynamics from the universal medium physics to all orders in perturbation theory in terms of gauge invariant operators. The EFT allows us to precisely define the domain of validity of different approximations and  to systematically go beyond leading order results in the literature through radiative corrections. For this observable, we show where the leading order GLV and BDMPS-Z results are valid and infer that higher order radiative corrections lead to both DGLAP and BFKL evolutions. We further show the impact of BFKL  resummation on the medium induced jet function for two point energy correlator. Crucially, the EFT approach enables us to evaluate the universality of the non-perturbative physics which is the key to predictive power in a strongly coupled medium.}


\begin{document}

\maketitle

\section{Introduction}

In the background of a strongly coupled medium such as the quark-gluon plasma (QGP) in heavy-ion collisions (HICs) or a nucleus in deep-inelastic scattering (DIS), a hard-scattering event can produce partons with energies significantly larger than the typical medium scales, such as the temperature ($T$)  or $\Lambda_{\rm QCD}$, making these high energy partons as natural X-rays for the medium. The subsequent parton cascade and fragmentation then lead to collimated sprays of particles known as jets. One would expect that the presence of the medium would modify the this fragmentation process. As a result, the final distribution of hadrons at the detector will carry the imprint of the microscopic/macroscopic properties of the medium created in high energy nuclear collisions. In the past decade there have been numerous studies for various hard probes for heavy ion colliders including jets~\cite{Connors:2017ptx,Tang:2020ame,Foka:2016zdb}. In particular, for jets, the initial production of the hard parton initiating the jet is theoretically well controlled, which makes them an excellent tool to probe the properties of the QGP in HICs ~\cite{ATLAS:2014ipv,ALICE:2015mjv,CMS:2016uxf,STAR:2005gfr,PHENIX:2004vcz,Vertesi:2024tdv,ALICE:2019whv,CMS:2018fof,CMS:2014jjt,ATLAS:2012tjt,Wiedemann:2009sh} and nuclei in electron-nucleus collisions~\cite{AbdulKhalek:2021gbh}. In this work, we primarily focus on HICs but analogous techniques can be applied to study cold nuclear matter effects in electron-nucleus collisions as well.

The direct observables such as inclusive jet production are sensitive to the energy loss in the medium. In contrast, the substructure observable being insensitive to the interactions between partons produced at the initial state interactions provides complimentary information about modifications of the internal structure of the jets in the medium. Therefore, jet substructure provides an excellent opportunity to study jet fragmentation within the QGP and hence medium induced modifications~\cite{Chien:2024uax,Barata:2023zqg,Budhraja:2023rgo, Cunqueiro:2023vxl,Casalderrey-Solana:2019ubu,Milhano:2017nzm,Caucal:2021cfb,Ehlers:2020piz,Li:2019dre,Chien:2016led}. In this regard, energy-energy correlators (EECs) have drawn tremendous attention due to their potential ability to offer a comprehensive view of jet substructure by measuring the correlations between the energies of the particles inside the jet~\cite{Chen:2020vvp,Bossi:2024qho,Andres:2023ymw,Andres:2023xwr,Kang:2023big,Budhraja:2024xiq,Andres:2022ovj,Chen:2024nfl,Liu:2024kqt,Holguin:2023bjf,Devereaux:2023vjz}. EECs have been studied to high precision for proton-proton (pp) and $e^+ e^-$ collisions where they offers a way to directly visualize emergent scales in QFT as kinks in the distribution curve \cite{Dixon:2019uzg,Komiske:2022enw}. Further, In pp, two and three particle energy correlation distributions have been measured to extract strong coupling constant, i.e., $\alpha_s(m_Z)$~\cite{CMS:2024mlf}. Moreover, higher point energy correlators which are less sensitive to soft radiations due to energy suppression have also been used to study the collinear substructure of the radiation~\cite{Yang:2024gcn,Yan:2022cye,Yang:2022tgm,Chicherin:2024ifn}.  Recently, using field theoretic techniques EEC has been proposed to study hadronization process in a model independent way~\cite{Chen:2024nfl}. Likewise, nucleon energy correlators have also been proposed to study microscopic features such as partons angular distribution, all order collinear splitting and internal transverse dynamics of nucleon in the lepton-nucleon deep inelastic scattering~\cite{Liu:2022wop}.

The computation of the EECs for the case of heavy ion collisions  is gaining interest due to its potential ability to resolve the color coherence effect characterized by the emergent resolution scales of the medium such as coherence angle, i.e., $\theta_c$. In particular, using both multiple scattering approach with BDMPS-Z and single scattering framework, i.e., GLV it has been shown that energy flow correlators can be used to study color coherence effect via medium modified splitting functions~\cite{Andres:2023xwr}. However, the corresponding effect is somewhat weaker in the single scattering limit~\cite{Andres:2023xwr}.  Further, using $\gamma$-triggered jets, it was also shown that at large angles EEC is  enhanced due to medium response arising from elastic scatterings. In addition, the small angle suppression is induced by jet energy loss and transverse momentum broadening leading to deviation from vacuum scaling behaviour~\cite{Yang:2023dwc}.  Using full splitting for $\gamma\to q\bar{q}$, at leading log (LL) accuracy it was shown that the energy loss effects can suppress 2-point energy correlator at large angles and may lead to enhancement at small angles. This conclusion is also supported by Monte Carlo simulation~\cite{Barata:2023bhh}. Recently, a more realistic description of EEC measurements on inclusive jet samples has been studied in Ref.~\cite{Andres:2024ksi} by including hydrodynamic effects, broadening corrections, selection bias and hadronization effects.  

All these result use theoretical approaches previously developed to understand jet propagation in quark gluon plasma \cite{Andres:2023xwr,Andres:2022ovj}. However, a systematically improvable approach that accounts for  medium physics in a factorized framework consistent with the vacuum result \cite{Dixon:2019uzg}  would significantly enhance our understanding of jet-medium interaction along with emergent scale dynamics. This is crucial, not only to be able to give accurate predictions but also to address a key question in heavy ion collisions: How do we isolate the non-perturbative physics from the perturbative to all orders in $\alpha_s$? This is important in order to establish universality of non-perturbative parameters across different jet observables. Moreover, with a systematic factorized approach, higher-order perturbative effects can be incorporated methodically, providing better theoretical control over computations.

Parton fragmentation in QGP by comparison to vacuum is a complex process which involves multiple manifest as well as emergent scales~\cite{Casalderrey-Solana:2011ule,Iancu:2014kga,Mehtar-Tani:2011lic,Caucal:2021cfb}.
In many phenomenological cases, all or some of these scales are hierarchically separated~\cite{Mehtar-Tani:2017web}. Hence a natural approach to analyze this system is through the lens of Effective Field Theory(EFT). The program of \textit{factorization} has been enormously successful in electron-proton (ep) and pp collisions where the universal non-perturbative physics encoded in well defined distribution functions of the hadron has been leveraged to make precise predictions. An open question in Heavy Ion collisions is whether this paradigm can be applied and could help us achieve an equal degree of accuracy in the light of high energy nuclear collision experiments at sPHENIX, ALICE, CMS etc.

In this work we use Soft collinear Effective Theory \cite{Bauer:2000yr,Bauer:2001ct,Bauer:2002nz} and its Glauber extension\cite{Rothstein:2016bsq} to address this problem in the context of the EEC observable. While we work with EEC, the factorization approach can be extended to other substructure observables as well.  The main purpose of adopting this methodology is to derive a framework for a complete calculation of jet observables which account for all possible vacuum and medium effects in a manner consistent with power counting. We note that this is different from previous approaches in literature~\cite{Ovanesyan:2011xy,Idilbi:2008vm,DEramo:2010wup} in the treatment of soft physics which controls the dynamics of the QGP medium and has important consequences for the structure of factorization and radiative corrections.   
An EFT approach offers a way to define the domain of validity of different approaches currently employed in literature and go beyond them in a systematically improvable manner. This was applied for describing dijet asymmetry in Refs.~\cite{Vaidya:2020cyi,Vaidya:2020lih,Vaidya:2021mly,Vaidya:2021vxu}.

The idea of the EFT is to exploit the hierarchy between scales to systematically define and match to EFTs at successively lower scales by integrating out heavier modes in a Wilsonian sense, finally completely isolating the physics at the non-perturbative scale. The resulting Renormalization group flow will allow us to resum large logarithms between the scales improving the accuracy of our prediction. We will show how the EFT captures all the previous results such as GLV, BDMPS-Z and more importantly, the regime where each is valid for describing the observable. At the same time, we can separate the universal physics of the medium from the observable dependent jet dynamics to all orders in perturbation theory through gauge invariant operator matrix elements. The factorization and the resultant renormalization group equations obeyed by these functions allows us to go beyond leading order (LO) results in a systematic way.

As stated earlier, another important goal of the EFT approach is to answer the key question:  is the non-perturbative physics universal thereby giving us predictive power across distinct jet observables in a strongly coupled medium? For jets in a medium, the answer depends strongly on the possible emergence of a \textit{perturbative} medium scale in a dense medium when the number of interactions of the medium with the jet is large. This is usually parameterized in literature through the jet quenching parameter $\hat q$~\cite{Baier:1996sk}. 

We would like to derive a precise definition for this scale in terms of field theoretic operator matrix elements, i.e to all orders in perturbation theory. This serves two purposes: first to check the universality of this parameter and second to be able to compute this object numerically. In this paper we take the first steps towards answering this important question.

The paper is organized as follows. We discuss various scales, both manifest and emergent that arise in jet-medium interaction dynamics as well as their hierarchy as a guide towards the EFT setup in Section \ref{sec:Scales}. We distinguish two regimes of the EFT and derive the factorization and leading order results  for each in Section \ref{sec:factone} and \ref{sec:factwo}. In Section \ref{sec:Qmed}, we discuss the universality or lack thereof of the non-perturbative physics in a strongly coupled medium through an emergent scale. Finally, we present our conclusions and future projections in Section \ref{sec:sum}.

\section{The observable, Physical scales and EFT setup}
\label{sec:Scales}

The differential cross-section for the two point energy correlators is defined as~\cite{Basham:1978bw} 
\bea
\frac{d\sigma}{d\chi} = \sum_{i,j} \int d\sigma \frac{E_iE_j}{Q^2} \delta \Big(\chi-\frac{1- \cos \theta_{ij}}{2}\Big),
\label{eq:eec}
\eea
where $E_i, E_j$ are the energies of two final state particles which are separated by an angle $\theta_{ij}$ and the energy scale $Q$ is defined through the total energy in the final state parton, i.e.,  $Q = \sum_{i} E_i$. We consider the collinear limit of the above observable and assume that $\chi\ll R$, where $R$ is jet radius.   Therefore, in that case $R$ dependence appears as power corrections and is suppressed by $\mathcal{O}(\chi/R^2)$ which we drop at leading order.
Thus, in the collinear region, we can simplify the definition given in Eq.~\ref{eq:eec} as
\bea
\frac{d\sigma}{d\chi} = \sum_{i,j} \int d\sigma \frac{E_iE_j}{Q^2} \delta\Big(\chi - \frac{\theta_{ij}^2}{4}\Big).
\label{eq:eec1}
\eea
In order to show the factorization and perform resummation, it is often more convenient to work with the cumulant, which is expressed as
\bea 
\Sigma(\chi_c) = \sum_{i,j} \int d\sigma \frac{E_iE_j}{Q^2} \Theta \Big( \chi_c - \frac{\theta_{ij}^2}{4}\Big).
\label{eq:eec2}
\eea
From here onwards we will refer to the differential observable as $\chi$ and to the cumulant as $\chi_c$.

For the case of jets in HICs, the physical process is that at the initial stages, the collision between two nuclei, creates an energetic parton in a ``hard interaction" which occurs at  scale $Q$. This parton then subsequently fragments both in the vacuum and the medium preferentially into energetic partons at angles of $\mathcal{O}(\sqrt{\chi} \ll 1 $). Without loss of generality, we set the direction of the initial parton to be along the $z$-axis and denote the light-like direction by vectors $n^{\mu}=(1,0,0,1)$ and $\bar{n}^{\mu}=(1,0,0,-1)$ so that $n^2=\bar{n}^2=0$ and $n\cdot\bar{n}=2$. In the light-cone co-ordinates, any four vector $p^{\mu}=(p^{-},p^{+},p_{\perp})$ with $p^{-}=\bar{n}\cdot p$ and $p^{+}=n\cdot p$ can be decomposed as
\begin{equation}
p^{\mu}=\bar{n}\cdot p\frac{n^{\mu}}{2}+n\cdot p\frac{\bar{n}^{\mu}}{2}+p_{\perp}^{\mu}.    
\end{equation}
The phase space for the parton shower in the collinear region is therefore populated by ``collinear" radiation that scales  as  
\begin{eqnarray}
p_c=Q(1,\lambda^2,\lambda),
\label{eq:Coll}
\end{eqnarray}
where $\lambda = \sqrt{\chi}$ is an expansion parameter of our EFT. For jet evolution in vacuum, this is the only region of phase space that we need and the corresponding factorization of the hard physics from the collinear was presented in \cite{Dixon:2019uzg} which we will reproduce in section \ref{sec:factone}.

Now we consider the scenario in which the jet produced in the hard interaction passes through the QGP medium populated by partons with energy $T \ll Q$. In most current heavy ion colliders the temperature  achievable is of the order of a few hundred MeV, making $T$ a non-perturbative scale. While traversing the medium, jet partons interact with thermal constituents through soft elastic collisions and inelastic processes such as radiation. In the single scattering limit, the jet parton typically receives a transverse kick of $\mathcal{O}(m_D \sim gT)$.  However, in a dense medium, jet undergoes multiple interactions with the thermal partons.  In the literature, this effect is typically parameterized by $\hat q$, which measures the average transverse momentum squared per unit length that the medium imparts to the jet parton. We use the phenomenological value of this parameter which is $\sim$ 1-2 GeV$^2$/fm. In a medium of size 5 fm, this corresponds to a total transverse momentum of 2-3 GeV. We will call this transverse momentum scale $Q_{\text{med}}$. Note that even though the scale $Q_{\rm med}$ originates in the multiple scattering scenario we will use this as a guide for the possible hierarchy of scales involved in the EECs and our hope would be to obtain a rigorous definition for this scale in the EFT framework in a self consistent manner. 

The full spectrum for the $\chi$ distribution, even in the collinear limit, ranges over several possible hierarchies of the scales involved. Since the scale $Q_{\text{med}}$ characterizes the medium, we consider the following two hierarchies,
\begin{itemize}
\item{}  Region I: $ Q\sqrt{\chi} \sim Q_{\text{med}}$ while  $Q\sqrt{\chi} \ll Q$. Here we need a single stage of matching between two widely separated scales, and the factorization for this scenario will be addressed in Section \ref{sec:factone}.
\item{} Region II: The scales Q, $ Q\sqrt{\chi}$  and $Q_{\text{med}}$ are well separated from each other i.e.,$Q\gg Q\sqrt{\chi}\gg Q_{\text{med}}$. This requires two stages of factorization. In stage I, we can separate physics at scale $Q$ from the IR physics at $Q\sqrt{\chi}$  and below.  Stage II involves a matching between the scales $Q\sqrt{\chi}$ and $Q_{\text{med}}$. The EFT for this case will be discussed in Section \ref{sec:factwo}

\end{itemize}
The reason for distinguishing these hierarchies is twofold: First, we want to exploit the hierarchies to simplify our calculation as much as possible for e.g., in region I, at leading order we require the full GLV \cite{Gyulassy:2000er,Ovanesyan:2011xy} result, while in region II we will only require the soft limit. Second, factorization of well separated scales avoids any ambiguity in the choice of the renormalization scale for the factorized functions. This systematic separation also enables us to isolate the non-perturbative physics from the perturbative.

In order to have perturbative control of the vacuum calculation,  we will always assume $ Q \sqrt{\chi}$ to be a perturbative scale and also $ Q \sqrt{\chi} \geq Q_{\text{med}}$, while as mentioned before, $T$ is non-perturbative which for illustration we assume to be  approximately $0.5$ GeV throughout this paper. Likewise, we will use a typical value of around 5 fm for the medium length. Another important scale for medium induced radiation is the coherence or formation time  $t_f \sim E/q_{\perp}^2$ where $E$ is energy of radiated gluon and $q_{\perp}$ is the transverse momentum. Typically, radiation with formation time larger than the medium length undergoes Landau-Pomeranchuk-Migdal (LPM) suppression~\cite{Migdal:1956tc}.
Furthermore, the medium can resolve two partons if the angle between them exceeds an emergent resolution angle, i.e., critical angle $\theta_c \sim \frac{1}{Q_{\text{med}}L}$. The typical value of the critical angle for above mentioned medium parameters therefore lies in the range $\mathcal{O}(10^{-2} - 10^{-1})$. Therefore, the two partons resolved by the medium act as independent sources of medium induced radiation so that the modification of the jet in the medium depends on its substructure. Below we discuss factorization for above two regions in more detail.


\section{Factorization for $Q\sqrt{\chi} \sim Q_{\text{med}}$}
\label{sec:factone}
We first examine the scenario where $Q\sqrt{\chi} \sim Q_{\text{med}}$. For a hard scale $Q \sim \mathcal{O}(10^2)$ GeV with $Q_{\text{med}} \sim$ 1-3 GeV, this hierarchy describes the  regime of $\chi \sim \mathcal{O}{(10^{-4} -10^{-3})} $. For the collinear mode defined in Eq.\ref{eq:Coll}, the parametric value of the formation time for this hierarchy reads as
\bea
t_f \sim \frac{Q}{Q^2_{\text{med}}} \sim \frac{1}{Q \chi}.
\eea
For $Q=200$ GeV,  $\hat{q}=1$ GeV$^2$fm$^{-1}$ and the medium length $L=5$ fm the typical value of the formation time is  $t_f=8$ fm. Note that $t_f$ here is larger than the length of the medium. Therefore, we can expect medium induced collinear radiation to suffer LPM suppression, reducing the impact of the medium on the observable in this regime. For jets of radius $R \sim 1$, the energy loss due to medium induced radiation that moves out of the jet is suppressed by $Q_{\text{med}}^2/Q^2$ and hence we ignore it at leading power \footnote{This will change for jets of smaller radius when energy loss effects will need to be accounted for.}.

In order to facilitate factorization, we work in a frame where the energy of the medium partons is $Q \sqrt{\chi}$. This requires us to work in a frame where the medium is boosted in the $\bar n$ direction by a factor $Q\sqrt{\chi}/T$. Under this boost, the collinear mode therefore scales as 
\bea 
p_c \sim \frac{T}{\sqrt{\chi}}\left(1,\delta^2,\delta \right),
\eea
where $ \delta = \frac{ Q\chi}{T} \sim  \frac{Q_{\text{med}}\sqrt{\chi}}{T}$ is the angle of deflection for jet partons after interacting with the medium. Note that for the hard scales of the order of a few hundreds of GeV, we get $\delta \ll 1 $. Therefore, $\delta$ is another expansion parameter of our EFT. We remind the readers that the first expansion parameter is $\sqrt{\chi}$.
Now, in this frame, the modes in the medium are soft and scales as 
\bea
\label{eq:soft}
p_s \sim  \frac{T}{\sqrt{\chi}}\left( \delta, \delta, \delta \right).
\eea
The interaction between the medium and jet partons is primarily governed by $t$-channel forward scattering. This interaction is mediated by Glauber gluons which are instanteneous exchanges of space-like gluons. In terms of the above two expansion parameters, Glauber modes scales as
\bea
p_G \sim  \frac{T}{\sqrt{\chi}}\left( \delta, \delta^2, \delta \right). 
\eea
Note that the Glauber exchange maintains the scaling of both the collinear and soft modes while exchanging transverse momentum of $\mathcal{O}(Q_{\text{med}})$.  

\begin{figure}[t]
\centering
\includegraphics[scale=0.7]{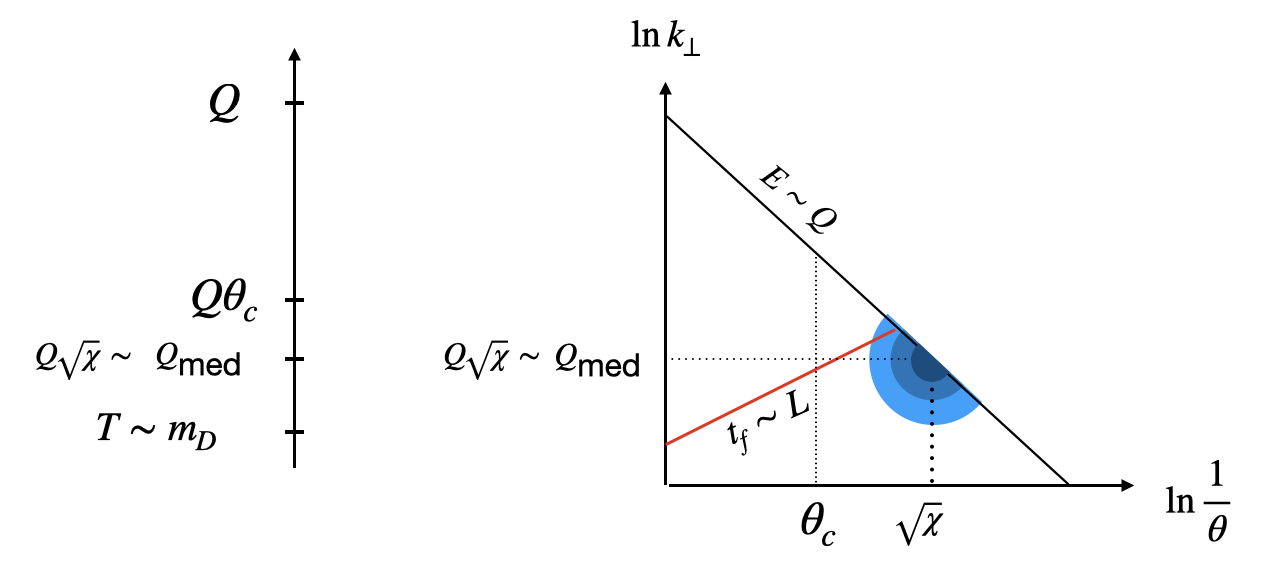}
\caption{{\bf Region I EFT}:The hierarchy of scales and the Lund plane for the region I EFT. The blue shaded region is the phase space of collinear radiation that contributes at leading power to the EEC in this regime. ~\label{fig:lund1}}
\end{figure}
To explicitly visualise the scales and the corresponding regions of the phase space, we can put all the above mentioned  information including the medium length and its temperature together in the form of a Lund plane shown in Fig. \ref{fig:lund1}. Here $k_{\perp}$ is transverse momentum of radiated gluon and $\theta$ is its angle from the jet direction. The shaded blue region indicates the collinear phase space emissions that dominate the measurement. As mentioned above, we see that these emissions have formation time larger than or equal to the medium length. Therefore, the radiation in this phase space cannot be resolved by the medium as a result of which the jet will act as a single coherent source of medium induced radiation.

\subsection{Factorization}

To simplify the computation for the factorization of EEC, we work with simple $e^{+}e^{-}$ initial state. We stress that this factorization can be extended to pp or PbPb collisions simply by incorporating appropriate initial state parton/nuclear distribution functions and correspondingly modifying the hard function, responsible for production of jet initiating parton.

We start by defining the factorized density matrix at time $t=0$ which is simply an $e^{+}$ $e^{-}$ state that will create the jet and a background QGP medium as
\begin{equation}
\rho(0)=|e^{+}e^{-}\rangle \langle e^+ e^- | \otimes \rho_M(0),
\end{equation}
where $\rho_M(0)$ is the initial density matrix of the medium. With this form of the initial density matrix we therefore assume that the initial partons involved in the hard interaction are disentangled from the QGP medium. Moreover, the medium density matrix is populated by the soft modes defined by their scaling in Eq.\ref{eq:soft}. As the system evolves, the interaction between the jet and the medium constituents can be encapsulated in the time dependent density matrix. In terms of total Hamiltonian, the time evolution of the density matrix reads as
\begin{equation}
\rho(t)=e^{-iHt}\rho(0)e^{iHt},    
\end{equation}
where the Hamiltonian $H$ is given by
\begin{equation}
H= H_{\rm QCD} + C(Q)l^{\mu}j_{\mu} \equiv  H_{\rm QCD}+\mathcal{O}_H.
\end{equation}
Here $j_\mu=\bar{\chi} \gamma_{\mu}\chi $ denotes the quark current which creates the $q \bar q$ pair, while $l_{\mu}$ represents the initial state lepton current.  Imposing the measurement on the density matrix as $t \rightarrow \infty$, we define the quantity 
\bea
\Sigma = \lim_{t\rightarrow \infty}\tr[\rho(t) \mathcal{M}].
\eea
where $\mathcal{M}$  denotes the measurement imposed on the final state particles, which at present is the EEC. We will relate this quantity to the differential cross-section at the later stage.  Plugging the Hamiltonian back in the time dependent density matrix and inserting the single hard vertex at both sides in the above equation, we obtain
\begin{equation}
\Sigma =|C(Q)|^2L_{\mu\nu}\lim_{t\to \infty}\int d^4xd^4ye^{iq\cdot(x-y)} \tr[e^{-iH_{\rm QCD}t}J^{\mu}(x)\rho(0)\mathcal{M}J^{\nu}(y)e^{iH_{\rm QCD}t}]
\label{evolution}
\end{equation}
where $q^{\mu} = p_e^{\mu}+ p^{\mu}_{\bar e}$ is the sum of the momenta of $e^+ e^-$ initial state. In the center of mass frame $q^{\mu} = (2Q, 0, 0, 0)$. 
Now at this stage we can perform an Operator product expansion(OPE) as $ q \rightarrow \infty$ to match the above equation onto the collinear jet function~\cite{Kang:2016mcy} in SCET. Let us note that this step involves integrating out the final state with $total$ large invariant mass ($\gg Q\sqrt{\chi}$), leaving behind only modes of virtuality $Q\sqrt{\chi}$ and lower. Finally, we can relate Eq.\ref{evolution} to the cross differential cross section  normalized by the born level cross section $\sigma_0$ and obtain
\begin{align}
\frac{1}{\sigma_0}\frac{d\sigma}{d\chi}=&\sum_{i\in \{q, \bar q, g\}}\int dx x^2 H_{i}(x,Q,\mu)J_i(xQ, \chi,\mu).
\label{eq:densitymat}
\end{align}
The terms appearing in the above factorized expression are interpreted as follows. The hard function $H_i(x)$ describes the process of initial hard interaction that creates a parton of species $i$ and carrying a fraction $x$ of the initial hard momentum $Q$. Therefore, the parton with energy $xQ$ initiates the jet. Moreover, the jet function $J_i$ captures jet dynamics which is mainly radiation pattern inside the jet. The jet function for the quark now reads as
\begin{equation}
J_q(xQ,\chi,\mu)= \frac{1}{2N_c}\sum_{X}\tr\Big[\rho_M \frac{\slashed{\bar n}}{2}e^{i  \int dt H_{S}(t) } \chi_{n} (0)\delta^2(\mathcal{P}_{\perp})\delta(\omega -\bar{n}\cdot \mathcal{P})\mathcal{M}|X\rangle  \langle X| e^{-i  \int dt H_{S}(t) }\bar{\chi}_{n}(0)\Big],
\label{eq:Sigma}    
\end{equation}
where $\omega=2xQ$ and $\mathcal{M}$ is the differential measurement function in the collinear limit defined as 
\begin{equation}
\mathcal{M}|X\rangle=\sum_{i,j \in X}\frac{p^-_ip^-_j}{\omega^2} \delta\Big(\chi-\frac{\theta_{ij}^2}{4}\Big)|X\rangle.  
\label{eq:measurement}
\end{equation}
Here $p^-= \bar n \cdot p$ is the large light-cone momentum component of the collinear partons.
Finally, the effective Hamiltonian in the evolution operator in Eq.~\ref{eq:Sigma} now is given by the Hamiltonian in Soft Collinear Effective Theory(SCET) 
\bea 
\label{eq:Heff}
H_S = H_n+ H_s+ H_G,
\eea
where $H_n$ is the standard SCET collinear Hamiltonian \cite{Bauer:2000yr,Bauer:2001ct,Bauer:2002nz}. $H_s$ is the full QCD Hamiltonian which controls the dynamics of the soft modes in the medium. 
Moreover, $H_G$ is the Glauber Hamiltonian \cite{Rothstein:2016bsq} that mediates forward scattering between the soft and collinear modes and which is given as 
\bea 
H_G = 8\pi \alpha_s\int d^3 x\sum_{i,j \in \{q,\bar q,g\}} \mathcal{O}_{ns}^{ij}(x).
\eea
Here $\mathcal{O}_{ns}^{ij}$ is Glauber interaction operator which  is written as a contact operator between collinear and soft currents as 
\begin{equation}
\mathcal{O}_{ns}^{qg}=O_n^{qA}\frac{1}{\mathcal{P}_{\perp}^2}O_s^{gA}, \,\,\, \mathcal{O}_{ns}^{qq}=O_n^{qA}\frac{1}{\mathcal{P}_{\perp}^2}O_s^{qA},    
\end{equation}
where $A$ stands for color indices and $\mathcal{P}_{\perp}$ operates on soft current to pull out Glauber gluon exchange momentum. Finally, the soft and collinear operators for quark and gluon are given as
\begin{align}
\mathcal{O}_{n}^{qA} &= \bar{\chi}_{ n}T^A\frac{\slashed{\bar{n}}}{2}\chi_{n}, 
 \  \   \    
\mathcal{O}_{s}^{qA}= \bar{\chi}_{s}T^A\frac{\slashed{n}}{2}\chi_{s}, 
 \   \   \    
\mathcal{O}_{s}^{gA}=  \frac{i}{2}f^{ACD}\mathcal{B}_{S \perp\mu}^C\frac{n}{2}\cdot(\mathcal{P}+\mathcal{P}^{\dagger})\mathcal{B}_{S \perp}^{D\mu}, 
\label{eq:SandC}
\end{align}
where $\chi_n/\chi_s$ are collinear/soft fermion fields and $\mathcal{B}_{S\perp \mu}$ is soft gluon field strength, which are built out of gauge invariant blocks in SCET defined as
\begin{align}
 & \chi_{n} = W_{n}^{\dagger}\xi_{n}, 
  \   \   \   \    
  W_{n} = \text{FT} \  {\bf P} \exp \Big\{ig\int_{-\infty}^0 {\rm d}s \, \bar{n}\cdot A_{n}(x+\bar{n}s)\Big\}  ,
  \nonumber\\
 & \chi_{S} = S_{n}^{\dagger}\xi_{S},  
   \   \   \   \    
   S_{n} = \text{FT} \ {\bf P} \exp \Big\{ig\int_{-\infty}^0 {\rm d}s\, n\cdot A_{S}(x+s n)\Big\}
   , \nonumber\\
  &\mathcal{B}_{n \perp}^{C\mu} T^C = \frac{1}{g}\Big[W_{n}^{\dagger}iD_{n \perp}^{\mu}W_{ n}\Big],    \  \  
 \mathcal{B}_{S\perp}^{C\mu} T^C = \frac{1}{g}\Big[S_{n}^{\dagger}iD_{S \perp}^{\mu} S_{n}\Big], 
 \label{eq:BBlock}
\end{align}
where FT stands for fourier transform and $W_n$ and $S_n$ are collinear and soft Wilson lines, respectively. Note that the collinear/soft quark and gluon fields are dressed with the Wilson lines making the operators gauge invariant.

The presence of the Glauber term in Eq.\ref{eq:Heff} causes non-commutativity among various Hamiltonian involved in the evolution operator. Therefore, we rearrange the evolution terms in Eq.~\ref{eq:Sigma} to distinctly isolate the Glauber component and re-write the quark jet function as
\begin{align}
J_q(xQ, \chi, \mu)=&\frac{1}{2N_c}\sum_{X}\tr\Big[\rho_M\frac{\slashed{\bar n}}{2} e^{i H_{n}+H_{s}t}\bar {\mathcal{T}}\Big\{e^{-i\int_0^t dt' H_{G,I}(t')} \chi_{n,I}(0) \Big\}\mathcal{M}|X\rangle \nonumber\\
&\langle X|\mathcal{T}\Big\{e^{-i\int_0^{t} dt' H_{G,I}(t')}\delta^2(\mathcal{P}_{\perp})\delta(2xQ-\bar{n}\cdot \mathcal{P})\bar{\chi}_{n,I}(0) \Big\}e^{-iH_{n}+H_{s}t}  \Big],   
\label{eq:Jet}
\end{align}
where the operators are now defined as
\begin{equation}
\mathcal{O}_{I}=e^{i(H_n+H_s)t}\mathcal{O}e^{-i(H_n+H_s)t}.    
\end{equation}
Since Glauber modes mediate interactions between collinear and soft modes, they explicitly break factorization and  we can only perform soft-collinear factorization order by order in the Glauber Hamiltonian. In order to do that we expand out the jet function in the Glauber Hamiltonian as
\bea
J_q(xQ, \chi, \mu) = \sum_{i=0}^{\infty} J_q^{(i)}(xQ,\chi, \mu)
\label{eq:jetglauber}
\eea
where $i=0$ corresponds to vacuum jet function, $i=1$ correspond to single scattering with the medium, and $i\geq 2$ represents multiple scatterings scenario. 


\subsubsection{Factorization in vacuum}
To obtain the vacuum results, we expand the evolution operator containing Glauber interactions in  Eq.~\ref{eq:densitymat} at leading order, i.e., no insertions of the Glauber Hamiltonian. In this case, the soft and collinear sectors are manifestly decoupled. We can then factorize the Hilbert space as $|X\rangle = |X_n\rangle |X_s\rangle$ which enables us to factorize the soft physics from the collinear one. Since the measurement of the EEC is acting on the collinear modes, the contribution from the soft mode is power suppressed. Thus, the soft function is completely inclusive and does not depend on the measurement. By using $\text{Tr}[\rho_M]= 1$, the soft function therefore is just an identity. This enables us to express the jet function as
\begin{eqnarray}
J_{q}^{(0)}(xQ,\chi,\mu)=\frac{1}{2N_c}\sum_{X_n} \tr\Big[\frac{\slashed{\bar n}}{2} \langle 0|\chi_{n}(0)\mathcal{M}|X_n\rangle \langle X_n|\delta(2xQ-\bar{n}\cdot \mathcal{P})\delta^2(\mathcal{P}_{\perp})\bar{\chi}_{n}(0)|0\rangle \Big],  
\label{eq:vacjet}
\end{eqnarray}
where as before $\mathcal{M}$ is measurement function defined in Eq.~\ref{eq:Jet}. Note that the above jet function is identical to the vacuum jet function obtained in Ref.\cite{Dixon:2019uzg}. Moreover, we explicitly verify the one loop result and the anomalous dimension in Section \ref{sec:one loop}.


\subsubsection{Factorization for single medium interaction}
For medium induced jet function we need to consider NLO terms in the evolution operator expansion. A single interaction with the medium at the amplitude squared level will involve making two insertions of the Glauber Hamiltonian, either on the same or opposite side of the cut. We follow the procedure outlined in \cite{Vaidya:2020lih} and factorize the soft physics from the collinear one. For a homogeneous medium of length $L$, this yields a medium induced jet function that can be written as 
\begin{align}
J_q^{(1)}(xQ,\chi;L)=L\int \frac{d^2k_{\perp}}{(2\pi)^2} {J}_{q,1}(xQ,\chi,k_{\perp},\mu,\nu;L)\otimes \mathcal{B}(k_{\perp},\mu,\nu),
\label{eq:FactJOne}
\end{align}
where $k_{\perp}$ is the Glauber momentum exchanged between the jet and the medium, $\mathcal{J}_{1}$ is the single interaction medium induced jet function, and $\mathcal{B}$ is medium function which can be obtained from the spectral function. We will discuss this the medium function in the later sections. Note that each function in Eq.~\ref{eq:FactJOne} depends on two renormalization scales, $\mu$, which is the virtuality scale and $\nu$, a rapidity scale \cite{Chiu:2011qc}. The medium function $\mathcal{B}$ does not depend on the observable and is therefore a universal function of the medium properties. Now the jet function $J(xQ,\chi,k_{\perp},\mu,\nu;L)$ gets contributions from both same and opposite side Glauber insertion. Therefore, the total jet function is given as 
\begin{equation}
J_{q,1}(xQ,\chi,k_{\perp};L)=J_{R}(xQ,\chi,k_{\perp};L)-J_{V}(xQ,\chi,k_{\perp};L), 
\label{eq:totaljet}   
\end{equation}
where $J_{R}$ represents jet function for Glauber insertions on the opposite side of the cut. Note that in the above equation we have suppressed $\mu$ and $\nu$ dependence to keep notations compact.  Similarly, $J_V$ denotes jet function for Glauber insertions on the same side of the cut. From here onwards we will call $J_R$ as real and $J_V$ as virtual jet function.  The real jet function is defined as as
\begin{align}
\label{eq:Jo}
&J_{R}(xQ,\chi,k_{\perp};L)=\frac{1}{2N_c}\frac{e^{-i\frac{L}{2}(\mathcal{P}^{A}_{+}-\mathcal{P}^{B}_{+})}}{k_{\perp}^2}\text{sinc}\Big[\frac{L}{2}(\mathcal{P}^{A}_{+}-\mathcal{P}^{B}_{+}) \Big]\nonumber\\ 
& \sum_{X}\tr\Big[\frac{\slashed{\bar{n}}}{2}\langle 0|\bar{\mathcal{T}}\Big\{e^{-i\int dt H_{n}(t)}\Big[\delta(\mathcal{P}^-)\delta^2(\mathcal{P}_{\perp}-k_{\perp})O_{n}^{qB}(0)\Big]\chi_{n}(0) \Big\} \mathcal{M}|X\rangle \nonumber\\
&\langle X|\mathcal{T}\Big\{e^{-i\int dt H_{n}(t)} \Big[\delta(\mathcal{P}^-)\delta^2(\mathcal{P}_{\perp}+k_{\perp})O_{n}^{qB}(0)\Big]\Big[\delta(\omega-\bar{n}\cdot\mathcal{P})\delta^2(\mathcal{P}^{\perp})\bar{\chi}_{n}(0)\Big]\Big\}|0\rangle \Big]\delta^{AB}, 
\end{align}
where $\omega=2xQ$, $A$ and $B$ are color indices, and the delta function denotes energy and transverse momentum conservation. Note that the transverse momentum delta functions acts on the operators within the same bracket. Moreover, $\mathcal{P}_A^+$, $\mathcal{P}_B^+$ are the operators that extract the $+$ component of the momenta of the collinear operators $O_n^A$ and $O_n^B$ respectively. We see that only the $-$ and $\perp$ components of momenta are conserved for Glauber gluon exchange while the $+$ component leads to a phase factor which is ultimately responsible for the LPM effect.  
Likewise, expanding the Glauber Hamiltonian up to second order at same side of the cut, $J_V$ takes the form 
\begin{align}
\label{eq:Js}
&J_{V}(xQ,\chi,k_{\perp};L)=\frac{1}{2N_c}\frac{1}{2}\frac{e^{-i\frac{L}{2}(\mathcal{P}^{A}_{+}+\mathcal{P}^{B}_{+})}}{k_{\perp}^2}\text{sinc}\Big[\frac{L}{2}(\mathcal{P}^{A}_{+}+\mathcal{P}^{B}_{+}) \Big] \nonumber \\
&\sum_{X} \tr\Big[\frac{\slashed{\bar{n}}}{2}\langle 0|\mathcal{\bar T}\Big\{e^{-i\int dt H_{n,I}(t)}\chi_{n}(0)\Big\}\mathcal{M}|X\rangle \langle X|\mathcal{T}\Big\{e^{-i\int dt H_{n}(t)}\Big[\delta^2(\vec{\mathcal{P}}_{\perp}+ \vec{k}_{\perp})\delta(\mathcal{P}^-)O_{n}^{A}(0)\Big]\nonumber \\ & \times \Big[\delta^2(\vec{\mathcal{P}}_{\perp}-\vec{k}_{\perp})\delta(\mathcal{P}^-)O_{n}^{B}(0)\Big]\Big[\delta^2(\mathcal{P}^{\perp})\delta(\omega-\bar{n}\cdot\mathcal{P})\bar{\chi}_{n}(0)\Big]\Big\}|0 \rangle\Big]\delta^{AB}+\, c.c, 
\end{align}
where again the delta functions act on the operators within the bracket. Finally, in terms of soft operators the medium correlator is defined as 
\bea
\mathcal{B}(k_{\perp})= \frac{1}{k_{\perp}^2}\frac{1}{N_c^2-1}\int \frac{dk^-}{2\pi}\int d^4r e^{-i k_{\perp} \cdot r^{\perp}+ik^-r^+}\text{Tr}\Big[e^{-i\int dt H_s(t) }O^A_{s}(r) \rho_M O^A_{s}(0)e^{i\int dt H_s(t) }\Big].\,\qquad
\label{eq:B}
\eea
 The operator $O_s^{A}$ acts as a local gauge invariant color source for Glauber gluons in the medium. We therefore see that the jet probes the correlation of color sources in the medium. This is reminiscent of the color density function $\rho^A$ that sources the background glauber field in the Color Glass Condensate(CGC) \cite{McLerran:1998nk}. We have now completely separated out the physics of the medium from the process dependent dynamics of the jet. 


\subsubsection{Factorization for arbitrary number of interactions}

We can continue the expansion to arbitrary orders in Glauber Hamiltonian in Eq.~\ref{eq:jetglauber} to  derive factorization formulas at each order. In order to do factorization, we will assume that the successive interaction of the jet and the medium happen with color uncorrelated partons in the medium. This is equivalent to the assumption that the mean free path of the jet partons is much longer than the color screening length $1/m_D$ in the medium. In this case we can factor out medium function and write it in terms of the same $\mathcal{B}(k_{\perp})$. In particular, expanding the Glauber Hamiltonian up to $\mathcal{O}(H_G^{2m})$ we get
\bea
J_q^{(m)}(xQ,\chi,\mu) = \frac{L^m}{m!} \Big[\prod_{i=1}^{m} \int \frac{d^2k_{i\perp}}{(2\pi)^3}\mathcal{B}(k_{i\perp},\mu,\nu)\Big]{J}_{q,m}(xQ,\chi,k_{1\perp}, . .. k_{m\perp},L,\mu,\nu) 
\label{eq:multiplescat}
\eea
which has $m$ copies of the medium function $\mathcal{B}$ defined in Eq.\ref{eq:B} with a single distinct medium induced jet function ${J}_m$ at each order. The full operator definition for ${J}_m$ is straightforward to extrapolate from ${J}_1$ but is cumbersome to write out in full. Likewise, it has more complicated dependence on $L$ that captures the LPM interference effects between $m$ interactions with the medium. 
The final form for the factorization in this region of the EFT takes the compact form 
\small
\bea
\label{eq:FactI}
\frac{1}{\sigma^0}\frac{d\sigma}{d\chi}=\!\! \int dx x^2 H_i(x,Q,\mu) \Big( J_i^{(0)}(xQ,\chi,\mu)\! +\! \sum_{m=1}^{\infty}\! \frac{L^m}{m!}\Big[\prod_{j=1}^{m} \int \frac{d^2k_{j\perp}}{(2\pi)^3}\mathcal{B}(k_{j\perp})\Big]{J}_{i,m}(xQ,\chi,k_{1\perp}, .. k_{m\perp};L) \Big)\nonumber 
\eea
\normalsize
where again we have suppressed scale dependence in the second term. All the universal medium physics is encoded in a single object $\mathcal{B}$ while all the jet dynamics including the measurement dependence is fully separated out into the jet function. We can see how this could be easily extended to other jet observables; the $\mathcal{B}$ functions would remain the same, so will the operators that define the jet function for a given parton species; only the measurement $\mathcal{M}$  inside the jet function would change. Therefore, the above factorized equation is very powerful since it has universal elements that would appear in any other jet observable. The fact that we fix the operator definitions of all the objects for \textit{all} observables means that the anomalous dimensions of all these functions in distinct observables are also universal and we will explicitly compute those in the next section. 
We note that in order to describe the observable in the regime of $Q\sqrt{\chi} \sim Q_{\text{med}}$ in a dense medium requires us to evaluate the jet function ${J}_{q,m}$ for arbitrary number of interactions, which is beyond the scope of this paper. As noted earlier, the LPM effect which is important in this regime tends to suppress the medium contribution and hence in this regime, it is expected that the vacuum result will dominate.

\subsection{One loop results}

\label{sec:one loop}
In this section we give the one loop result for all the functions described in the previous section that appear in the factorization formula Eq.\ref{eq:FactI}. Let us first start with vacuum jet function.
The quark jet function in vacuum in terms of gauge invariant SCET operator is given as Eq.\ref{eq:vacjet}.
\begin{eqnarray}
J_{q}^{(0)}(xQ,\chi,\mu)=\frac{1}{2N_c}\sum_{X_n} \tr\Big[\frac{\slashed{\bar n}}{2} \langle 0|\chi_{n}(0)\mathcal{M}|X_n\rangle \langle X_n|\delta(\omega-\bar{n}\cdot \mathcal{P})\delta^2(\mathcal{P}_{\perp})\bar{\chi}_{n}(0)|0\rangle \Big],  
\label{eq:qjet}
\end{eqnarray}
with the measurement function same as in Eq.~\ref{eq:measurement}
\begin{equation}
\mathcal{M}=\sum_{i,j \in X_n}\frac{p^-_i p^-_j}{(2xQ)^2}\delta\Big(\chi-\frac{\theta_{ij}^2}{4}\Big), 
\end{equation}
where $p^-_i,\, p^-_j$ represents the large light-cone component of the final state partons and  $\theta_{ij}$ is their angular separation. Note that we have already taken the collinear limit of the observable. Likewise, the summation over  $i, j$ also includes the case when EEC detectors are placed on the same particle, i.e., $i=j$. Similarly, for gluon jet, the jet function is given as
\begin{equation}
\!J_g^{(0)}(xQ,\chi,\mu)\!=\!\frac{\omega}{2(N_c^2-1)}\tr\left[\langle 0|\mathcal{B}_{n\perp\mu}(0)\mathcal{M} |X\rangle \langle X| \delta(\omega-\bar{n}\cdot \mathcal{P})\delta^2(\mathcal{P}_{\perp})\mathcal{B}_{n\perp}^{\mu}(0)|0\rangle \right]. 
\label{eq:gjet}
\end{equation}
Now we start computing the quark jet function $J_q^{0}$. At leading order, the energy weights in Eq.~\ref{eq:qjet} are just one and the jet function turns out to be similar to the standard semi-inclusive jet function. Therefore, we obtain
\begin{equation}
J_{q,\text{LO}}^{(0)}=\delta(\chi).
\end{equation}
\begin{figure}[h!]
\centering
\includegraphics[scale=0.7]{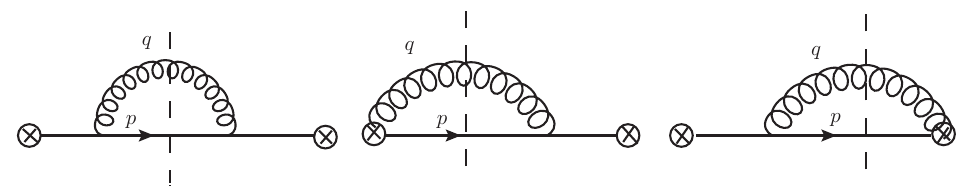}
\caption{Feynman diagrams contributing to vacuum jet function for quark.}
\label{fig:vacNLO}
\end{figure}
Now let us turn to the detailed calculation of next-to-leading order (NLO) quark jet function. While the result has already been computed in literature \cite{Dixon:2019uzg}, we present it here for completeness. The Feynman diagrams contributing to NLO quark jet function are shown in Figure~\ref{fig:vacNLO} where the incoming quark of light-cone momentum $\omega$ splits into a gluon of momenta $q^{\mu}=(q^{-},q^{+},q_{\perp})$ and a quark with momenta $p^{\mu}=(p^{-},p^{+},p_{\perp})$. The virtual contributions lead to a scaleless integral which in dimensional regularization vanishes. Thus, adding all the real contributions we get
\begin{align}
J_{q,\text{NLO}}^{(0)}(\omega,\chi,\mu)&=\frac{2g^2\delta_{AB}}{2N_c}\int \pphase\int\qphase\frac{1}{q_{\perp}^2\omega}\bigg(\frac{2p^-\omega+(q^-)^2}{q^-} \bigg)\nonumber\\
&\left[((q^-)^2+(p^-)^2)\delta(\chi)+2q^{-}p^{-}\delta\Big(\chi-\frac{\theta_{ij}^2}{4}\Big)\right]\delta^2(\vec{q}_{\perp}+\vec{p}_{\perp})\delta(\omega-q^--p^-).
\label{eq:jetmeas}
\end{align}
Using the on-shell and transverse momentum conserving delta functions, one can perform the trivial momentum integration. Therefore, Eq.~\ref{eq:jetmeas} acquires the form
\begin{align}
J_{q,\text{NLO}}^{(0)}(\omega,\chi, \mu)=&\frac{2\alpha_s C_F}{\pi}\frac{(\mu^2e^{\gamma_E})^{\epsilon}}{\Gamma[1-\epsilon]}\int_0^1 dz \hat{P}_{gq}(z) \int \frac{d^2q_{\perp}}{q_{\perp}^{2+2\epsilon}}\nonumber\\ 
&\left[z^2\delta(\chi)+(1-z)^2\delta(\chi)+2z^3(1-z)^3\omega^2\delta\left(q_{\perp}^2-[z(1-z)\omega]^2\chi\right) \right]\nonumber \\
&=\frac{2\alpha_s C_F}{\pi\Gamma[1-\epsilon]}\int_0^1 dz\left(\frac{\mu^2e^{\gamma_E}}{\omega^2 z(1-z)}\right)^{\epsilon}\hat{P}_{gq}(z)z(1-z)\frac{1}{\chi^{1+\epsilon}}
\label{eq:nlojet}
\end{align}
where $d=2-2\epsilon$ and $z=q^-/\omega$ is energy fraction of gluon in the final state. $\hat{P}_{gq}(z)$ is given as
\begin{equation}
\hat{P}_{gq}(z)=\frac{1+(1-z)^2}{z}.    
\end{equation}
 In cumulant space this yields the anomalous dimension for the jet function 
\bea
\frac{d}{ d\ln \mu^2}J^{(0)}_q(\omega,\chi_c,\mu) = \frac{\alpha_sC_F}{\pi}\int dz z^2 \left( P_{qq}(z)J_q^{(0)}(z\omega,\chi_c,\mu)+P_{gq}(z)J_g^{(0)}(z\omega,\chi_c,\mu)\right) 
\eea
inducing a mixing into the gluon jet function
where $P_{ij}$ are the regularized vacuum splitting functions which are given as 
\bea 
& P_{qq} = \frac{1+z^2}{(1-z)_+} +\delta(1-z)\frac{3}{2}\nonumber\\
& P_{gq} =  \frac{1+(1-z)^2}{z}.
\eea

Finally plugging back the above splitting function in Eq.~\ref{eq:nlojet} and  performing $z$ integration, the complete NLO jet function can be written as
\begin{align}
J^{(0)}_{q,\text{NLO}}(\omega,\chi)&=\delta(\chi)+\frac{\alpha_s C_F}{\pi}\left(-\frac{3}{2\epsilon}\delta(\chi)+\frac{3}{2}\left[\frac{1}{\chi}\right]_{+}-\frac{3}{2}\delta(\chi)\ln\left(\frac{\mu^2}{\omega^2}\right)-\frac{19}{3}\delta(\chi)+\mathcal{O}(\epsilon) \right).   
\end{align}
Combining the plus distribution term and the delta function, we can further write
\begin{equation}
\frac{3}{2}\left[\frac{1}{\chi}\right]_{+}-\frac{3}{2}\delta(\chi)\ln\left(\frac{\mu^2}{\omega^2}\right)=\frac{3}{2}\left[\frac{\mu^2}{\omega^2\chi} \right]_{+},    
\end{equation}
which provides jet scale $\omega\sqrt{\chi}$ for the renormalization group (RG) evaluation. Likewise computing the gluon jet function, we arrive at the full RG evolution of the jet function 
\bea
\frac{d}{ d\ln \mu^2}J^{(0)}_i(\omega,\chi_c,\mu) = \frac{\alpha_sC_F}{\pi}\int dz z^2  J_j^{(0)}(z\omega,\chi_c,\mu)P_{ji}(z). 
\label{eq:vacJAD}
\eea

The one loop result for the hard function $H_i$ (Eq.\ref{eq:FactI}) can be found in \cite{Dixon:2019uzg}. The anomalous dimension for the hard function can be inferred from that of the jet function by demanding RG scale invariance of the cross section. 
\bea
\frac{d}{ d\ln \mu^2}H_i(x\omega,\mu) = -\frac{\alpha_sC_F}{\pi}\int_x^1 \frac{dz}{z}   P_{ij}(z)H_j(x/z\omega,\mu). 
\eea

\subsubsection{Single medium interaction}

Here we give the one loop results and anomalous dimensions for the medium induced jet function and also evaluate the medium correlator resulting from a single interaction with the medium.

\subsection{The medium correlator} 
\begin{figure}[h!]
\centering
\includegraphics[scale=0.61]{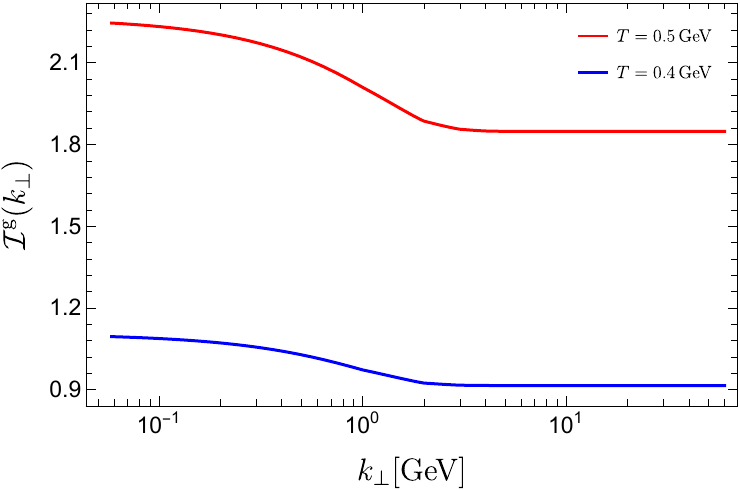}
\includegraphics[scale=0.61]{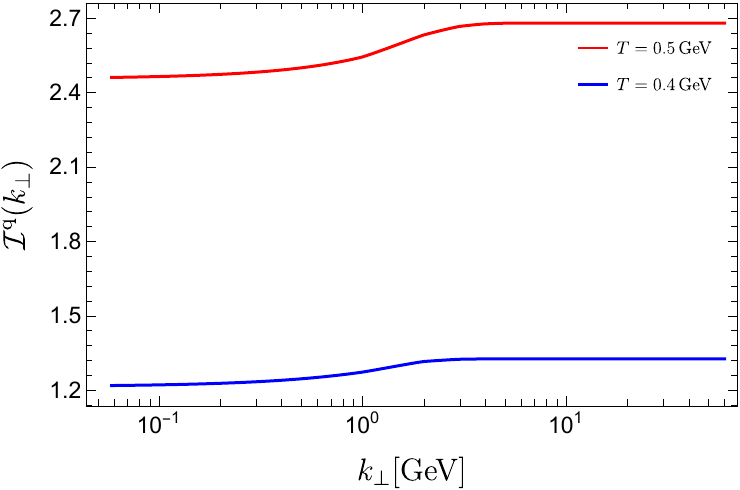}
\caption{The LO medium correlator for Quark and Gluon degrees of freedom.}
\label{fig:medfn}
\end{figure}

The function $\mathcal{B}$ defined in Eq.\ref{eq:B} is identical to the one defined in \cite{Vaidya:2020cyi}. An explicit computation of this function in a quark thermal bath was performed in \cite{Vaidya:2021vxu} along with the one loop radiative corrections. Here we also show the explicit computation of this function at tree level for thermal gluons in Appendix~\ref{sec:med}.
In terms of Wightman correlator for thermal quarks and gluons tree level medium correlator can be written as 
\bea
\mathcal{B}_{\text{LO}}(k_{\perp}) = D_{>}^g(k)+D_{>}^q(k)
\label{eq:bfunction}
\eea
where $D_{>}(k)=(1+f(k_0))\rho(k)$ is a Wightman correlator in the thermal medium and $\rho(k)$ is spectral function which gets contributions both from quark and gluon soft operators evaluated in Appendix~\ref{sec:med}. Evaluating the spectral function in the imaginary time formalism and plugging it back in Eq.~\ref{eq:bfunction} we get
\begin{equation}
\mathcal{B}_{\text{LO}}(k_{\perp})=(8\pi\alpha_s)^2\left(\frac{2\pi N_c^2}{16k_{\perp}^2}\mathcal{I}^g(k_{\perp})+\frac{2\pi N_f}{k_{\perp}^2} \mathcal{I}^q(k_{\perp}) \right)   
\end{equation}
where the first term arises from gluon contributions and the second one from quarks in the medium. For gluons, the function $\mathcal{I}^g(k_{\perp})$ is
\begin{align}
\mathcal{I}^g(k_{\perp})=&\frac{1}{2\pi}\int\frac{dq^-d^2q_{\perp}}{(2\pi)^3}\frac{q_{\perp}^2}{(q^{-})^2}f\left(\frac{q^-}{2}+\frac{q_{\perp}^2}{2q^{-}} \right)\bigg[1+f\left(\frac{k^-+q^-}{2}+\frac{q_{\perp}^2}{2q^-} \right)\bigg].    
\end{align}
Here $f$ is Bose-Einstein distribution function. Note that the distribution functions lead to Bose-Einstein enhancement which is expected for the case of gluons.  The $k^{-}$ component of Glauber momentum reads as
\begin{equation}
k^-=-q^-+\frac{q^-(\vec{k}_{\perp}+\vec{q}_{\perp})^2}{q_{\perp}^2}.    
\end{equation}
Similarly, for quark operators in the thermal medium the function $\mathcal{I}^q$ acquires the form 
\begin{align}
\mathcal{I}^q(k_{\perp})&=\frac{1}{2\pi}\int\frac{dq^-d^2q_{\perp}}{(2\pi)^3}\frac{q_{\perp}^2}{(q^-)^2}\tilde{f}\left(\frac{q^-}{2}+\frac{q_{\perp}^2}{2q^{-}} \right)\bigg[1-\tilde{f}\left(\frac{k^-+q^-}{2}+\frac{q_{\perp}^2}{2q^-} \right)\bigg]   
\end{align}
where $\Tilde{f}$ is Fermi-Dirac distribution function. Note that the distribution functions in the above equation lead to Pauli blocking which is expected for the case of fermions. In Figure~\ref{fig:medfn}, we show the variation of $\mathcal{I}^g(k_{\perp})$ and $\mathcal{I}^g(k_{\perp})$ as function of Glauber momentum $k_{\perp}$ for $T=0.4$ GeV and $\alpha_s=0.3$. Note that both the function shave somewhat weaker dependence on the Glauber momentum.

It was shown in \cite{Rothstein:2016bsq,Vaidya:2021vxu} that the medium function obeys two renormalization group equations, one in  virtuality $\mu$ and one in rapidity $\nu$. 
\begin{align}
\label{eq:BRG}
&\frac{d  \mathcal{B}(k_{\perp};\nu;\mu)}{d\ln \nu} 
= \frac{\alpha_s N_c}{\pi^2} \int d^2u_{\perp} \Bigg[\frac{\mathcal{B}(u_{\perp};\nu;\mu)}{(\vec{u}_{\perp}-\vec{k}_{\perp})^2} -\frac{k_{\perp}^2\mathcal{ B}(k_{\perp};\nu;\mu) }{2u_{\perp}^2(\vec{u}_{\perp}-\vec{k}_{\perp})^2}\Bigg]\nonumber\\
&\frac{d  \mathcal{B}(k_{\perp};\nu;\mu)}{d\ln \mu} = -\frac{\alpha_s \beta_0}{\pi}\mathcal{B}(k_{\perp};\nu;\mu).
\end{align}
The RG equation in the rapidity scale is just the Balitsky–Fadin– Kuraev–Lipatov (BFKL) \cite{Kuraev:1977fs,Balitsky:1978ic}  equation.
Since the $\mathcal{B}$ function only depends on the scale $k_{\perp}$, the natural scales for minimizing the logarithms in both $\mu$ and $\nu$ for this function is $k_{\perp} \sim Q_{\text{med}}$.

\subsection{The medium induced jet function}
Now we compute medium modified quark jet function ${J}_{q,1}$ for EECs. As mentioned earlier, we get real and virtual terms for Glauber insertions defined in Eq.~\ref{eq:totaljet}. The corresponding real and virtual NLO Feynman diagram for $J_R$ and $J_V$ are evaluated in Appendix~\ref{sec:feynjetmed} in more detail. Here we will combine all the contributing terms and give results for total one loop jet function.

\begin{figure}[t]
\centering
\includegraphics[scale=0.3]{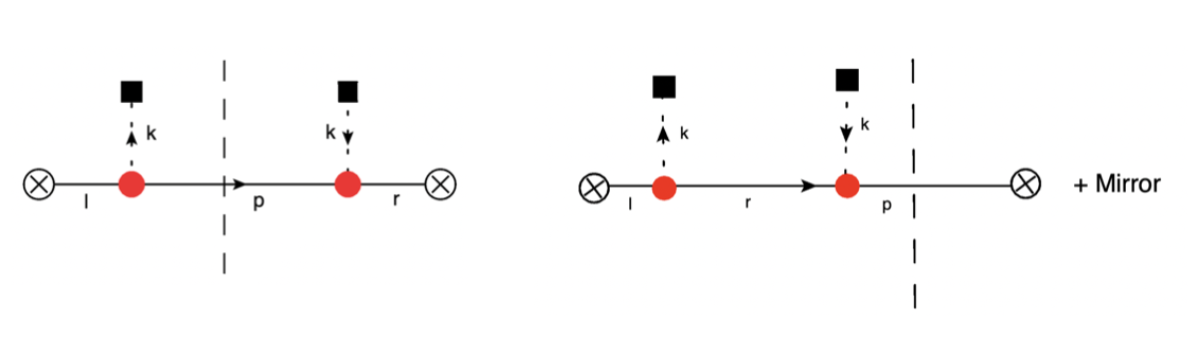}
\caption{LO contribution to the medium modified jet function.} 
\label{fig:tree}
\end{figure}

At leading order, we only have a single high energy quark propagating through the medium which interact with medium partons through soft elastic scatterings.  The corresponding diagrams are shown in Figure.~\ref{fig:tree}.  Without loss of generality, we can assume that the initial transverse momentum of this quark is zero.  
Since we have only one parton the energy weight is one and the measurement contribution is a delta function. Therefore, the LO jet function for Figure~\ref{fig:tree}(a) reads as       
\begin{align}
J_{R,\text{LO}}(\omega,\chi)=&-\delta_{AB}\int \frac{d^4p}{(2\pi)^4}\delta(p^2)\delta(p^{-}-\omega)\delta^2(\vec{p}_{\perp}-\vec{k}_{\perp})\int \frac{d l^{+}}{2\pi}\int \frac{d r^{+}}{2\pi}\frac{\bar{n}\cdot l}{l^2+i\epsilon} \frac{\bar{n}\cdot r}{r^2-i\epsilon}\bar{n}\cdot p\nonumber\\
& e^{-i(\frac{L}{2}(l^{+}-r^{+}))} \sinc\bigg[\frac{L}{2}(l_{+}-r_{+}) \bigg] \delta(\chi). 
\label{eq:lojet1}
\end{align}
 Note that with the Glauber exchange the plus component of light-cone momentum is not conserved which reflects in form of \sinc\, function. With the simple contour integrations over $r^{+}$ and $l^{+}$ and remaining integrations using delta functions, we obtain
\begin{eqnarray}
 J_{R,\text{LO}}&=&-4\delta_{AB}\delta(\chi).
\end{eqnarray}
 Following the same steps as in Eq.~\ref{eq:lojet1}, for Figure~\ref{fig:tree}(b) we get 
\begin{eqnarray}
J_{V,\text{LO}}&=&-4\delta_{AB}\delta(\chi)
\end{eqnarray}
As shown in Eq.~\ref{eq:totaljet}, the total contribution to the jet function is $J_{R}-J_{V}$ and therefore the LO modifications to the jet function vanishes. This just means that the medium has no effect on the observable at tree level.

At NLO we have an energetic quark passing through the medium that splits into a quark and gluon. Similar to the case of LO, we need to consider two cases, i.e., Glauber insertions on the same and opposite side of the cut. However, as mentioned earlier for each case we will have both real and virtual gluon emission contributions as well. A complete list of non zero Feynman diagrams and corresponding matrix element is shown in Appendix.~\ref{sec:realop}, \ref{sec:realsame}, \ref{sec:virtualop} and \ref{sec:virtualsame}. We have checked that the medium does not induce any new UV divergences. In order to compute the jet function we work with the differential cross-section and write the jet function as
\begin{align}
{J}_{1,\text{NLO}}(\chi,\omega,k_{\perp})&=\frac{g^2C_F}{(2\pi)^3k_{\perp}^2}\int \frac{dz}{z}\int d^2q_{\perp}\bigg[\Big(|\mathcal{M}|_{RR}^2\{\delta(\chi-\theta_1^2)-\delta(\chi)\}-|\mathcal{M}|_{VR}^2\{\delta(\chi-\theta_2^2)\nonumber\\
&-\delta(\chi)\} \Big) 2z(1-z)+(|\mathcal{M}|_{RR}^2+|\mathcal{M}|_{VR}^2-|\mathcal{M}|_{VR}^2-|\mathcal{M}|_{VV}^2)\delta(\chi)\bigg]
\label{eq:fulljet}
\end{align}
where $\theta_1^2=\frac{k_{\perp}^2 z^2-2z\vec{k}_{\perp}\cdot\vec{q}_{\perp}+q_{\perp}^2}{[z(1-z)\omega]^2}$ and $\theta_2^2=\frac{q_{\perp}^2}{[z(1-z)\omega]^2}$. Furthermore, $\mathcal{M}_{RR}/\mathcal{M}_{RV}$ are integrands for real/virtual diagrams for glauber insertions on the same side of the cut. $\mathcal{M}_{RR}$ can be obtained by adding all the diagrams discussed in Appendix~\ref{sec:realop}.  Similarly, $\mathcal{M}_{RV}$ can be obtained adding all the diagrams discussed in Appendix~\ref{sec:virtualop}. Moreover,  $\mathcal{M}_{VR}/\mathcal{M}_{VV}$ are integrands for real/virtual diagrams for glauber insertions on the same side of the cut which can be obtained by adding all the contributions given in Appendix~\ref{sec:realsame} and Appendix~\ref{sec:virtualsame}. In the second line of the above equation, due to the inclusive nature, the rapidity and UV divergences get cancelled between real and virtual terms for both same and opposite side Glauber insertion diagrams.  Finally, including all the color factors the fixed order medium modified EEC NLO jet function for a finite homogeneous medium of length $L$ acquires the form
\begin{align}
\label{eq:GLV}
{J}_{1,\text{NLO}}(\chi,\omega,k_{\perp})&=\frac{\bar{\alpha}C_F\omega^2}{2\pi^2k_{\perp}^2}\int dz z^3(1-z)^3\int d^2q_{\perp}\bigg[\bigg\{-\frac{2N_c(\vec{q}_{\perp}\cdot\vec{\kappa}_{\perp})}{q_{\perp}^2\kappa_{\perp}^2}f(z)\bigg(1-\frac{\sin \omega_1}{\omega_1}-\frac{\sin\omega_2}{\omega_2} \nonumber\\
&\hspace{-1cm}+\frac{\sin(\omega_2-\omega_1)}{\omega_2-\omega_1} \bigg)-\frac{4N_c(1-z)^2}{\kappa_{\perp}^2Q_{\perp}^2}\bigg(\frac{\vec{q}_{\perp}\cdot\vec{\kappa}_{\perp}}{z}+\frac{\kappa^2+\vec{q}_{\perp}\cdot\vec{\kappa}_{\perp}}{2(1-z)}+\frac{\vec{k}\cdot\vec{\kappa}_{\perp}}{2}+\frac{\kappa_{\perp}^2z}{2(1-z)^2} \bigg)\nonumber\\
&\hspace{-1cm}\left(1-\frac{\sin\omega_1}{\omega_1} \right)+\frac{4N_c f(z)}{\kappa_{\perp}^2}\left(1-\frac{\sin\omega_1}{\omega_1} \right)+\frac{4C_Fz}{q_{\perp}^2}\left(1-\frac{\sin\omega_1}{\omega_1} \right)+\frac{2}{3}\frac{z(1-z)^2}{q_{\perp}^2Q_{\perp}^2}\nonumber\\
&\hspace{-1cm}\left(\frac{\vec{q}_{\perp}\cdot\vec{\kappa}_{\perp}}{(1-z)^2}+\frac{\vec{k}_{\perp}\cdot\vec{\kappa}_{\perp}}{1-z} \right)\left(1-\frac{\sin\omega_1}{\omega_1} \right)-2C_F\frac{z(1-z)^2}{Q_{\perp}^2}\left(k_{\perp}^2+\frac{\kappa_{\perp}^2}{(1-z)^2}+\frac{\vec{k}_{\perp}\cdot\vec{\kappa}_{\perp}}{(1-z)} \right)\nonumber\\
&\hspace{-1cm}+\frac{4C_F(1-z)}{q_{\perp}^2z}\bigg(1-\frac{\sin\omega_1}{\omega_1}\bigg)-\frac{2}{3}\frac{(1-z)}{zQ_{\perp}^2}\frac{\sin\omega_1}{\omega_1}+\frac{2N_c(1-z)}{zq_{\perp}^2}\left(1-\frac{\sin\omega_1}{\omega_1} \right)\nonumber\\
&\hspace{-1cm}+\frac{2N_c(1-z)}{zQ_{\perp}^2} \bigg\}\bigg(\frac{\delta(q_{\perp}-q_{0})}{2|q_0-k_{\perp}z\cos\theta|}-\delta(\chi) \bigg)-\bigg\{\frac{4C_F(1-z)}{q_{\perp}^2z}-\frac{2N_c(\vec{q}_{\perp}\cdot\vec{\kappa}_{\perp})}{q_{\perp}^2\kappa_{\perp}^2}f(z)\nonumber\\
&\hspace{-2.5cm}\left(\frac{\sin(\omega_2-\omega_1)}{\omega_2-\omega_1}-\frac{\sin\omega_1}{\omega_1} \right)+\frac{2N_c}{q_{\perp}^2}f(z)\left(1-\frac{\sin\omega_1}{\omega_1} \right)\bigg\}\bigg(\delta(q_{\perp}^2-[z(1-z)\omega]^2\chi)-\delta(\chi)\bigg)\bigg]
\end{align}
where $\bar{\alpha}=\frac{g^2}{4\pi}$ is Glauber and jet coupling constant, $q_0=zk_{\perp}\cos\theta+\sqrt{\chi(z(1-z)\omega)^2-z^2k_{\perp}^2\sin^2\theta}$ and $Q_{\perp}^2=\omega(\kappa_{\perp}^2q^{-}+q_{\perp}^2p^{-})-k_{\perp}^2p^{-}q^{-}$. To keep the notations short we have also defined
\begin{equation}
\omega_1=\frac{L\kappa_{\perp}^2}{z(1-z)\omega}, \hspace{0.5cm}\,   \omega_2=\frac{Lq_{\perp}^2}{z(1-z)\omega} \hspace{0.2cm}\,\text{and}\, \hspace{0.2cm}\, f(z)=\frac{2+z^2-2z}{z}. 
\end{equation}
We have verified numerically that this is identical to the full GLV\cite{Ovanesyan:2011xy} result. 
Given the anomalous dimension of the medium correlator in Eq.\ref{eq:BRG}, by RG consistency in Eq.\ref{eq:FactJOne} we therefore require that the medium induced jet function also obey an RG equation in both $\mu $ and $\nu$ 
\begin{align}
\label{eq:JRG}
 &\frac{d{J}_{1,i}(xQ, \chi,k_{\perp};\nu;\mu)}{d\ln \nu} 
 = -\frac{\alpha_s N_c}{\pi^2} \int d^2u_{\perp} \Bigg[\frac{{J}_{1,i}(xQ, \chi,u_{\perp};\nu;\mu)}{(\vec{u}_{\perp}-\vec{k}_{\perp})^2} -\frac{k_{\perp}^2{J}_{1,i}(xQ, \chi,k_{\perp};\nu;\mu) }{2u_{\perp}^2(\vec{u}_{\perp}-\vec{k}_{\perp})^2}\Bigg]\nonumber\\
 &\frac{d \vec{{J}}_{1,i}(Q, \chi,k_{\perp};\nu;\mu)}{d\ln \mu^2} = \int_0^1 dy y^2 \vec{{J}}_{1,j}(y^2Q, \chi,k_{\perp};\nu;\mu) P_{ji}(y,\mu),\
\end{align}
The equation in $\mu $ is just the time-like DGLAP evolution equation. Since the leading order result for the jet function only appears at one loop, the UV and rapidity divergences will only be visible next-to-next-to-leading-order (NNLO). 
The NLO jet function therefore sets the initial conditions to solve the DGLAP and BFKL equation in rapidity. This demonstrates the power of the EFT framework which allows us to understand higher order effects by construction.


\subsection{One loop Jet function for arbitrary number of interactions}

The factorization for arbitrary number of interactions of the jet with the medium is given in Eq.\ref{eq:multiplescat}. The one loop computation for arbitrary number of interactions is a challenging one and is beyond the scope of this paper.   A simpler regime where such a computation is possible is when the radiation is soft; this is the BDMPS-Z limit. However, the soft limit is not valid in this regime of the EFT where $Q\sqrt{\chi} \sim Q_{\text{med}}$. We can , however take the soft limit when $Q\sqrt{\chi} \gg Q_{\text{med}}$. The factorization for this regime will be covered in Section \ref{sec:factwo}.
We can however, still make quantitative statements about the higher order radiative corrections simply by using RG consistency and show that we require that the medium induced jet function to obey the BKFL equation in all the $k_{i\perp}$ momenta as
\small
\bea
 \frac{d\mathcal{J}^{(m)}(xQ, \chi,k_{1\perp}, ...k_{m\perp};\nu;\mu)}{d\ln \nu} 
 &=& -\sum_{i=1}^{m}\frac{\alpha_s N_c}{\pi^2} \int d^2u_{\perp} \Bigg[\frac{\mathcal{J}^{(m)}(xQ, \chi,k_{1\perp}...k_{i-1\perp},u_{\perp}, k_{i+1\perp} .. k_{m\perp};\nu;\mu)}{(\vec{u}_{\perp}\vec{k}_{i\perp})^2}\nonumber\\
 & - &\frac{k_{\perp}^2\mathcal{J}^{(m)}(xQ, \chi,k_{1\perp},k_{2\perp}, ...k_{m\perp};\nu;\mu) }{2u_{\perp}^2(\vec{u}_{\perp}-\vec{k}_{\perp})^2}\Bigg]
\eea
\normalsize
The natural scale $\nu$ which will minimize the rapidity logs is $xQ$. The BFKL resummation between the jet and the medium function will therefore resum $\ln Q_{\text{med}}/(xQ) \sim \ln \sqrt{\chi}$. For small values of $\sqrt{\chi}$ where this EFT is valid, this resummation can be substantial.

Likewise, this functions also obeys the time-like DGLAP evolution equation in the scale $\mu$.
\bea
\frac{d}{d \ln \mu^2}\mathcal{J}_{i,m}(xQ, \chi, k_{1\perp},...,k_{m\perp})=  \frac{\alpha_sC_F}{\pi}\int dz z^2  \mathcal{J}_{j,m}(xQ, \chi, k_{1\perp},...,k_{m\perp})P_{ji}(z)
\eea
\begin{figure}[t]
\centering
\includegraphics[scale=0.6]{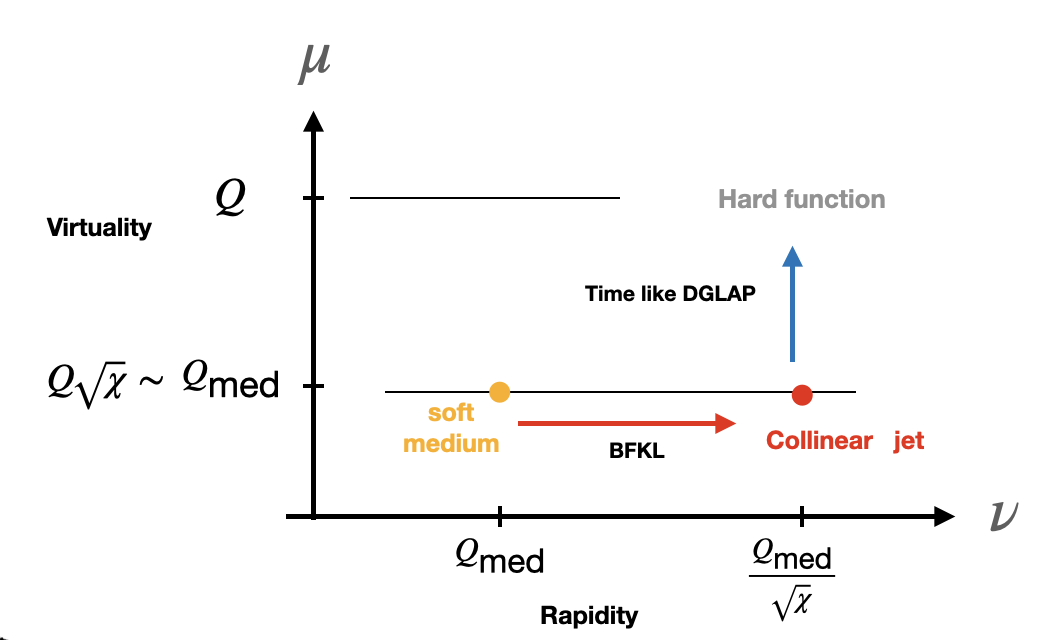}
\caption{Renormalization group flow for factorized functions in the regime $Q_{\text{med}} \sim Q\sqrt{\chi}$ } 
\label{fig:RGEOne}
\end{figure}
 We can then summarize the RG evolution between the various modes in Fig. \ref{fig:RGEOne}.

\section{Factorization for $Q \gg Q\sqrt{\chi} \gg  Q_{\text{med}}$}
\label{sec:factwo}
We now discuss the scenario where three scales Q, $Q\sqrt{\chi}$ and $Q_{\text{med}}$ are are well separated from each other. This is still within the collinear limit of the EEC but with a higher value of $\chi$ compared to the region I EFT. As discussed earlier, in this case we have a two stage EFT: The first of which is defined at the scale $Q \sqrt{\chi}$ and below. Subsequently, we match this EFT the scale $Q_{\text{med}}$.  The details of the nature of the jet medium interaction and the EFT modes needed in this regime will be presented in a companion letter that discusses the case of inclusive jet production. Here we apply this for the case of the EEC. As before we work in a frame where medium partons have energy $Q_{\rm med}$. In the same frame, hard collinear mode scales as 
\bea 
p_{hc} \sim Q\frac{T}{Q_{\text{med}}}\left( 1,\delta^2, \delta \right) 
\eea
with $\delta = \frac{\sqrt{\chi}Q_{\text{med}}}{T}$. This mode therefore has a virtuality $p_c^2 \sim Q^2\chi$. On the other hand, 
the medium partons as before scale as a soft mode 
\bea
p_s \sim Q_{\text{med}}(1,1,1)
\eea
which sits at the virtuality $p_s^2 \sim Q_{\text{med}}^2 \ll Q^2\chi$. The interaction between the soft partons and the hard collinear mode is therefore suppressed by $\mathcal{O}(Q_{\text{med}}^2/Q^2\chi)$ since they are widely separated in virtuality. Therefore the hard collinear mode describes only the vacuum evolution of the jet.
The jet-medium interaction and as also the medium induced radiation is described by a new radiation mode which we call the collinear-soft mode which scales as
\bea
p_{cs} \sim \left( \frac{Q_{\text{med}}}{\sqrt{\chi}},  \ Q_{\text{med}}\sqrt{\chi}, Q_{\text{med}}\right).
\eea
The scaling for this mode can be easily determined by demanding that it has a virtuality $Q_{\text{med}}^2$, with an angle of emission $\sqrt{\chi}$ so that it contributes to the measurement. Again for the case when the three scales $Q, Q\sqrt{\chi}$ and $Q_{\text{med}}$ are widely (or equally) separated from each other, we also have  $Q_{\text{med}}/\sqrt{\chi} \sim Q\sqrt{\chi}$ so that the energy of this mode is much smaller than the energy of the jet but much larger than the medium partons. Consequently the contribution to the EEC from this mode is suppressed by a factor $\sqrt{\chi}$. Thus, in this regime, the leading  contribution from medium interaction is parametrically power suppressed compared to the vacuum result. However, we may expect that in a dense medium, with multiple interactions, this contribution may be enhanced and become comparable to that from vacuum evolution.
\begin{figure}[t]
\centering
\includegraphics[scale=0.5]{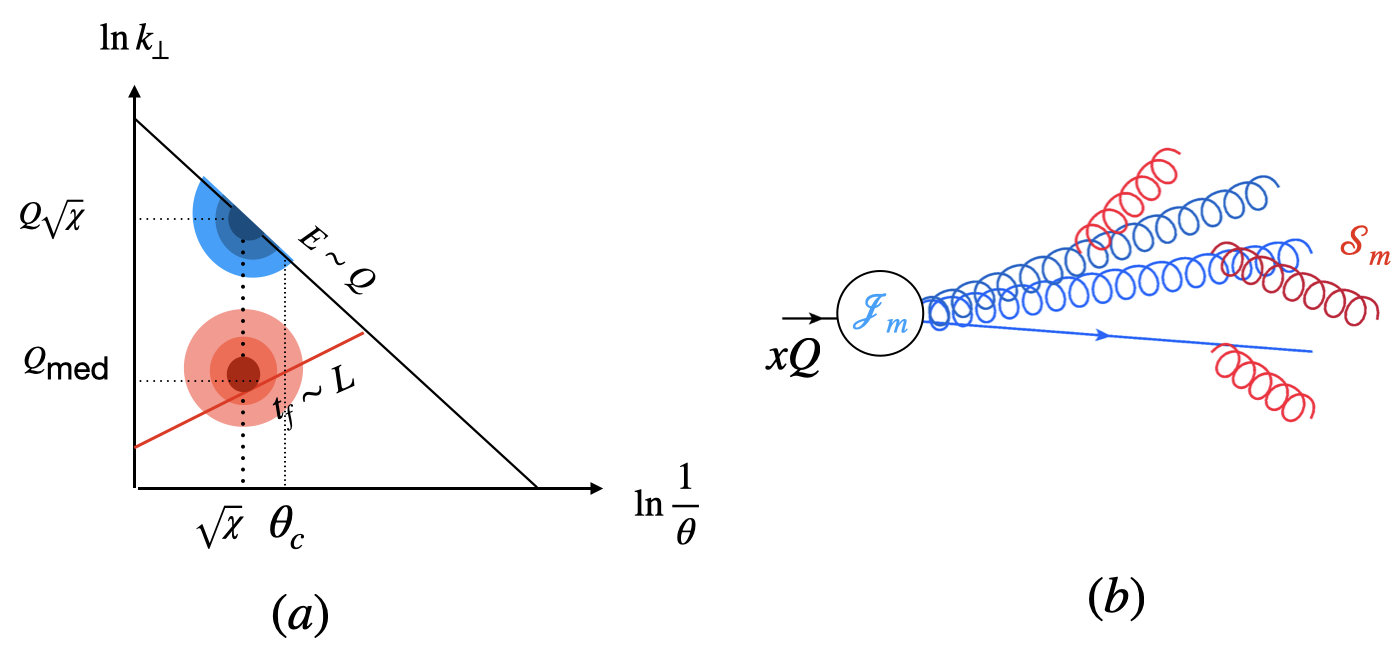}
\caption{Lund plane illustrating the measurement and kinematic constraints. Illustration for a jet with energy 200 GeV, with $L= 5$ fm, $\sqrt{\chi}=0.1$ and $\theta_c = 0.03$.} ~\label{fig:lund2}
\end{figure}

The phase space contributing to the measurement therefore is encoded in two regions as shown in the Lund plane Fig. \ref{fig:lund2}. The blue region corresponds to the hard-collinear physics while the red is populated by collinear soft(CS) radiation.
We see that for this regime  $\sqrt{\chi}$ is typically larger than the critical angle $\theta_c$ so that the medium can resolve multiple independent sources of medium induced collinear soft radiation.
 
Here we directly present the form of the factorization at stage II where we  refactorize the both the hard and hard collinear physics from the IR physics at the  scale $Q_{\text{med}}$. 

\small
\begin{align}
\label{eq:JSFact}
&\frac{1}{\sigma^0}\frac{d\sigma}{d\chi} = 
\int dx x^2 H_i(xQ, \mu) \Bigg[  J_i^{(0)}(xQ,\chi, \mu)+ \sum_{m=1}^{\infty} \sum_{j=1}^{m} \mathcal{J}^j_{i\rightarrow m}(\{\underline{m}\},\theta_c, xQ, \mu)\otimes_{\theta}\, \, {\cal S}_{m,j} (\{\underline{m}\} , \chi ,\mu) \Bigg]\nonumber\\ 
&+ {\cal O} \left( \frac{Q_{\rm med}^2}{Q^2\chi}\right)+{\cal O} (\chi)
\end{align}
\normalsize
The hard function is identical to the case of vacuum factorization. The first term $J_i^{(0)}(xQ,\chi, \mu)$ implements the measurement only on the hard-collinear partons and is identical to the vacuum jet function.  The function $\mathcal{J}^j_{i \rightarrow m}$ describes the production of \text{resolved} $m$ high energy $E \sim xQ$ prongs from the initial parton $i$. In this case the measurement acts on one hard-collinear, namely one of the resolved  partons $j$ and one collinear soft parton. Note that the contribution to the  EEC from two CS partons is power suppressed by a factor of $\chi$ due to their small energy and hence is not included at this order in the expansion.  
 The function $S_{m,j}$ describes the production of collinear soft radiation and its interaction with the medium. The two functions have a convolution in the angle of the $m$ high energy partons that source the collinear soft radiation. The result is written as a series over arbitrary number of resolved hard prongs.
 Therefore, all the medium physics is completely captured through the functions $\mathcal{S}_m^j$.
 Here, the collinear-soft(CS) functions are defined as 
\begin{equation} 
{\cal S}^{m,j}(\{\underline{m}\},\epsilon) \equiv    \text{Tr}\Big[U_m(n_m)...U_{1}(n_1)U_0(\bar n)\rho_M U^\dag_0(\bar n)U_1^\dag(n_1)...U_m^\dag(n_m)\mathcal{M}^j \Big] \,,
\label{eq:collsoft}
\end{equation}
Here $n_1$  . .. $n_m$ are the light-like directions of the $m$ high energy $E\sim Q$ partons.

We see that this refactorization looks similar to that encountered in the context of non-global logarithms~\cite{Dasgupta:2001sh} in Ref.~\cite{Becher:2015hka,Banfi:2002hw} but is more involved since we are probing the substructure of the jet. The function $\mathcal{S}_m$ is written in terms of Collinear soft Wilson line defined as 
\begin{align}
U(n) &=\text{P}\exp \Big[ig\int_0^{\infty} ds n \cdot A_{cs}(ns)\Big]
\label{eq:CollSoft}
\end{align}
The measurement $\mathcal{M}^j$ is defined as 
\bea
\sum_k E_k \Theta \left( \chi - \frac{\theta_{jk}^2}{4}\right)
\eea
where $E_k$ is the energy of the collinear soft parton and  $\theta_{jk}$ is its angle with the $j^{\text{th}}$ hard collinear parton.

The functions $\mathcal{J}_{i\rightarrow m}$, which are the perturbative matching co-efficients start at ${\cal O}(\alpha_s^{m-1})$ and therefore computing them requires successively higher order loop calculations. In this paper, we consider the first term is the series, namely the single hard prong. The analysis for subsequent terms will be left for future work.
 
\subsection{The single subjet}

The first term which gives rise to medium induced effects in Eq.~\ref{eq:JSFact} is given by 
\begin{align}
\mathcal{J}_{i\rightarrow 1}(\theta_c, xQ, \mu) \, {\cal S}_{1} (\chi,\mu)  \, 
\end{align}
$\mathcal{J}_{i\rightarrow 1}( \theta_c, xQ, \mu)$ is the matching co-efficient which we can fix to be 1 at leading order from a one loop matching. The collinear soft function $\mathcal{S}_1$ is evaluated with a SCET Hamiltonian  
\small
\begin{align}
\label{eq:H2}
\int dt H_{\text{I}} &= \int dt \left(H_{cs}+H_{s}+ H_{cs;s}\right) + \int d s \mathcal{O}_{cs;s}(sn).
\end{align}
\normalsize
Here $H_{cs}$ is the standard SCET collinear Hamiltonian and $H_{s}$ describes the soft dynamics of the medium partons and is just the full QCD Hamiltonian. $H_{cs;s}$ describes the scattering of the collinear-soft gluon off the soft medium through a Glauber exchange. All of this is identical to the Hamiltonian encountered for the Region I of the EFT. The new ingredient is  the medium-induced radiation to all orders in $\alpha_s$ by a hard-collinear parton in the direction $n$ and is described in the last term of Eq. \ref{eq:H2} as an  operator along the world-line of the hard parton.  For $\mathcal{S}_m$, we will have $m$ such operators, one for each hard collinear parton. This is described in terms of the  operator $O_{cs}$ which is written in terms of an interaction operator which was derived in Ref.~\cite{Rothstein:2016bsq}.  
\bea
\mathcal{O}_{cs;s}(sn) = \int d^2q_{\perp} \frac{1}{q_{\perp}^2}\Big[O^{BA}_{cs}\frac{1}{\mathcal{P}_{\perp}^2}\mathcal{O}_{s}^A\Big](sn,q_{\perp}) T^B
\eea
where 
\bea 
O^{BA}_{cs}(x) &=& 8\pi \alpha_s\Bigg[\mathcal{P}_{\perp}^{\mu}S_n^TW_{n}\mathcal{P}_{\perp \mu} -\mathcal{P}_{\mu}^{\perp}g \mathcal{\tilde B}_{S\perp}^{n\mu}S_n^TW_n -\nonumber \\
&& S_n^TW_ng \mathcal{\tilde B}^{n \mu}_{\perp}\mathcal{P}_{\mu}^{\perp}- g \mathcal{\tilde B}_{S \perp}^{n\mu}S_n^TW_ng\mathcal{\tilde B}^{n}_{\perp \mu}-\frac{n_{\mu}\bar n_{\nu}}{2}S_n^T ig  G^{\mu\nu}W_n\Bigg]^{BA}
\label{eq:Ocs}
\eea
$\mathcal{O}_s^{A} =\sum_{j\in\{q,\bar q, g\}}\mathcal{O}_s^{j,A}$ is the soft medium operator identical to the one that appears in the region I EFT defined in Eq.\ref{eq:SandC} that acts as source for the Glaubers. The $O_{cs}$ operator is again built out of gauge invariant SCET operators whose definitions are provided in Eq.\ref{eq:BBlock} while $ G^{\mu \nu}$ is the gluon field strength tensor.
 
\begin{figure}
\centering
\includegraphics[width=\linewidth]{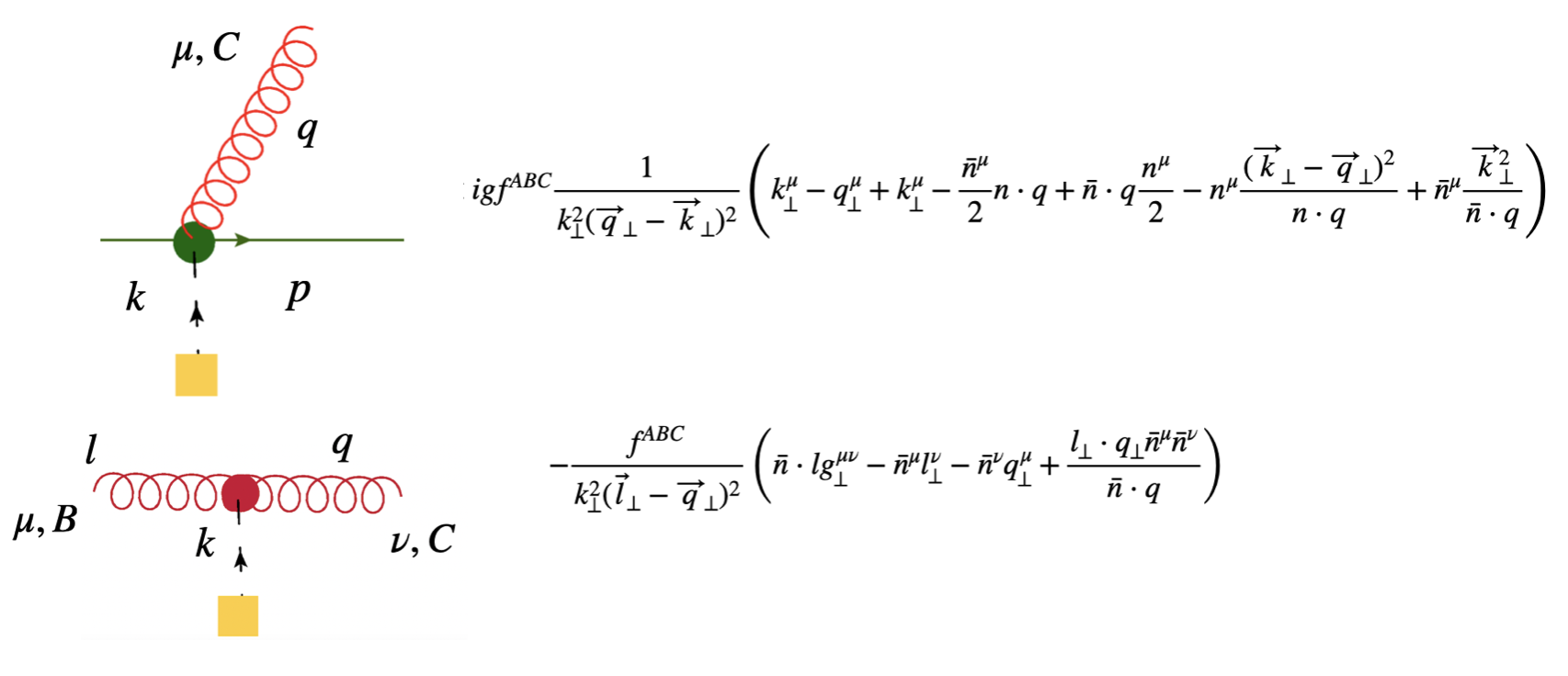}
\caption{The leading feynman rules  from medium induced radiation operator $O_{cs}$ and scattering of CS gluon off the medium. The red gluon is a $CS$ gluon. The green  line represents the world-line of the hard quark. The dashed line is the Glauber propagator sourced by the medium~\label{LIP}.}
\end{figure}

At one loop, the operator $O_{cs}$ reproduces the Lipatov vertex. 
For this paper, we are computing only the one-loop result so we present the relevant Feynman rules derived from the two interaction operators $H_{cs;s}$ and $O_{cs}^{BA}$ which is the medium-induced CS radiation through the Lipatov vertex and the interaction of a CS gluon with the medium as shown in Fig.~\ref{LIP}. 

The nature of the observable is such that the entirety of vacuum evolution is contained in the hard collinear matching co-efficient $\mathcal{J}_{i\rightarrow 1}$ while the collinear soft function contains all the medium induced jet physics and is UV finite. This implies, by consistency of RG evolution,
\begin{align}
   & \frac{d}{d \ln \mu^2} \mathcal{J}_{i\rightarrow 1}(\chi_c,Q,\mu) = \int dx x^2 \mathcal{J}_{j\rightarrow 1}(\chi_c,xQ,\mu)P_{ji}(x)\nonumber \\
   &  \frac{d}{d \ln \mu^2}\mathcal{S}_1(\chi_c, \mu) =0
\end{align}

\subsubsection{ Factorization of the medium physics.} 
At this stage, the functions $\mathcal{S}_1$ depends on both the properties of the jet and the medium through the collinear soft and soft modes sourced by the medium density matrix $\rho_M$. The two modes cannot be decoupled to all orders in a simple manner due to the non-eikonal nature of the glauber exchange. Instead, the factorization of the universal physics associated with the medium from the jet requires us to expand out the action Eq.~\ref{eq:H2} and separate the soft physics of the medium from the collinear-soft jet radiation order by order in the number of interactions between the jet and the medium. For the single subjet case, we have ${\cal S}_{1,1} = \sum_{i=0}^{\infty}{\cal S}_{1}^{(i)}$ (Eq.~\ref{eq:soft-funct}) where the summation is over the number of interactions, i,e, glauber exchanges of the jet with the medium soft partons. Here, ${\cal S}_{1}^{(0)}$ is the vacuum contribution, which for this observable is just identity. At ${\cal O}(n)$ with $n>0$, for a homogeneous medium of size $L$, we can write 
\small
\begin{align}\label{eq:FctAl}
&{\cal S}_{1}^{(n)}(\chi) = \frac{\left(|C_{G}|^{2}L\right)^n}{n!}\Bigg[\prod_{i=1}^{n}\int \frac{d^2k_{i\perp}}{(2\pi)^3}\mathcal{B}(k_{i\perp},\mu,\nu)\Bigg]
{\bf S}_{1}^{(n)}(\chi,L ;k_{1\perp}, \ldots, k_{n\perp},\nu)\,.
\end{align}
\normalsize
Here $C_G = 8\pi \alpha_s(\mu)$. 
This expression contains $n$ copies of the medium correlator $\mathcal{B}$, which is defined as
\bea
\mathcal{B}(k_{\perp})= \frac{1}{k_{\perp}^2}\frac{1}{N_c^2-1}\int \frac{dk^-}{2\pi}\int d^4r e^{-i k_{\perp} \cdot r^{\perp}+ik^-r^+}\text{Tr}\Big[e^{-i\int dt H_s(t) }O^A_{s}(r) \rho_M O^A_{s}(0)e^{i\int dt H_s(t) }\Big].\,\qquad
\label{eq:B2}
\eea
which is identical to the definition of the correlator for the regime $Q\sqrt{\chi} \sim Q_{\text{med}}$ defined in \ref{eq:B}.
This result is valid when the mean free path of the jet $\lambda_{\text{mfp}}$ is much larger than the color screening length $1/m_D$ in the medium. As before, the function $\mathcal{B}$ obeys the BFKL evolution equation Eq.\ref{eq:BRG} as  in the rapidity renormalization scale $\nu$, along with $\alpha_s$ running in the scale $\mu$.This function depends \textit{only} on universal microscopic properties of the medium. The function ${\bf S}_{1}^{(n)}$ in Eq.~(\ref{eq:FctAl}) obeys a BFKL equation in all its $k_{i\perp}$ arguments but with an opposite sign to maintain renormalization group (RG) consistency in the evolution associated with the scale $\nu$. 
\bea
\label{eq:CS_RG}
 \frac{d{\bf S}_{1}^{(n)}( \chi_c,L;k_{1\perp}, ...k_{m\perp};\nu)}{d\ln \nu} 
 &=& -\sum_{i=1}^{n}\frac{\alpha_s N_c}{\pi^2} \int d^2u_{\perp} \Bigg[\frac{{\bf S}_{1}^{(n)}( \chi_c,L;k_{1\perp}...k_{i-1\perp},u_{\perp}, k_{i+1\perp} .. k_{m\perp};\nu)}{(\vec{u}_{\perp}\vec{k}_{i\perp})^2}\nonumber\\
 & - &\frac{k_{\perp}^2{\bf S}_{1}^{(n)}( \chi_c,L;k_{1\perp},k_{2\perp}, ...k_{m\perp};\nu) }{2u_{\perp}^2(\vec{u}_{\perp}-\vec{k}_{\perp})^2}\Bigg]\nonumber
 \eea
 \bea
 \frac{d{\bf S}_{1}^{(n)}( \chi_c,L;k_{1\perp}, ...k_{m\perp};\nu)}{d\ln \mu} =0 
\eea

\subsubsection{Single interaction medium induced collinear soft function}

\begin{figure}
\centering
\includegraphics[width=0.75\linewidth]{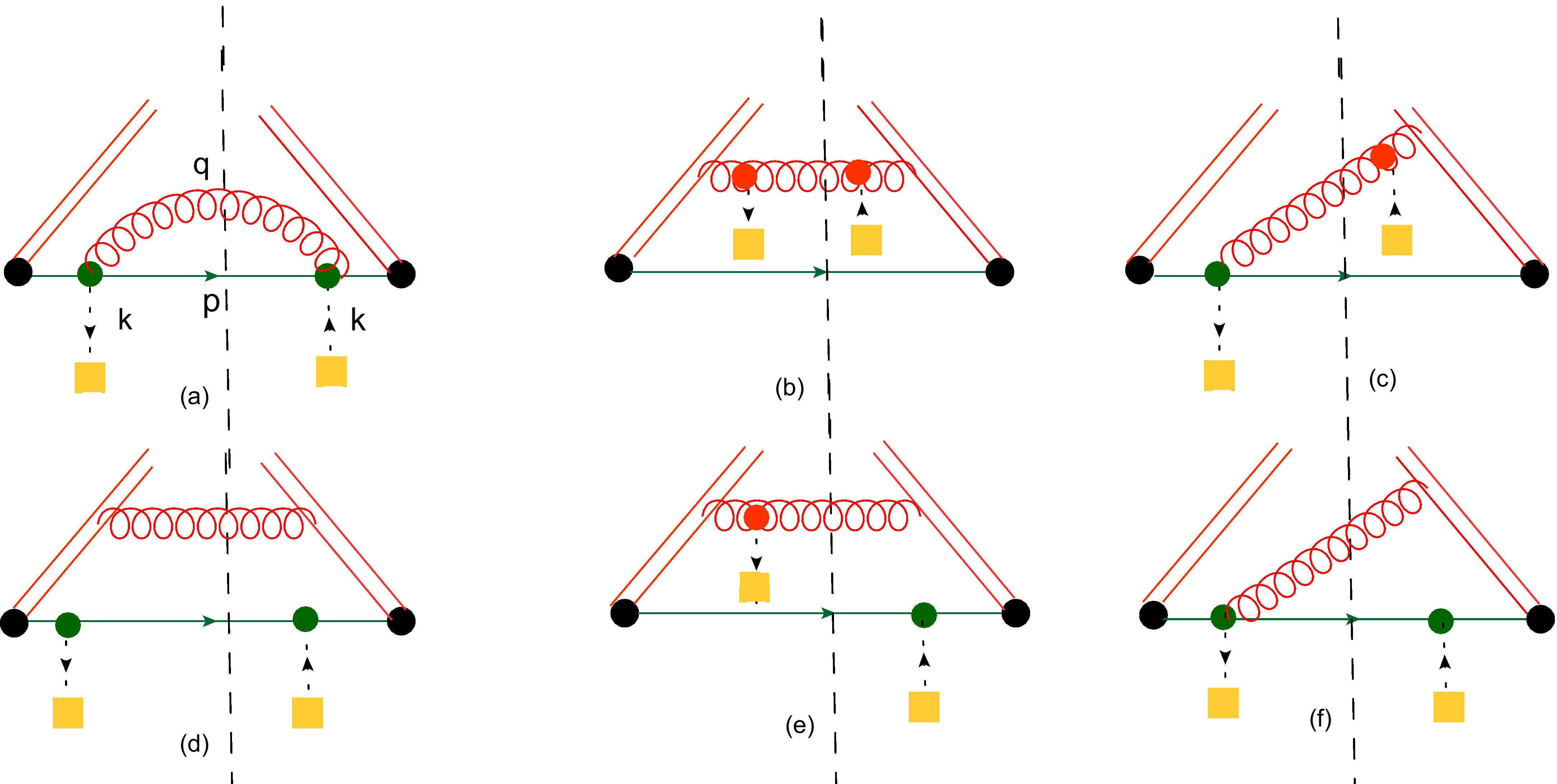}
\caption{Real emission diagrams at the one-loop level with Glauber insertions on the opposite sides of the cut. The red parallel lines indicate $U(n)$ Wilson lines. The green dot is the insertion of the $O_{cs;s}$ operator.}
\label{Loop2RR}
\end{figure}

Now we give here the expressions for the medium induced collinear soft function for a single interaction with the medium i.e, the function ${\bf S}_{1}^{(1)}$. 
We can define the collinear soft function from real and virtual functions in the same manner as  in Eq.~\ref{eq:totaljet}. 
\begin{align}
\label{eq:CSDef}
{\bf{S}}^{(1)}_{1}(\chi,\omega,k_{\perp})&={\bf{S}}_{R}(\chi,\omega,k_{\perp})-{\bf{S}}_{V}(\chi,\omega,k_{\perp}).  
\end{align}
where ${\bf{S}}_{R}$ corresponds to glauber exchange on the opposite sides of the cut while ${\bf{S}}_{V}$ stands for same side glauber exchange. 
The operator for these functions are defined using two types of glauber operators, namely the collinear operator $O_n^A$ and the medium induced radiation operator $O^{BA}_{cs}$ Eq.\ref{eq:Ocs} and Eq.\ref{eq:SandC}. Therefore, for ${\bf{S}}_{R}$ we can  then write 
\begin{align}
\label{eq:CSoftI}
{\bf S}_{R} &=\frac{1}{2N_c}\frac{e^{-i\frac{L}{2}(\mathcal{P}^{A}_{+}-\mathcal{P}^{B}_{+})}}{k_{\perp}^2}\text{sinc}\Big[\frac{L}{2}(\mathcal{P}^{A}_{+}-\mathcal{P}^{B}_{+}) \Big]\nonumber\\ 
&\hspace{-0.5cm} \sum_{X}\tr\Big[\langle 0|\bar{\mathcal{T}}\Big\{e^{-i\int dt H_{n}(t)}\Big[\delta(\mathcal{P}^-)\delta^2(\mathcal{P}_{\perp}-k_{\perp})O_{n}^{qA}(0)+ \frac{1}{(\vec{\mathcal{P}}_{\perp}-\vec{k}_{\perp})^2}O_{cs}^{CA}(0)T^C\Big]U^{\dagger}(n)U^{\dagger}(\bar n) \Big\} \mathcal{M}|X\rangle \nonumber\\
&\langle X|\mathcal{T}\Big\{e^{-i\int dt H_{n}(t)} \Big[\delta(\mathcal{P}^-)\delta^2(\mathcal{P}_{\perp}+k_{\perp})O_{n}^{qB}(0)+ \frac{1}{(\vec{\mathcal{P}}_{\perp}-\vec{k}_{\perp})^2}O_{cs}^{CB}(0)T^C\Big]U(n)U(\bar n)\Big\}|0\rangle \Big]\delta^{AB}, 
\end{align}
 and then likewise for ${\bf S}_V$.$\vec{\mathcal{P}}_{\perp}$ is that extracts out the transverse momentum of the operator that it acts on.
In Fig.\ref{Loop2RR} we show a subset of diagrams that goes into the one loop computation, namely real gluon emission with opposite side glauber exchanges.
We see that the gluon is sourced from either the vacuum Wilson line $U(n)$ \footnote{The $U(\bar n)$ Wilson line does not lead to any contribution atleast at leading order so we do not show it in the Feynman diagrams. } which can then scatter off the medium through vertex (b) in Fig.\ref{LIP} or is medium induced through the Lipatov vertex (a) in Fig.\ref{LIP}. 
We will have a similar set of diagrams for virtual gluon emissions and then repeated for glauber exchanges on the same side. 
At leading order, we expect the result to be just the soft limit of the jet function defined in region I EFT. This is because the matching coefficient $\mathcal{J}_{q\rightarrow 1}$ at leading order is one. Moreover, at higher orders  the collinear soft function can be obtained from Eq.~\ref{eq:CSoftI}. Combining the real and virtual diagrams along with the measurements, we get
\begin{align}
{\bf{S}}_{1}^{(1)}(\chi,\omega,k_{\perp})&=({\bf{S}}_{RR}-{\bf{S}}_{VR})\,2z\Big[\delta\Big(\chi-\frac{q_{\perp}^2}{z^2\omega^2}\Big)-\delta(\chi)\Big]
\label{eq:softS}
\end{align}
where

\begin{align}
{\bf{S}}_{RR}-{\bf{S}}_{VR}=&8N_cg^2\int \frac{dq^-}{(2\pi)^3k_{\perp}^2}\int d^2q_{\perp}\left(\frac{1}{q^-\kappa^2}-\frac{\vec{q}_{\perp}\cdot\vec{\kappa}_{\perp}}{q_{\perp}^2\kappa_{\perp}^2q^-} \right)\Big[1-\frac{q^{-}}{\kappa_{\perp}^2L}\sin\Big(\frac{L\kappa_{\perp}^2}{q^{-}} \Big)  \Big] \nonumber\\
&=8N_cg^2\int \frac{dq^-}{(2\pi)^3}\int d^2q_{\perp}\frac{\vec{q}_{\perp}\cdot \vec{k}_{\perp}}{q_{\perp}^2\kappa_{\perp}^2q^-}\Big(1-\frac{q^{-}}{\kappa_{\perp}^2L}\sin\Big[\frac{L\kappa_{\perp}^2}{q^{-}} \Big]\Big).
\end{align}
Therefore the collinear soft function 
\begin{align}
\label{eq:SoftGLV}
{\bf{S}}_{1}^{(1)}(\chi,\omega,k_{\perp})=&\frac{2(N_c^2-1)\bar{\alpha}}{\pi^2k_{\perp}^2}\int \frac{dz}{z}\int d^2q_{\perp}\frac{\vec{q}_{\perp}\cdot \vec{k}_{\perp}}{q_{\perp}^2\kappa_{\perp}^2}\Big[1-\frac{z\omega}{\kappa_{\perp}^2L}\sin\Big(\frac{L\kappa_{\perp}^2}{z\omega}\Big) \Big]\nonumber\\
&2z^3\omega^2\left[\delta(q_{\perp}^2-z^2\omega^2 \chi)-\delta(\chi)\right] 
\end{align}
where $\bar{\alpha}=\frac{g^2}{4\pi}$ is the coupling between Glauber gluon and collinear jet parton. Note that the above expression coincides with the soft limit of GLV results obtained in Ref.~\cite{Gyulassy:2000er}.

\subsubsection{Towards  all order in interactions}
\label{sec:BDMPS-Z}
If the medium is dense, then multiple interactions between jet and medium become important, and in that case at NLO we need to sum over arbitrary number of Glauber exchanges with a single collinear-soft gluon emission. This is essentially the BDMPS-Z\cite{Baier:1994bd,Baier:1996kr} regime. We can see that computation for any fixed number of interactions will require one insertion of either the vacuum Wilson line gluon or the Lipatov vertex with multiple scatterings off the medium. We can then follow exactly the same procedure as the original derivation \cite{Baier:1996kr} which uses an iterative procedure to obtain an all orders expression. The difference in our case is the copies of the $\mathcal{B}$ function that carry the all order dynamics of the medium partons and are carried along in the calculation . This calculation will be the subject of an upcoming work. The most important aspect of multiple scattering would be the emergence of the $Q_{\text{med}}$ scale which, if sufficiently separated from T would precipitate another step of matching. We discuss this in more detail in Section \ref{sec:Qmed}.

\begin{figure}
\centering
\includegraphics[width=0.75\linewidth]{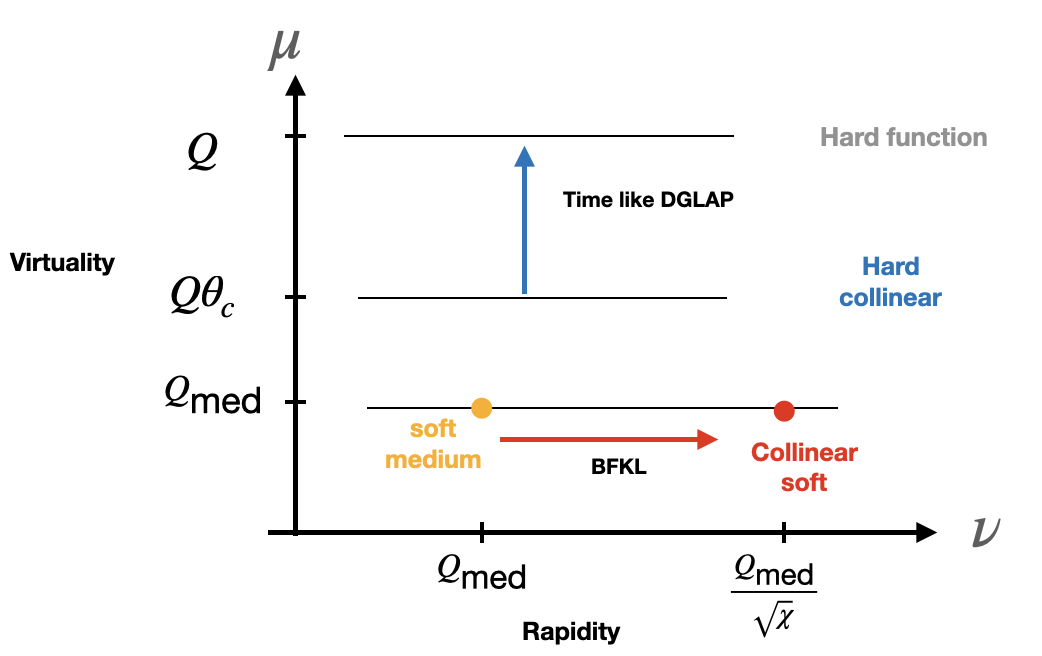}
\caption{Renormalization Group evolution for a single subjet}
\label{RG2}
\end{figure}

We can then summarize the EFT in this regime, writing out the complete factorization formula as 

\begin{align}
\label{eq:FactFin}
\frac{1}{\sigma^0}\frac{d\sigma}{d\chi}&= \int dx x^2 H_i(x,Q,\mu) \Bigg \{ J_i^{(0)}(\chi,\mu) + \sum_{m=1}^{\infty} \sum_{j=1}^m \mathcal{J}^j_{i\rightarrow m}(\theta_c, xQ,\mu)  \nonumber \\
& \otimes_{\{\theta_1....\theta_m\}} \Bigg[\sum_{n=1}^{\infty} \frac{L^n}{n!}\Big[\prod_{l=1}^{n} \int \frac{d^2k_{l\perp}}{(2\pi)^3}\mathcal{B}(k_{l\perp},\mu,\nu)\Big]{\bf S}^{(n)}_{m,j}(\chi,k_{1\perp}, . .. k_{n\perp},L,\mu,\nu)\Bigg] \Bigg\}
\end{align}
The function ${\bf S}^{(n)}_{m,j}$ refers to the the collinear soft function for $m$ hard prongs with $n$ interactions with the medium. The subscript j refers to the measurement acting on the $j^{th}$ prong and one collinear soft parton.
The RG evolution between the various modes for medium modification of the measurement is shown in Fig. \ref{RG2}.

\subsection{Impact of BFKL resummation}
Although a full phenomenological analysis is beyond the scope of this paper, we assess here the impact of the BFKL evolution induced due to the medium evolution of the jet. 
We implement this evolution by solving the BFKL equation between the collinear soft and medium correlator Eq.~\ref{eq:B}. This evolution is in rapidity and happens at a single virtuality $\mu \sim Q_{\text{med}}$.
 While the exact rapidity scale of the collinear soft function in our case will appear at NNLO, we parametrically take this scale to be $Q_{\text{med}}/\sqrt{\chi}$. Likewise, the natural rapidity scale for the medium function  is $Q_{\text{med}}$ so that the BFKL evolution will resum $\alpha_s^n(Q_{\text{med}}) \ln^n 1/\sqrt{\chi}$.

\begin{figure}[h!]
	\centering
	\includegraphics[scale=0.62]{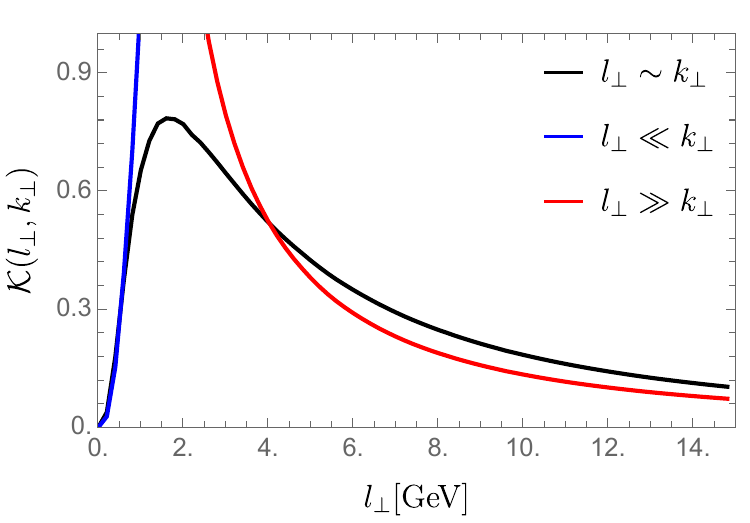}
	\includegraphics[scale=0.61]{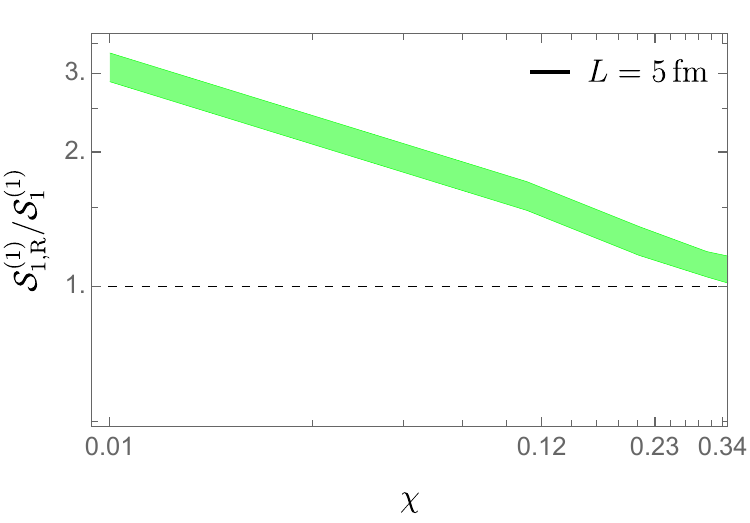}
	\caption{Left: The resummed BFKL kernel as a function of the transverse momentum for three regimes with $\chi=0.01$. Right: The ratio of the  BFKL resummed medium induced jet function to the corresponding NLO jet function for medium length $L=5$ fm and $T=0.4$ GeV. The band corresponds to varying the scale $\mu\sim Q_{\rm med}\in (2-3)$ GeV.}
	\label{fig:bfkl}
\end{figure}

As seen the previous section, the tree level contribution of the medium to the cross section,i.e. $\mathcal{O}(\alpha_s^2)$  result from exchange of a single Glauber vanishes for this observable. The soft limit of the GLV result appears at $O(\alpha_s^3)$ while BFKL logs appear at $\mathcal{O}(\alpha_s^4)$ . If we count $\alpha_s \ln \sqrt{\chi}$ as $\mathcal{O}(1)$, then it appears that the contribution from the medium only starts at $\mathcal{O}(\alpha_s^3)$, i,e, N3LL. However, this is misleading for two reasons: The $\alpha_s$ for BFKL resummation is evaluated at the scale $Q_{\text{med}}$ where it is much larger compared to that for the vacuum result. Likewise, in a dense medium with multiple interactions, the leading contribution for the medium can easily become $\mathcal{O}(1)$ as has been seen in experiments, which could elevate BFKL resummation to the same level as the LL result. Hence, for any phenomenologically rigorous prediction, it becomes imperative to include BFKL resummation effects. The collinear soft function obeys BKFL  equation given as
\begin{equation}
\nu\frac{d{\bf S}^{(1)}_1(k_{\perp},\nu)}{d\nu}=-\frac{\alpha_s(\mu) N_c}{\pi^2}\int d^2l_{\perp}\bigg[\frac{{\bf S}^{(1)}_1(l_{\perp},\nu)}{(\vec{l}_{\perp}-\vec{k}_{\perp})^2}-\frac{k_{\perp}^2{\bf S}^{(1)}_1(k_{\perp},\nu)}{2l_{\perp}^2(\vec{l}_{\perp}-\vec{k}_{\perp})^2} \bigg], 
\label{eq:bfkl}
\end{equation}
where $\nu\sim Q_{\text{med}}/\sqrt{\chi}$ is natural scale for the collinear soft function. In order to solve above equation we define BFKL kernel 
\begin{equation}
\int d^2l_{\perp}K_{\rm BFKL}(\vec{l}_{\perp},\vec{k}_{\perp}) {\bf S}_1^{(1)}(\vec{l}_{\perp}) = \frac{1}{\pi} \int \frac{d^2l_{\perp}}{(\vec{l}_{\perp}-\vec{k}_{\perp})^2}\Bigg[{\bf S}_1^{(1)}(l_{\perp},\nu)-\frac{k_{\perp}^2}{2l_{\perp}^2}{\bf S}_1^{(1)}(\vec{k}_{\perp},\nu)\Bigg].    
\end{equation}
In order to solve above equation we follow the prescription given in Ref.~\cite{Kovchegov:2012mbw}. The BFKL kernel has the eigenfunction of the form $f(\vec{l}_{\perp})=l_{\perp}^{2\gamma-1}e^{i n\phi_l}$ where $\phi_l$ is azimuthal angle with $n$ being an integer and $\gamma$ an arbitrary complex number. With this ansatz we can find the eigenvalue
\begin{equation}
\int d^2l_{\perp}K_{\rm BFKL}(\vec{l}_{\perp},\vec{k}_{\perp})l_{\perp}^{2(\gamma-1)}e^{in\phi_l}= \chi(n,\gamma)k_{\perp}^{2(\gamma-1)}e^{i n\phi_k},
\end{equation}
where 
\begin{equation}
\chi(n,\gamma)=2 \psi(1)-\psi\left(\gamma+\frac{|n|}{2}\right)-\psi\left(1-\gamma+\frac{|n|}{2}\right),
\end{equation}
with $\psi$ being the Polygamma function. The result for the eigenvalue is valid for $0< {\rm Re}(\gamma)<1$. With this general solution of BFKL equation we can expand out the collinear soft function in the terms of eigenfunctions
\begin{equation}
{\bf S}^{(1)}_1(\vec{l}_{\perp},\nu)=\sum_{n=-\infty}^{\infty} \int_{a-i\infty}^{a+i\infty}\frac{d\gamma}{2\pi i}\mathcal{C}_{n,\gamma}(\nu) l_{\perp}^{2(\gamma-1)}e^{in\phi_l},
\end{equation}
where $\mathcal{C}_{n,\gamma}$ is unknown function which we need find using fixed order NLO collinear soft function as boundary condition. Now we can plug in the expanded jet function in the collinear-soft function evolution equation Eq.~\ref{eq:bfkl} to obtain
\begin{equation}
\nu \frac{d}{d\nu}{\bf S}^{(1)}_1(k_{\perp},\nu)=-\frac{\alpha_s(\mu) N_c}{\pi}\sum_{n=-\infty}^{\infty}\int_{a-i\infty}^{a+i\infty}\frac{d\gamma}{2\pi i} \chi(n,\gamma)\mathcal{C}_{n,\gamma}(\nu)k_{\perp}^{2(\gamma-1)}e^{in\phi_k},    
\end{equation}
where $\mathcal{C}_{n,\gamma}$ is obtained by solving 
\begin{equation}
\nu \frac{d}{d\nu}\mathcal{C}_{n,\gamma}(\mu, \nu) = -\frac{\alpha_s(\mu) N_c}{\pi}\chi(n,\gamma)\mathcal{C}_{n,\gamma}(\nu),   
\end{equation}
from some known scale $\nu_{0}$ which in our case is $Q_{\text{med}}/\sqrt{\chi}$ to final scale $\nu_f$ which is $k_{\perp} \sim Q_{\text{med}}$. The corresponding solution for $\mathcal{C}_{n,\gamma}$ is given as
\begin{equation}
\mathcal{C}_{n,\gamma}(\mu,\nu_f) = \mathcal{C}_{n,\gamma}(\mu, \nu_0)e^{ -\frac{\alpha_s(\mu) N_c}{\pi}\chi(n,\gamma)\ln \frac{\nu_f}{\nu_0}},  \end{equation}
where the scale $\mu$ enters through coupling only. With this solution of $\mathcal{C}_{n,\gamma}$, we can re-write the collinear soft function as
\begin{equation}
{\bf S}^{(1)}_1(k_{\perp},\mu,\nu_f)=\sum_{n=-\infty}^{\infty} \int_{a-\infty}^{a+i\infty}\frac{d\gamma}{2\pi i}\mathcal{C}_{n,\gamma}(\mu,\nu_0) e^{ -\frac{\alpha_s(\mu) N_c}{\pi}\chi(n,\gamma)\ln \frac{\nu_f}{\nu_0}}k_{\perp}^{2(\gamma-1)}e^{in\phi_k}.
\label{eq:medjetrsm}
\end{equation}
For complete resummed CS function we  need to compute $\mathcal{C}_{n,\gamma}(\mu,\nu_0)$ which we can do in the following way. We can see that at the scale $\nu_f=\nu_0$  the CS function reduces to NLO fixed order result.  In Eq.\ref{eq:medjetrsm} we multiply both side by a factor $e^{-im\phi_k}k_{\perp}^{2\alpha^*-1}$ and use the the orthogonality relation
\begin{align}
\int d^2k_{\perp}e^{-im\phi_k}k_{\perp}^{2\alpha^*-1}k_{\perp}^{2\gamma-1}e^{in\phi_k}=&2\pi\delta_{mn}\int dr e^{(2\alpha_R+2\gamma_R-2)r}e^{i(-2\alpha_I+2\gamma_I)r}\nonumber\\
&=2\pi^2\delta_{mn}\delta(\alpha_I-\gamma_I),
\end{align}
where in the second line we have used $\alpha_R=\alpha_{\gamma}=1/2$. Thus the coefficient $\mathcal{C}$ is given as
\begin{equation}
\mathcal{C}_{n,\gamma}(\mu,\nu_0)=\int d^2l_{\perp}{\bf S}^{(1)}_1(l_{\perp},\mu,\nu_0)e^{-im\phi_l}l_{\perp}^{2\alpha^*-1}.    
\end{equation}
Therefore, the resummed CS function takes the form
\begin{equation}
{\bf S}^{(1)}_{1,{\rm R}}(k_{\perp},\mu,\nu_f)\!=\!\!\!\!\sum_{n=-\infty}^{\infty} \int_{\frac{1}{2}-i\infty}^{\frac{1}{2}+i\infty}\frac{d\gamma}{2\pi i}\!\!\!\int d^2l_{\perp}e^{i (n\phi_k-m\phi_l)} k_{\perp}^{2(\gamma-1)}l_{\perp}^{2(\gamma^*-1)}{\bf S}^{(1)}_1(l_{\perp},\mu,\nu_0)e^{ -\frac{\alpha_s(\mu) N_c}{\pi}\chi(n,\gamma)\ln \frac{\nu_f}{\nu_0}}   
\end{equation}
We can use $\gamma=1/2+i\nu$ to rewrite above equation as
\begin{equation}
{\bf S}^{(1)}_{1,{\rm R}}(k_{\perp},\mu,\nu_f)=\int d^2l_{\perp}{\bf S}^{(1)}_1(l_{\perp},\mu,\nu_0)\int\frac{d\nu}{2\pi}k_{\perp}^{-1+2i\nu}l_{\perp}^{-1-2i\nu}e^{in(\phi_k-\phi_l)}e^{-\frac{\alpha_s(\mu)N_c}{\pi}\chi(n,r)\log\frac{\nu_f}{\nu_0}}.  
\label{eq:resumS}
\end{equation}
Now we solve above equation analytically in the following three regime
\begin{itemize}
\item For $k_\perp\sim l_{\perp}$ Eq.~\ref{eq:resumS} can be written as
\begin{align}
{\bf S}^{(1)}_{1,{\rm R}}(k_{\perp})=&\frac{1}{\pi k_{\perp}}\sqrt{\frac{\pi}{14\zeta(3)\bar{\alpha}Y}}e^{(a_p-1)Y}\int d^2l_{\perp} \frac{{\bf S}_1^{(1)}(l_{\perp})}{l_{\perp}} e^{-\frac{\log^2(k_{\perp}/l_{\perp})}{14\zeta(3)\bar{\alpha}Y}}\nonumber\\
&=\int d^2l_{\perp}\mathcal{K}(l_{\perp},k_{\perp}),
\label{eq:lsamek}
\end{align}
where $a_p=1+\frac{4 \alpha_s N_c}{\pi}\ln{2}$ and $Y=\log(\nu_0/k_{\perp})$ with $\nu_0 = Q_{\text{med}}/\sqrt{\chi}$. The transverse momentum scale $k_{\perp} \sim Q_{\text{med}}$ so that $Y \sim \ln 1/\sqrt{\chi}$.
\item In the limit $k_{\perp}\gg l_{\perp}$ Eq.~\ref{eq:resumS} acquires the form
\begin{equation}
{\bf S}^{(1)}_{1,{\rm R}}(k_{\perp})= \frac{(\bar{\alpha}Y)^{1/4}}{\pi^{\frac{1}{2}}}\int \frac{d^2l_{\perp}{\bf S}_1^{(1)}(l_{\perp})}{l_{\perp}^2\ln^{3/4}(k_{\perp}^2/l_{\perp}^2)}e^{2\sqrt{\bar{\alpha}Y\ln(k_{\perp}^2/l_{\perp}^2)}},
\label{eq:kggl}
\end{equation}
where $\bar{\alpha}=\frac{\alpha_s N_c}{\pi}$ and $Y$ is same the one defined in the last equation.
\item Finally with $k_{\perp}\ll l_{\perp}$ Eq.~\ref{eq:resumS} can be simplified to 
\begin{equation}
{\bf S}^{(1)}_{1,{\rm R}}(k_{\perp})=\frac{(\bar{\alpha}Y)^{1/4}}{2\pi^{1/2}k_{\perp}^2}\int \frac{d^2l_{\perp}{\bf S}^{(1)}_1(l_{\perp})}{\ln^{3/4}(l_{\perp}^2/k_{\perp}^2)}e^{2\sqrt{\bar{\alpha}Y\ln(l_{\perp}^2/k_{\perp}^2)}}. \label{eq:lggk}
\end{equation}
\end{itemize}
In Figure~\ref{fig:bfkl}(left), we plot the integrands appearing in the resummed collinear soft function in Eqs.~\ref{eq:lsamek}, \ref{eq:kggl} and \ref{eq:lggk} as a function of $l_{\perp}$ for Glauber momentum $k_{\perp}=2.5$ GeV and various values of $\chi$. Note that the solution in Eq.~\ref{eq:lsamek} is valid in the intermediate range where both Eq.~\ref{eq:kggl} and Eq.~\ref{eq:lggk} blows up. However, even in the range $k_{\perp}\gg l_{\perp}$ and $k_{\perp}\ll l_{\perp}$, Eq.~\ref{eq:lsamek} gives qualitatively reasonable solutions. We stress that for accurate phenomenology one should use an interpolation between these three regimes. Since we are only interested in the effect of BFKL resummation qualitatively we will use Eq.~\ref{eq:lsamek} for all three regimes. 

Now with this solution at hand we plot the ratio of resummed collinear soft function that include $\mathcal{B}$ and fixed order collinear soft function, i.e., $\mathcal{S}_{1}^{(1)}$ in Figure~\ref{fig:bfkl}(right). Here we take one loop QCD coupling constant
\begin{equation}
\alpha_s(\mu)=  \frac{\alpha_s(m_Z)}{1+\frac{\alpha_s(m_z)\beta_0}{2\pi}\log\Big[\frac{\mu}{m_Z} \Big]},  
\end{equation}
where 
\begin{equation}
 \beta_0=\frac{11 N_c}{3}-\frac{2 N_f}{3}.   
\end{equation}
We take the scale $\mu=Q_{\rm med}\sim \sqrt{\hat{q}L}, m_Z=90$ GeV and $\alpha_s(m_Z)=0.11$. For the band in Figure~\ref{fig:bfkl} we vary $Q_{\rm med}\in (2-3)$ GeV which qualitatively corresponds to $\hat{q}\in (1-2)$ GeV$^2$fm$^{-1}$. 
We note that as anticipated BFKL resummation is relevant in the small $\chi$ limit and merges to its fixed order counterpart at  high $\chi$ limit. The large impact of  BFKL evolution can be undertood as the growth of the logarithms $(a_p-1)^n\ln^n 1/\sqrt{\chi}$ with decreasing $\chi$. For $\chi=0.01$, $\exp[(a_p-1) \ln(1/\sqrt{\chi})]\sim 3.5$.

\section{The scale $Q_{\text{med}}$}
\label{sec:Qmed}

\begin{figure}
\centering
\includegraphics[width=0.75\linewidth]{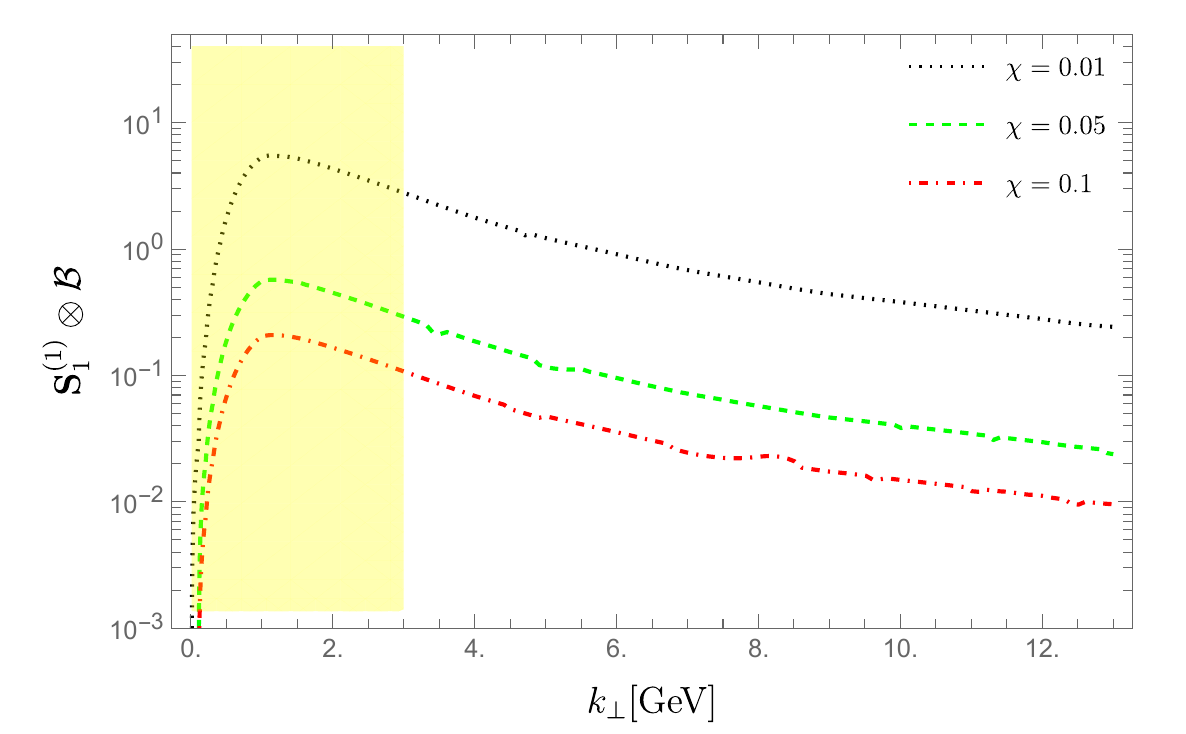}
\caption{Variation of ${\bf S}_{1}^{(1)}(k_{\perp},\chi)\otimes \mathcal{B}(k_{\perp})$ in Eq.\ref{eq:qmedscale} as a function of the Glauber transverse momentum $k_{\perp}$. For all values of $\chi$,  $m_D=0.8$ GeV. The yellow band shows the region dominated by non-perturbative physics. }
\label{fig:integrand}
\end{figure}
So far we have not precisely defined  the scale $Q_{\text{med}}$ which can be thought of as an intrinsic transverse momentum scale, expected to emerge through jet and medium interaction. For this, let us consider the regime $Q\sqrt{\chi} \gg Q_{\text{med}}$, where the cross section considering only a single subjet can be written as 
\small
\bea 
\Sigma(\chi) =  \int dx x^2 H_i(x,\mu) \Big[J_i^{(0)}(\chi,\mu) + \mathcal{J}_{i \rightarrow 1}(\theta_c, xQ,\mu) \left(L \int d^2k_{\perp}{\bf{S}}^{(1)}_{1}(\chi,\omega,k_{\perp},\nu)\mathcal{B}(k_{\perp},\mu,\nu) + ...\right)\Big],
\label{eq:qmedscale}
\eea
\normalsize
where the ellipses $. ..$ indicate higher order terms in the number of interactions with the medium.
If the medium is dilute, we can truncate this series to just a single interaction term. In that case the typical momentum transfer in a single kick is of the order of Debye mass, i.e, $k_{\perp}\sim m_D$. This is due to the the Glauber propagator $\sim 1/(k_{\perp}^2+ m_D^2)$ which sets the scale for the exchange momentum.  Note that in current colliders, the scale $m_D$ is a non-perturbative scale. Thus, in this scenario we expect $Q_{\text{med}}\! \sim\! m_D\! \sim\! T$. This can be observed from the $k_{\perp}$ dependence of the integrand ${\bf{S}}^{(1)}_{1}(\chi,\omega,k_{\perp})\mathcal{B}(k_{\perp})$ which we plot as a function of $k_{\perp}$ in Fig.\ref{fig:integrand}. Therefore we conclude that for a dilute medium, the non-perturbative physics, is not universal, i.e., it depends on both the properties of the medium and jet dynamics through measurements imposed on the collinear soft function ${\bf S}_1$. This also holds when we include BFKL resummation as shown in Fig.\ref{fig:bfkl}. 

For a dense medium, jet quenching mechanism is dominated by multiple scatterings~\cite{Baier:1996kr} which appears as higher order terms in Eq.~\ref{eq:qmedscale}. In that case, the jet partons can accrue larger transverse momentum which can be a perturbative scale. In literature \cite{Kumar:2020wvb}, this scale is parameterized by the jet quenching parameter $\hat q$ which is defined as the average transverse momentum squared  gained per unit medium length. Therefore, we expect that the scale $Q_{\text{med}} \sim \sqrt{\hat q L}$.

Our goal is to be able to derive an explicit operator definition for this parameter within our EFT framework.
In this paper, we develop the necessary tools to compute the result for all orders in jet-medium interactions through Eq. \ref{eq:JSFact}. 
At leading order, we therefore expect to recover an equivalent of the BDMPS-Z result, an exercise which we will carry out in an upcoming paper. The EFT will then enable us to go beyond BDMPS-Z formalism by employing RG evolution through the BFKL equation. This will allow us to examine the possible emergence of a $Q_{\text{med}}$ scale as the typical value of $k_{\perp}$ transferred to the jet, as well as its modification due to radiative corrections. 

We can then outline the next steps required for a phenomenological analysis of this observable. Once a \textit{perturbative} emergent scale $Q_{\text{med}}$ has been identified for multiple interactions, if it is significantly larger than the medium temperature $T$ further matching is necessary to separate the scale $Q_{\text{med}}$ from the scale $T \sim m_D$. This will fully isolate the non-perturbative physics, completing our factorization. At that stage, we will be able to compare with data and extract the universal non-perturbative structure function of the medium.

\section{Summary and outlook}
\label{sec:sum}
In this paper, we develop a comprehensive Effective Field Theory framework for the two point energy correlators in the collinear limit for HICs. Starting from the QCD action, we systematically match to EFTs at lower virtuality separating out the physics at widely separated scales. We examine two distinct hierarchies for the observable and formulate a factorization expression for each regime, valid to all orders in perturbation theory for the complete jet observable. We incorporate both vacuum and medium dynamics, providing gauge invariant operator definitions for the factorized functions. Further, we compute these functions to LO and also provide the corresponding anomalous dimensions, which incorporate higher order radiative corrections that can be resummed systematically. In particular, we recover previous LO results in literature, such as single scattering limit in GLV formalism within our framework and observe that the radiative corrections leads to a BFKL evolution. Below we briefly summarize the results in this paper.

For the region $Q\sqrt{\chi} \sim Q_{\text{med}}$ we first present a factorized expression including multiple scattering dynamics for differential cross-section of two-point energy correlators. With this, at LO we recover vacuum jet function shown in Eq.\ref{eq:vacjet} along with the corresponding anomalous dimensions. Further, for medium induced radiation we provide an operator definition for medium modified jet function within the  single scattering scenario along with the one loop anomalous dimension given in Eq.\ref{eq:JRG} which is inferred to obey both BFKL and DGLAP evolution by RG consistency. In the factorized approach we further provide the operator definition for the medium correlators, necessary to compute differential distribution for the observable. These correlators encodes the measurement independent universal physics of the medium and only depends on the medium parameters such as length and temperature. We discuss the corresponding anomalous dimensions in Eq.\ref{eq:BRG}. Finally, we perform the one loop calculation for the single interaction medium modified jet function and recover full GLV results for the quark jet function. This leading order result sets the initial conditions for the resummations of DGLAP and BFKL logarithms that account for higher order radiative corrections.


We perform a similar analysis for the case when the scales are widely separated, i.e., $Q\sqrt{\chi} \gg Q_{\text{med}}$ (see Section \ref{sec:factwo}). Given the substantial scale separation, we refactorize the jet function and perform an additional matching by integrating out physics at high virtuality, i.e., $Q\sqrt{\chi}$, through the matching coefficients. The medium induced radiation are now incorporated into the collinear-soft function defined in Eq.~\ref{eq:collsoft}. Moreover, in this step the medium correlators remains same as that of in $Q\sqrt{\chi} \sim Q_{\text{med}}$ case.   
Next,  we provide a complete factorized expression for all order of Glauber interactions with the medium, expressed in terms of multi-pronged collinear soft functions that includes all order radiations.  Using this we further compute medium induced NLO collinear-soft function, which matches the soft limit of GLV result. In Figure~\ref{fig:bfkl}, we demonstrate that for single scattering the dominant contribution to the medium induced jet function arises from the non-perturbative regime. Since the medium function remains unchanged, we infer that the collinear-soft function also obeys the BFKL evolution equation. Finally, in Figure~\ref{fig:bfkl}(right), we show for the first time the impact of BFKL resummation on two-point energy correlator distributions and conclude that in the small $\chi$ region the resummation enhances the distribution. Our this result goes beyond the current leading order results that exist in the literature. Moreover, in Section \ref{sec:BDMPS-Z}, we sketch the procedure through which an equivalent of the BDMPS-Z result can be obtained in our EFT framework and leave a detailed calculation for the future. 


Our framework allows to improve the description of jets in HICs in the following ways,
\begin{itemize}
\item The systematic treatment of EFTs allows us to separate out the perturbative physics from the non-perturbative to all orders in $\alpha_s$ and probe the universality of the strongly coupled medium.
This is crucial in order to have predictive power across different observables.
We conclude that for a dilute medium the non-perturbative physics is \textit{not} universal across different jet observables.
For a dense medium, we sketch the procedure that would allow us to recover at leading order the BDMPS-Z result and identify an emergent medium induced scale related to the jet quenching parameter $\hat q$. If this scale is sufficiently separated from the medium temperature, then it would require us to do another step of matching to completely isolate the non-perturbative physics at the scale of the temperature. In that case it might be possible to have an all order factorization with a universal non-perturbative component that can be defined and extracted from experiment/computed on the lattice. This possibility will be explored in a future work.
\item{} The EFT framework also allows us to systematically go beyond current results for the factorized functions defined at the perturbative scale. For instance, with the current factorization procedure, we can see that by demanding consistency of Renormalization Group evolution, we can infer that the radiative corrections to the GLV result give BFKL evolution. This is a generic feature of factorization that we can exploit; generally one of the functions in the factorization formula will be easier to compute than the other, for instance, we see a BFKL log in the medium function at one loop, while to do the same in the jet function requires us to go to two loops. A clear  separation of scales enables us to resum large logarithms systematically by performing a renormalization group running between functions that that depend on well-separated scales. The decoupling of the factorized functions also allows us to go to higher orders in perturbation theory by computing each function \textit{independently} as a series in $\alpha_s$.
\end{itemize}

To apply our framework to the phenomenology of jets in heavy-ion collisions (HICs), it will be necessary to extend our current approach for dense media which will be reported in the future publications.  In this paper we have limited our analysis to single scattering regime with a one prong configuration. Hence, in this setup another key step is to compute the collinear-soft function for multi-prong jets which will enable us to incorporate color coherence dynamics in a self-consistent manner. Additionally, this will also allow us to determine the matching co-efficients for the collinear-soft functions. 

An important extension of this EFT framework, especially, relevant for ongoing sPHENIX experiment is to incorporate heavy quark jets in this formalism. This can be addressed in a similar manner as done in Refs.~\cite{Makris:2018npl,Lee:2019lge} through a hybrid SCET- Heavy Quark Effective Theory(HQET) framework. Once the medium dynamics are well understood within a factorized framework, adapting this approach to heavy quark jets will be a natural extension.

\begin{acknowledgments}
We thank Yacine-Mehtar Tani and Felix Ringer for valuable discussions. V.V and B.S are supported by startup funds from the University of South Dakota.
\end{acknowledgments}

\appendix

\section{Wightman correlator in thermal medium}\label{sec:med}
Now we discuss the medium function that gets contribution from both thermal quarks and gluons. We first consider the case where glauber modes are generated by gluons in the thermal medium. We will use imaginary time formalism to compute Wightman correlator. In terms of SCET operators, the correlator in momentum space is given as
\begin{equation}
D^{AB}_{E}(K)=\int_0^\beta d\tau \int d^3x\, e^{iK\cdot X} \langle\frac{1}{\mathcal{P}_{\perp}^2} O_s^{g_nA}(X)\frac{1}{\mathcal{P}_{\perp}^2}O_s^{g_nB}(0)\rangle, 
\label{eq:wightman}
\end{equation}
where $\beta=1/T$ and $X(K)$ are position (momentum) vectors in  Euclidean space. Position vector $X=(\tau=it,\vec{x})$ and momentum vector $K=(k_m,\vec{k})$ with $K\cdot X=k_m\tau+\vec{k}\cdot\vec{x}$. Moreover, $k_m=2m\pi T$ is bosonic Matsubara frequency where $m$ is an integer.  The soft gluon operator $O_s^{gA}$ is given by
\begin{eqnarray}
O_s^{g_nB}=8\pi\alpha_s \left[\frac{i}{2}f^{BCD}\mathcal{B}_{s\perp,\mu}^{nC}\frac{n}{2}\cdot (\mathcal{P}+\mathcal{P}^{\dagger})\mathcal{B}_{s\perp}^{nD\mu} \right],    
\end{eqnarray}
where the superscripts $n$ represents that the soft gluon operators are dressed with soft Wilson lines that depends on the direction of collinear parton. Thus the Wilson lines make the operators gauge invariant. The operator $\mathcal{B}^n_{s\perp,\,\mu}$ is defined as
\begin{equation}
\mathcal{B}^n_{s\perp,\,\mu}=\frac{1}{g}[S_n^\dagger i D_{s\perp}^{\mu}S_n].  \end{equation}
Plugging the soft gluon operator back in Eq.~\ref{eq:wightman}, we obtain
\begin{align}
D_E^{AB}=&-\frac{(8\pi\alpha_s)^2}{4}\int_0^{\beta}d\tau \int d^3x \,T\sum_{m}T\sum_{n}\int\frac{d^3p}{(2\pi)^3}\int\frac{d^3q}{(2\pi)^3} e^{(K-P+Q)\cdot X}\nonumber\\
&\frac{1}{(\vec{p}_{\perp}-\vec{q}_{\perp})^4}\frac{e^{-ip_n\tau}}{p_n^2+E_2^2}\frac{e^{iq_m\tau}}{q_m^2+E_1^2}\tr[f^{ACD}f^{BC'D'}T^CT^DT^{C'}T^{D'}],
\end{align}
where $E_2=|\vec{p}|$ and $E_1=|\vec{q}|$ and trace is over colors. Now we can perform $x$ integration which gives $\delta^3(\vec{k}-\vec{p}+\vec{q})$. Using this delta function to perform $p$ integration, we get
\begin{align}
D_E^{AB}=&\frac{(8\pi\alpha_s)^2}{k_{\perp}^4}\frac{N_c^2}{32}\int_0^{\beta}e^{k_0\tau}d\tau \,\int\frac{d^3q}{(2\pi)^3} T\sum_{n}\frac{e^{-ip_n\tau}}{p_n^2+E_2^2}T\sum_{m}\frac{e^{iq_m\tau}}{q_m^2+E_1^2},    
\end{align}
where $p_n=2n\pi T$ and $q_m=2m\pi T$ are bosonic Matsubara frequencies. Now we can simplify above equation by summing over Matsubara frequencies 
\begin{equation}
 T\sum_{n}\frac{e^{-ip_n\tau}}{p_n^2+E_2^2}=\frac{1}{2E_2}\left[(1+f(E_2))e^{-E_2\tau}+f(E_2)e^{E_2\tau} \right],   
\end{equation}
where $f(E)=(e^{\beta E}-1)^{-1}$ is Bose-Einstein distribution function. With the frequency summation we can perform $\tau$ integration to obtain Euclidean correlator. Finally, with the Euclidean correlator we can compute spectral function via analytic continuation
\begin{equation}
\rho_{AB}(k)=-i(D_E^{AB}(-i(k_0+i0^{+}),\vec{k})-D_E^{AB}(-i(k_0-i0^{+}),\vec{k})).
\label{eq:spec}
\end{equation}
Plugging the Euclidean correlator in Eq.\ref{eq:spec}, we obtain
\begin{align}
\rho_{AB}(k)=&\frac{(8\pi\alpha_s)^2}{k_{\perp}^4}\frac{N_c^2}{32}\int\frac{d^3q}{(2\pi)^3}\frac{(n\cdot q)(n\cdot (q-k))}{4E_1E_2} \bigg[(f(E_1)-f(E_2))[\delta(k_0+E_1-E_2)\nonumber\\
&-\delta(k_0+E_2-E_1)]+(1+f(E_1)+f(E_2))[\delta(k_0-E_1-E_2)-\delta(k_0+E_1+E_2)]\bigg].    
\end{align}
In the above equation, the four delta functions represent four different processes. The first two delta functions correspond to scattering processes that include incoming soft gluon and outgoing soft gluon. Morevoer, the last two delta functions represent the scattering processes that involve either two incoming soft gluon or two outgoing soft gluon. In our case only first two delta functions contribute at leading order. 
Now with the spectral function we can compute the function $S_{AB}(k)$ which is Wightman correlator in real time and is given as
\begin{align}
D_{>}(k)=&(1+f(k_0))\rho_{AB}(k).
\end{align}
Simplifying the above equation we get
\begin{align}
D^g_{>}(k)=&\frac{1}{k_{\perp}^4}\int \frac{d^3q}{(2\pi)^3}\frac{(n\cdot q)(n\cdot (q-k))}{4E_1E_2}\bigg[f(E_1)(1+f(E_2))\delta(k_0+E_1-E_2)\nonumber\\
&-f(E_2)(1+f(E_2))\delta(k_0+E_2-E_1)\bigg], 
\label{eq:wightcorr}
\end{align}
where the term $1+f(E)$ represents Bose-Einstein enhancement factor. To perform the integration and systematically do power counting we need to express everything in light-cone coordinates. To this end we introduce a new integration variable $l$ and rewrite Eq.~\ref{eq:wightcorr} as
\begin{align}
D^g_{>}(k)=&\frac{(8\pi\alpha_s)^2}{k_{\perp}^4}\frac{4\pi N_c^2}{32}\int \frac{d^4q}{(2\pi)^3}\int d^4l \delta(l^2)\delta(q^2)f(q_0)(1+f(l_0))\delta^4(k+q-l)(n\cdot q)(n\cdot(q-k)) \nonumber\\
&=\frac{(8\pi\alpha_s)^2}{k_{\perp}^4}\frac{4\pi N_c^2}{32}\mathcal{I}^g(k).
\end{align}
After performing $l$ integration using the delta function, we get
\begin{align}
\mathcal{I}^g(k)=&\int \frac{dq^{+}dq^{-}d^2q_{\perp}}{(2\pi)^3}\delta(q^+q^--q_{\perp}^2)\delta((k+q)^+(k+q)^--(\vec{k}_{\perp}+\vec{q}_{\perp})^2)f\left(\frac{q^++q^-}{2}\right)\nonumber\\
&(1+f\left(\frac{k^++q^++k^-+q^-}{2}\right))(n\cdot q)(n\cdot(q-k)).    
\end{align}
Since $q$ scales as $(\lambda,\lambda,\lambda)$ we can drop $k^+$ component and after performing $q^+$ integration we get
\begin{align}
\mathcal{I}^g(k^-,k_{\perp})=&\int\frac{dq^{-}d^2q_{\perp}}{(2\pi)^3}\frac{1}{q^-}\delta\left(\frac{q_{\perp}^2}{q^-}(k^-+q^-)-(\vec{k}_{\perp}+\vec{q}_{\perp})^2 \right)f\left(\frac{q^-}{2}+\frac{q_{\perp}^2}{2q^{-}} \right) \nonumber\\
&\bigg[1+f\left(\frac{k^-+q^-}{2}+\frac{q_{\perp}^2}{2q^-} \right) \bigg]\frac{q_{\perp}^4}{(q^-)^2}.
\end{align}
For EEC jet function we also need to integrate over $k^-$ which now we can do using the delta function to get
\begin{align}
\mathcal{I}^g(k_{\perp})=&\frac{1}{2\pi}\int\frac{dq^-d^2q_{\perp}}{(2\pi)^3}\frac{1}{q_{\perp}^2}f\left(\frac{q^-}{2}+\frac{q_{\perp}^2}{2q^{-}} \right)\bigg[1+f\left(\frac{k^-+q^-}{2}+\frac{q_{\perp}^2}{2q^-} \right)\bigg]\frac{q_{\perp}^4}{(q^-)^2},    
\end{align}
where
\begin{equation}
k^-=-q^-+\frac{q^-(\vec{k}_{\perp}+\vec{q}_{\perp})^2}{q_{\perp}^2}.    
\end{equation}
Similarly for quark operators in the thermal medium the Wightman correlator is
\begin{equation}
D_{>}^q(k_{\perp})=\frac{(8\pi\alpha_s)^2 2\pi}{k_{\perp}^4} \mathcal{I}^q(k_{\perp})   
\end{equation}
where the function $\mathcal{I}^q(k_{\perp})$ is 
\begin{align}
\mathcal{I}^q(k_{\perp})&=\frac{1}{2\pi}\int\frac{dq^-d^2q_{\perp}}{(2\pi)^3}\frac{q_{\perp}^2}{(q^-)^2}\tilde{f}\left(\frac{q^-}{2}+\frac{q_{\perp}^2}{2q^{-}} \right)\bigg[1-\tilde{f}\left(\frac{k^-+q^-}{2}+\frac{q_{\perp}^2}{2q^-} \right)\bigg].   
\end{align}

\section{Feynman diagrams for medium jet function}\label{sec:feynjetmed}

\subsection{Real diagrams with insertions on the opposite side of the cut}\label{sec:realop}

Here we list out a complete set of the real and the virtual diagrams for a quark initiated jet that are required to compute EEC.  These diagrams can be systematically generated through the evolution operator by expanding it order by order in the interaction and Glauber Hamiltonian. We only consider the diagrams leading to finite matrix elements.  We first provide expressions for matrix elements with Glauber insertions on the opposite side of the cut. Let us first consider diagrams with collinear gluon emission from collinear quark. These diagrams arise from the insertions of collinear interaction terms in the evolution operator. In this case there are total six diagrams one of which we discuss one by one below. 

First we evaluate the diagram shown in Figure~\ref{fig:lag1}. Using the Feynman rules for collinear-collinear and Glauber-collinear interactions we get 
\begin{align}
&B3=-4g^2N_c\frac{\delta_{AB}}{2}\int \pphase\int \qphase \delta(\omega-p^{-}-q^{-})\delta^2(\vec{p}_{\perp}+\vec{q}_{\perp}-\vec{k}_{\perp})\nonumber\\
&\times \Big(\frac{\vec{q}_{\perp}\cdot (\vec{q}_{\perp}-\vec{k}_{\perp})}{q^{-}}+\frac{\vec{q}_{\perp}\cdot (\vec{q}_{\perp}-\vec{k}_{\perp})}{p^{-}}\Big)\int\frac{dl^{+}}{2\pi}e^{-iL/2(l^{+}-q^{+})}\int\frac{dr^{+}}{2\pi}e^{-iL/2(p^{+}-r^{+})}\nonumber\\
&\times \Big[\frac{p^{-}}{l^{+}+p^{+}+i\epsilon}\frac{1}{r^{+}-\frac{q_{\perp}^2}{p^{-}}-i\epsilon}\frac{1}{r^{+}+q^{+}-i\epsilon}\frac{1}{l^{+}q^{-}-(\vec{q}_{\perp}-\vec{k}_{\perp})^2+i\epsilon}\Big]\sinc\Big[\frac{L}{2}(l^{+}+p^{+}-q^{+}-r^{+}) \Big].
\label{realo1}
\end{align}
where $\omega$ is the energy of quark coming from the hard vertex which is same as $xQ$. The last two delta functions represents energy and transverse momentum conservation. First we perform $q^+$ and $p^+$ integrations using on-shell delta functions for each and then $p^-$ integration using the energy delta function. Next we perform the contour integration for $l^{+}$ and $r^{+}$ integrations. Note that both $l_{+}$ and $r_{+}$ have two poles in lower and upper half planes we therefore get total four contributions from these integrations. 
Performing $r_{+},l_{+}$ contour integration and simplifying phase space terms, contribution to the jet function can be written in a simple form that reads as

\begin{align}
\begin{minipage}{3cm}\includegraphics[width=\textwidth]{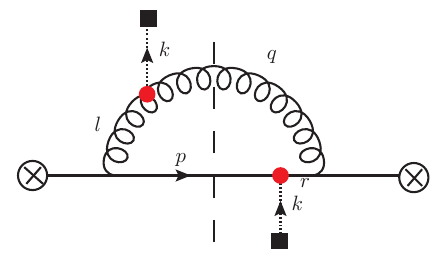} \end{minipage}+c.c=&-4g^2N_c\delta_{AB}\int \frac{dq^-}{(2\pi)^3} \int  \frac{d^2q_{\perp}}{q_{\perp}^2} \frac{(p^-)^2}{\omega^2{\kappa}_{\perp}^2}\mathcal{F}(q^{-},\vec{q}_{\perp},k_{\perp})\nonumber\\
&\Big[1-F\Big(\frac{L{\kappa}_{\perp}^2\omega}{2p^-q^-}\Big)-F\Big(\frac{L{q}_{\perp}^2\omega}{2p^-q^-}\Big)+F\Big(\frac{L\omega(q_{\perp}^2-\kappa_{\perp}^2)}{2p^-q^-}\Big)\Big]\mathcal{M}_{RO},
\label{fig:lag1}
\end{align}
where $p^{-}=\omega-q^{-}$ and
\begin{equation}
\mathcal{F}(q^{-},\vec{q}_{\perp},k_{\perp})=\frac{\vec{q}_{\perp}\cdot \vec{\kappa}_{\perp}}{q^-}+\frac{\vec{q}_{\perp}\cdot\vec{\kappa}_{\perp}}{p^-}+\frac{(\vec{q}_{\perp}\cdot\vec{\kappa}_{\perp})q^-}{2(p^-)^2},  \end{equation}
and the function $F$ is 
\begin{equation}
F(y)=\cos(y)\sinc(y).    
\end{equation}
For shorthand notation we define $\vec{\kappa}_{\perp}=\vec{q}_{\perp}-\vec{k}_{\perp}$. Further $\mathcal{M}_{RO}$ is measurement function for real diagrams appearing in Glauber insertions on the opposite side of the cut which for small angles defined as 
\bea
\mathcal{M}_{RO} = \frac{(p^-)^2+ (q^-)^2}{(xQ)^2} \delta(\chi)+\frac{2 p^-q^-}{(xQ)^2}\delta \Big( \chi- \Big(\frac{\vec{q}_{\perp}}{q^-}+\frac{\vec{\kappa}_{\perp}}{p^-}\Big)^2 \Big).
\eea
The first  two terms in the measurement represents the case when EEC is measured on the same parton leading to $\delta(\chi)$ contribution. The last term denotes the case when EEC is measured on both the final state partons separated by some angle.

Next we consider the diagram where Glauber insertion is on collinear gluon at one side and on collinear quark on the other side shown in Figure~\ref{eq:realo2}. Following the same prescription as the in the previous one the matrix elements  including the complex conjugate reads as 
\begin{align}
\begin{minipage}{3cm}\includegraphics[width=\textwidth]{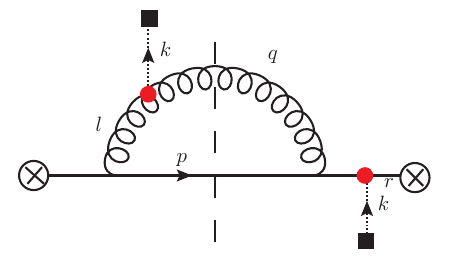} \end{minipage}+c.c=& -4g^2N_c \delta_{CD}\int \frac{dq^{-}}{(2\pi)^3} \int  d^2q_{\perp}\Big(1-F\Big(\frac{L\kappa_{\perp}^2\omega}{2p^{-}q^{-}} \Big) \Big)\nonumber\\
&\times\frac{(p^{-})^2\mathcal{A}(q^{-},\vec{q}_{\perp},\vec{k}_{\perp})}{\kappa_{\perp}^2[\omega(\kappa_{\perp}^2q^{-}+q_{\perp}^2p^{-})-k_{\perp}^2p^{-}q^{-}]}\mathcal{M}_{RO},
\label{eq:realo2}
\end{align}
where the function $\mathcal{A}$ is 
\begin{equation}
\mathcal{A}(q^{-},\vec{q}_{\perp},k_{\perp})=  \frac{\vec{q}_{\perp}\cdot\vec{\kappa}_{\perp}}{q^{-}}+\frac{\kappa_{\perp}^2+\vec{\kappa}\cdot\vec{q}_{\perp}}{2p^-}+\frac{\vec{k}_{\perp}\cdot\vec{\kappa}_{\perp}}{2\omega}+\frac{\kappa_{\perp}^2q^-}{2(p^-)^2}.  
\end{equation}
Note that in the soft limit, i.e., $q^-\to 0$, the dominant contribution comes from the first term in the above equation. However, for stage I factorization of EEC we need to keep subleading terms as well.

Next we consider the diagrams where both side the Glauber insertions are on the collinear gluon shown in Figure~\ref{fig:feyn3}. Note that this diagram has no conjugate term. Thus, the matrix element is given as
\begin{align}
\begin{minipage}{3.5cm}\includegraphics[width=\textwidth]{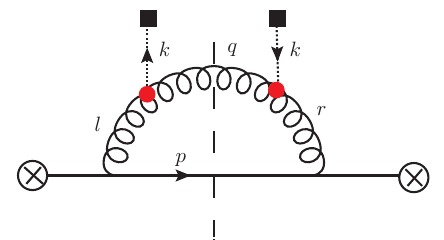} \end{minipage}=&\, 8g^2N_c\delta_{CD}\int \frac{dq^{-}}{(2\pi)^3} \int  d^2q_{\perp}\frac{(p^{-})^2}{q^{-}\kappa_{\perp}^2\omega^2}\Big(1+\frac{q^{-}}{p^{-}}+\frac{(q^-)^2}{2(p^-)^2} \Big)\nonumber\\
&\times\Big(1-F\Big(\frac{L\kappa_{\perp}^2\omega}{2p^{-}q^{-}} \Big) \Big)\mathcal{M}_{RO},
\label{fig:feyn3}
\end{align}
where the second and the third terms are subleading contributions. It is worth mentioning that the sum of above three diagrams does not contain any UV divergence. Finally, adding all these three diagrams in $L\to \infty$ limit we get
\begin{align}
\text{B}2+\text{B}6++\text{B}8&=\frac{(N_c^2-1)\bar{\alpha}}{\pi^2}\int dq^{-}\int d^2q_{\perp}\bigg[\frac{(p^{-})^2}{q^{-}\kappa_{\perp}^2\omega^2}\left(2-\frac{\vec{q}_{\perp}\cdot\vec{\kappa}_{\perp}}{q_{\perp}^2}\right)\bigg(1+\frac{q^{-}}{p^{-}}+\frac{(q^{-})^2}{2(p^{-})^2} \bigg)\nonumber\\
&-\frac{(p^{-})^2\mathcal{A}(q^{-},\vec{q}_{\perp},\vec{k}_{\perp})}{\kappa_{\perp}^2[\omega\kappa_{\perp}^2q^{-}+q_{\perp}^2p^{-}\omega-k_{\perp}^2p^{-}q^{-}]} \bigg]\mathcal{M}_{RO}. 
\end{align}
Here we have summed over color indices and also included $1/2N_c$ factor present in the jet function definition. Next we consider the diagrams with collinear gluon emission from interaction insertion and both the Glauber insertions are on the collinear quark. In this case there are three diagrams and the corresponding matrix elements including the complex conjugate ones are 
\begin{align}
\begin{minipage}{2.5cm}\includegraphics[width=\textwidth]{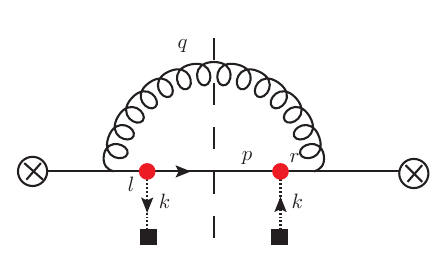} \end{minipage}=&4g^2C_F\delta_{CD}\int\frac{dq^-}{(2\pi)^3}\int d^2q_{\perp}\frac{q^-}{q_{\perp}^2\omega^2}\Big(1-F\Big(\frac{Lq_{\perp}^2\omega}{2p^-q^-} \Big) \Big)\mathcal{M}_{RO},\\
\begin{minipage}{2.5cm}\includegraphics[width=\textwidth]{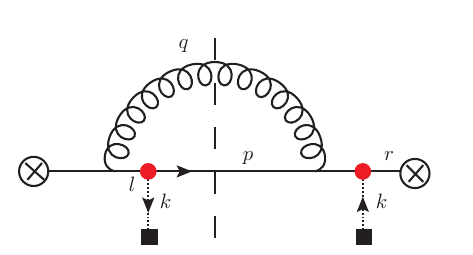} \end{minipage}+{\rm c.c}=&\,- 4g^2\left(C_F-\frac{N_c}{2}\right)\delta_{CD}\int \frac{dq^{-}}{(2\pi)^3} \int d^2q_{\perp}\frac{(p^{-})^2q^{-}}{q_{\perp}^2[\omega(q_{\perp}^2p^{-}+\kappa_{\perp}^2q^{-})-k_{\perp}^2p^{-}q^{-}]}\nonumber\\
&\Big(\frac{\vec{q}\cdot\vec{\kappa}_{\perp}}{(p^-)^2}+\frac{\vec{k}_{\perp}\cdot\vec{\kappa}_{\perp}}{p^-\omega}\Big)\Big[1-F\Big(\frac{Lq_{\perp}^2\omega}{2p^{-}q^{-}}\Big) \Big]\mathcal{M}_{RO},\\
\begin{minipage}{2.5cm}\includegraphics[width=\textwidth]{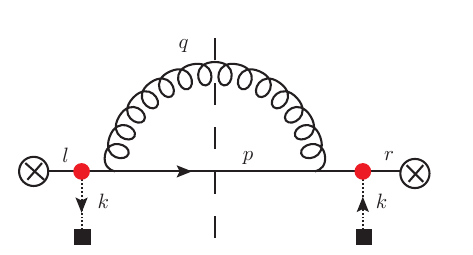} \end{minipage}=&\,-2g^2C_F\delta_{CD}\int \frac{dq^{-}}{(2\pi)^3} \int d^2q_{\perp} \frac{(p^{-})^2q^{-}\omega^2}{[\omega(\kappa_{\perp}^2q^{-}+q_{\perp}^2p^{-})-k_{\perp}^2p^{-}q^{-}]^2}\nonumber\\
&\bigg(\frac{k_{\perp}^2}{\omega^2}+\frac{\kappa_{\perp}^2}{(p^-)^2}+\frac{\vec{k}_{\perp}\cdot\vec{\kappa}_{\perp}}{p^-\omega}\bigg)\mathcal{M}_{RO}, 
\end{align}
where $C_F=4/3$. Note that all the above diagrams are subleading and the corresponding contribution vanishes in the soft limit. In the limit $L\to\infty$ the total contribution from the above diagrams is
\begin{align}
\text{B}10+\text{B}11+\text{B}12&=\frac{(N_c^2-1)\bar{\alpha}}{N_c\pi^2}\int dq^{-}\int d^2q_{\perp}\Big[\frac{q^{-}C_F}{q_{\perp}^2\omega^2}-\Big(C_F-\frac{N_c}{2}\Big)\Big(\frac{\vec{q_{\perp}}\cdot\vec{\kappa}_{\perp}}{(p^{-})^2}+\frac{\vec{k}_{\perp}\cdot\vec{\kappa}_{\perp}}{p^{-}\omega}\Big)\nonumber\\
&-\frac{C_F}{2}\frac{(p^{-})^2q^{-}\omega^2}{[\omega(\kappa_{\perp}^2q^{-}+q_{\perp}^2p^{-})-k_{\perp}^2p^{-}q^{-}]^2}\bigg(\frac{k_{\perp}^2}{\omega^2}+\frac{\kappa^2}{(p^-)^2}+\frac{\vec{k}_{\perp}\cdot\vec{\kappa}_{\perp}}{p^-\omega}\bigg)\Big]\mathcal{M}_{RO}.    
\end{align}
Next we consider the diagrams with Wilson lines originating from hard vertex. Diagrams with Glauber insertions on the collinear gluon vanishes due to $n^2=\bar{n}^2=0$. Therefore, the finite contributions come from Glauber insertions on quark only. The matrix elements of the non vanishing diagrams along with the mirror diagrams is
\begin{align}
\begin{minipage}{3cm}\includegraphics[width=\textwidth]{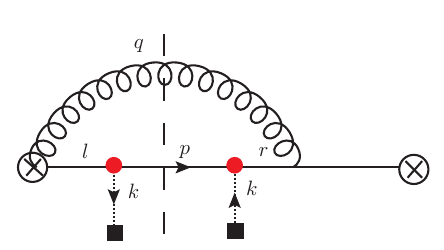} \end{minipage}+{\rm c.c}=&\, 4g^2C_F\delta_{CD}\int \frac{dq_{-}}{(2\pi)^3} \int  d^2q_{\perp} \frac{p^{-}}{q^{-} q_{\perp}^2\omega}\bigg(1-F\bigg(\frac{Lq_{\perp}^2\omega}{2p^{-}q^{-}} \bigg)\bigg)\mathcal{M}_{RO}\label{eq:hardwilson},\,\\
\begin{minipage}{3cm}\includegraphics[width=\textwidth]{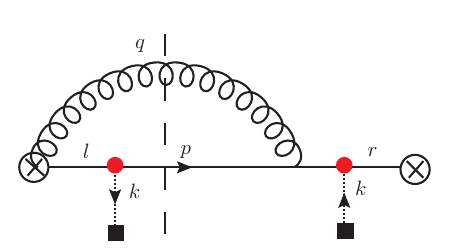} \end{minipage}+{\rm c.c}=&\, 4g^2\bigg(C_F-\frac{N_c}{2}\bigg)\delta_{CD}\int \frac{dq^{-}}{(2\pi)^3} \int  d^2q_{\perp} \frac{p^{-}}{q^{-}[\omega(\kappa_{\perp}^2q^{-}+q_{\perp}^2p^{-})-k_{\perp}^2p^-q^-]}\nonumber\\
&F\bigg(\frac{Lq_{\perp}^2\omega}{2 p^{-}q^{-}} \bigg)\mathcal{M}_{RO}.
\end{align}
Note that only Eq.~\ref{eq:hardwilson} contributes in $L\to \infty$ limit. Finally, we consider the diagrams wit Wilson lines originating from the Glauber vertex shown in Figures~\ref{fig:feyn9} and \ref{fig:feyn10}.  The correaponding matrix elements including the complex conjugate ones is  
\begin{align}
\begin{minipage}{3cm}\includegraphics[width=\textwidth]{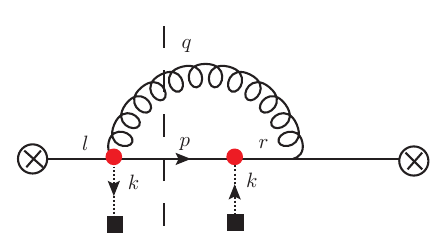} \end{minipage}+{\rm c.c}=& 2g^2N_c\delta_{CD}\int \frac{dq^{-}}{(2\pi)^3}\int d^2q_{\perp}\frac{p^{-}}{q_{\perp}^2q^{-}\omega}\bigg(1-F\left(\frac{Lq^2_{\perp}\omega}{2p^{-}q^{-}} \right), \bigg)\mathcal{M}_{RO}\label{fig:feyn9}\\
\begin{minipage}{3cm}\includegraphics[width=\textwidth]{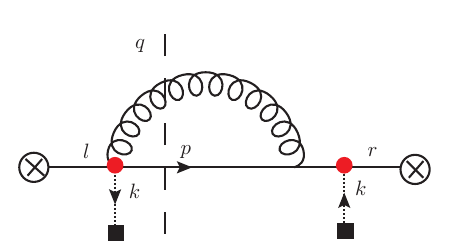} \end{minipage}+{\rm c.c}=&2g^2 N_c\delta_{CD}\int \frac{dq^{-}}{(2\pi)^3}\int d^2q_{\perp}\bigg[\frac{p^{-}\omega}{q^{-}[\omega(\kappa_{\perp}^2q^{-}+q_{\perp}^2p^{-})-k_{\perp}^2p^{-}q^{-}]}\bigg]\mathcal{M}_{RO}.\,
\label{fig:feyn10}
\end{align}
In the $L\to \infty$ limit the contribution from the above two diagrams is
\begin{equation}
\text{B}16+\text{B}17=\frac{N_c(N_c^2-1)\bar{\alpha}}{2\pi^2}\int dq^{-}\int d2q_{\perp}\bigg[\frac{p^-}{q_{\perp}^2q^-\omega}+\frac{p^{-}\omega}{q^{-}[\omega(\kappa_{\perp}^2q^{-}+q_{\perp}^2p^{-})-k_{\perp}^2p^{-}q^{-}]} \bigg]\mathcal{M}_{RO}.   
\end{equation}
The net contribution to the jet function from the above diagrams in the soft and $L \rightarrow \infty $ limit is 
\bea 
J_{RR} &=& \frac{C_F\bar{\alpha}}{2\pi^2} \int \frac{dq^-}{q^-}\int d^2q_{\perp}\Bigg[ \frac{C_F}{q_{\perp}^2}+ \frac{N_c}{q_{\perp}^2}+ \frac{2N_c}{\kappa_{\perp}^2}-\frac{2N_c \vec{q}_{\perp}\cdot \vec{\kappa}_{\perp}}{q_{\perp}^2\kappa_{\perp}^2}\Bigg]\mathcal{M}_{RO}\nonumber\\
&= &\frac{C_F\bar{\alpha}}{2\pi^2} \int \frac{dq^-}{q^-}\int d^2q_{\perp}\Bigg[ \frac{C_F}{q_{\perp}^2}+ \frac{N_c}{\kappa_{\perp}^2}+\frac{N_c k_{\perp}^2}{q_{\perp}^2\kappa_{\perp}^2}\Bigg]\mathcal{M}_{RO}.
\eea

\subsection{Virtual diagrams with insertions on the opposite side of the cut}\label{sec:virtualop}
Now we consider the virtual diagrams with Glauber insertions on the opposite side of the cut. For this the measurement acting on the final state parton takes the form 
\bea
\mathcal{M}_{VO} = \delta(\chi).
\eea
We first consider the diagrams with collinear gluon from the  Wilson lines originating from the hard vertex. For this case there are two diagrams along with the complex conjugate. The corresponding matrix elements after adding the complex conjugate diagrams are
\begin{align}
\begin{minipage}{3cm}\includegraphics[width=\textwidth]{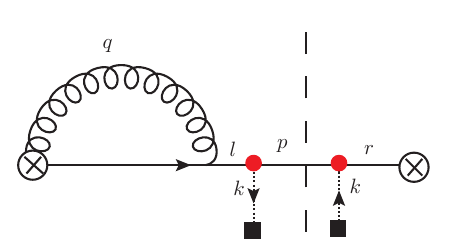} \end{minipage}+{\rm c.c.}=&\,4g^2C_F\delta_{CD}\int \frac{dq^{-}}{(2\pi)^3}\int d^2q_{\perp}\frac{p^{-}}{q_{\perp}^2q^{-}\omega}\bigg[-1+F\left(\frac{Lq_{\perp}^2\omega}{2p^{-}q^{-}} \right)\bigg]\mathcal{M}_{VO}, \\
\begin{minipage}{3cm}\includegraphics[width=\textwidth]{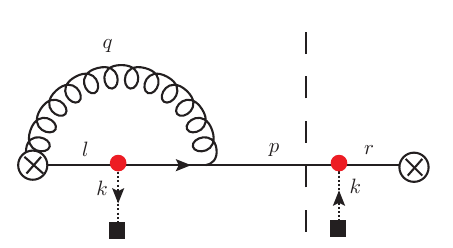} \end{minipage}+{\rm c.c.}=&\,4g^2\delta_{CD}\left(C_F-\frac{N_c}{2}\right)\int \frac{dq^{-}}{(2\pi)^3}\int d^2q_{\perp}\nonumber\\
&\times\frac{\omega p^-}{q^{-}[p^{-}q^{-}{k}_{\perp}^2-\omega(q^{-}\kappa_{\perp}^2+q_{\perp}^2p^{-})]}F\bigg(\frac{Lq_{\perp}^2\omega}{2p^{-}q^{-}} \bigg)\mathcal{M}_{VO}, \,
\end{align}
where same as previous case $\vec{\kappa}_{\perp}=\vec{q}_{\perp}-\vec{k}_{\perp}$ and $p^-=\omega-q^-$. Note that while these relations are true only for real diagrams, in order to keep the expressions compact we will use these definitions in the virtual diagrams as well. We stress that this is only to keep the expressions in compact form. The second diagram above does not contribute to the observable for infinitely large medium. 
Next we consider virtual diagrams with Wilson line originating from Glauber vertex shown in Figures~\ref{fig:feyn13} and \ref{fig:feyn14}. These diagrams give dominant contribution in the soft limit. The corresponding matrix elements including the mirror diagrams are given by
\begin{align}
\begin{minipage}{3cm}\includegraphics[width=\textwidth]{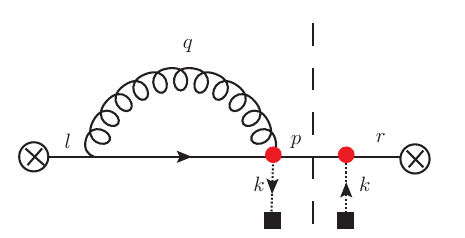} \end{minipage}+{\rm c.c.}=&\,-2g^2N_c\delta_{CD}\int \frac{dq^{-}}{(2\pi)^3}\int d^2q_{\perp}\frac{p^-}{q_{\perp}^2q^{-}\omega}\left(1-F\left(\frac{Lq_{\perp}^2\omega}{2p^{-}q^{-}} \right)  \right)\mathcal{M}_{VO},\label{fig:feyn13}  \\
\begin{minipage}{3cm}\includegraphics[width=\textwidth]{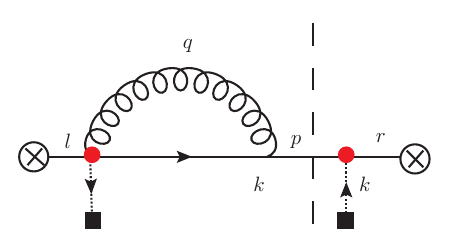} \end{minipage}+{\rm c.c.}=&\,2g^2N_c\delta_{CD}\int \frac{dq^{-}}{(2\pi)^3}\int d^2q_{\perp} \frac{p^{-}\omega}{q^{-}[k_{\perp}^2q^{-}p^{-}-\omega(q_{\perp}^2p^-+\kappa_{\perp}^2q^-)]}\mathcal{M}_{VO}.\label{fig:feyn14} 
\end{align}
In the $L\to \infty$ limit the contribution from these two diagrams are
\begin{equation}
\text{B}23+\text{B}24=\frac{(N_c^2-1)\bar{\alpha}}{2\pi^2}\int dq^{-}\int d^2q_{\perp}\bigg[-\frac{p^-}{q_{\perp}^2q^-\omega}+\frac{p^{-}\omega}{q^{-}[k_{\perp}^2q^{-}p^{-}-\omega(q_{\perp}^2p^-+\kappa_{\perp}^2q^-)]}\bigg]\mathcal{M}_{VO}   
\end{equation}
Finally, we consider diagrams with collinear gluon emission from the Lagrangian insertion and the corresponding diagram is shown in Figure~\ref{fig:feyn15}. The matrix elements including the complex conjugate reads as 
\begin{align}
\begin{minipage}{3cm}\includegraphics[width=\textwidth]{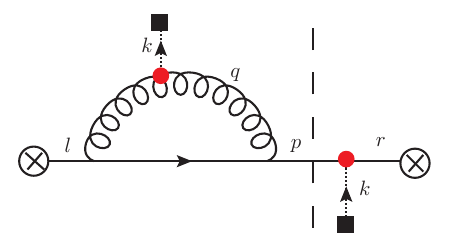} \end{minipage}+{\rm c.c.}=&4g^2 N_c\delta_{CD}\int \frac{dq^{-}}{(2\pi)^3} \int d^2q_{\perp}\bigg(1-F\bigg(\frac{L}{2}\frac{\kappa_{\perp}^2\omega}{p^{-}q^{-}} \bigg) \bigg) \nonumber\\
&\times \frac{(p^{-})^2\mathcal{B}(q^-,q_{\perp},k_{\perp})}{\kappa_{\perp}^2[\omega(q_{\perp}^2p^{-}+\kappa_{\perp}^2q_{-})-k_{\perp}^2q_{-}p^{-}]}\mathcal{M}_{VO}, 
\label{fig:feyn15}
\end{align}
where the function $\mathcal{B}$ is 
\begin{align}
\mathcal{B}(q^-,q_{\perp},k_{\perp})=&\frac{\vec{q}_{\perp}\cdot\vec{\kappa}_{\perp}}{q_{-}}+\frac{\kappa_{\perp}^2}{2p^{-}}+\frac{\vec{q}_{\perp}\cdot\vec{\kappa}_{\perp}}{2p^-}+\frac{\kappa_{\perp}^2q^-}{2(p^-)^2}.
\end{align}
In the soft limit the dominant contribution comes from the first term.

Next we consider the diagrams with collinear gluon emission from collinear insertion and both the Glauber insertions are on the quark. In this case there are three diagrams and the corresponding matrix elements are given as 
\begin{align}
\begin{minipage}{3cm}\includegraphics[width=\textwidth]{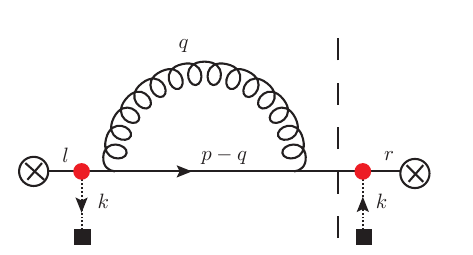} \end{minipage}+{\rm c.c.}=&-4g^2C_F\delta_{CD}\int \frac{dq^{-}}{(2\pi)^3}\int d^2q_{\perp}\left(\frac{\kappa_{\perp}^2}{(p^-)^2}+\frac{k_{\perp}^2}{\omega^2}+\frac{2\vec{k}_{\perp}\cdot\vec{\kappa}_{\perp}}{\omega p^-}\right)\nonumber\\
&\frac{q^-\omega^2(p^{-})^2}{[\omega(q_{\perp}^2p^-+\kappa_{\perp}^2q^-)\omega-k_{\perp}^2q^-p^-]^2}\mathcal{M}_{VO}, \\
\begin{minipage}{3cm}\includegraphics[width=\textwidth]{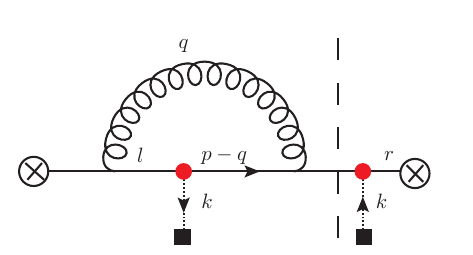} \end{minipage}+{\rm c.c.}=&4g^2\left(C_F-\frac{N_c}{2}\right)\int \frac{dq_{-}}{(2\pi)^3}\int d^2q_{\perp}\bigg(\frac{\vec{k}_{\perp}\cdot\vec{\kappa}_{\perp}}{\omega p^-}+\frac{\vec{q}_{\perp}\cdot\vec{\kappa}_{\perp}}{(p^-)^2}\bigg)\nonumber\\
&\frac{1}{q_{\perp}^2}\frac{q^-(p^{-})^2}{\omega(\kappa_{\perp}^2q^{-}+q_{\perp}^2p^{-})-k_{\perp}^2p^{-}q^{-}}\left(1-F\bigg(\frac{Lq_{\perp}^2\omega}{2q^- p^-}\bigg)\right)\mathcal{M}_{VO},  \\
\begin{minipage}{3cm}\includegraphics[width=\textwidth]{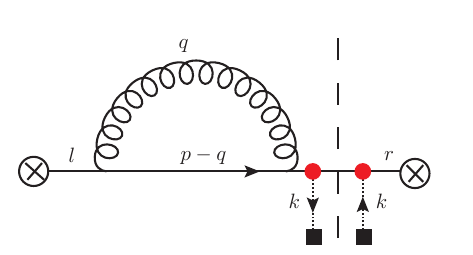} \end{minipage}+{\rm c.c.}=&-4g^2C_F\int\frac{dq^-}{(2\pi)^3}\int d^2q_{\perp}\frac{q^-}{q_{\perp}^2\omega^2}\bigg(1-F\bigg(\frac{Lq_{\perp}^2\omega}{2p^-q^-} \bigg) \bigg)\mathcal{M}_{VO}. 
\end{align}
Note that in the soft limit the contribution from the above diagrams vanishes. Again if we consider all terms in the soft limit as $L \rightarrow \infty$, we get 
\bea
J_{VO}  &=& 4g^2\delta_{CD} \int \frac{dq^-}{q^-}\int d^2q_{\perp}\Bigg[ -\frac{C_F}{q_{\perp}^2}- \frac{N_c}{q_{\perp}^2}+\frac{N_c \vec{q}_{\perp}\cdot \vec{\kappa}_{\perp}}{q_{\perp}^2\kappa_{\perp}^2}\Bigg]\mathcal{M}_{VO}\nonumber\\
&=& 4g^2\delta_{CD} \int \frac{dq^-}{q^-}\int d^2q_{\perp}\Bigg[ -\frac{C_F}{q_{\perp}^2}- \frac{N_c k_{\perp}^2}{2q_{\perp}^2\kappa_{\perp}^2}\Bigg]\mathcal{M}_{VO}.
\eea
where we have shifted $\vec{q}_{\perp} -\vec{k}_{\perp} \rightarrow \vec{q}_{\perp}$ in one of the terms since the measurement is just $\delta(\chi)$ and so remains unaffected.

Next we consider diagrams with Glauber insertions on the same side of the cut. As discussed earlier the total contribution to the jet function comes from the difference of real and opposite side Glauber insertions.
\subsection{Real diagrams with insertions on the same side of the cut}\label{sec:realsame}
Now we discuss real diagrams for the case of Glauber insertions on the same side of the cut. We first consider diagrams with collinear gluon emission from the Wilson line originating from the hard vertex. In this case there are three diagrams that give dominant contribution in the soft limit. The corresponding matrix elements for these diagrams including the complex conjugate ones are
\begin{align}
\begin{minipage}{3cm}\includegraphics[width=\textwidth]{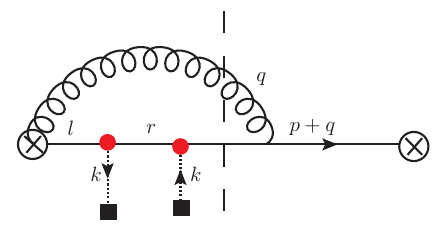} \end{minipage}+{\rm c.c.}=&2g^2C_F\delta_{CD}\int\frac{dq^{-}}{(2\pi)^3}\int d^2q_{\perp}\frac{p^{-}}{q^{-}q_{\perp}^2\omega}\mathcal{M}_{RS},\,\\
\begin{minipage}{3cm}\includegraphics[width=\textwidth]{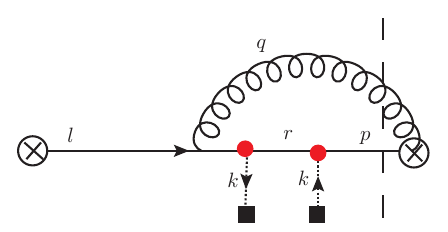} \end{minipage}+{\rm c.c.}=&2g^2C_F\delta_{CD}\int\frac{dq^{-}}{(2\pi)^3}\int d^2q_{\perp}\frac{p^{-}}{q_{\perp}^2q^{-}\omega}\left[1-F\left(\frac{Lq_{\perp}^2\omega}{2q^{-}p^{-}} \right) \right]\mathcal{M}_{RS}\label{eq:srwil2},\, \\
\begin{minipage}{3cm}\includegraphics[width=\textwidth]{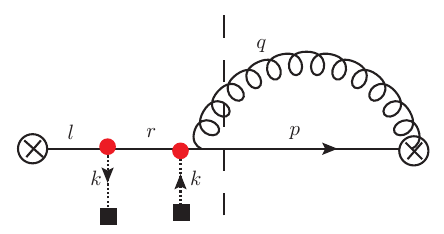} \end{minipage}+{\rm c.c.}=&2g^2C_F\delta_{CD}\int\frac{dq^{-}}{(2\pi)^3}\int d^2q_{\perp}\frac{p^{-}}{q^2_{\perp}q^{-}\omega}F\left(\frac{Lq_{\perp}^2\omega}{2q^{-}p^{-}} \right)\mathcal{M}_{RS},\label{eq:srwil3}
\end{align}
where the measurement function for this case reads as
\bea
\mathcal{M}_{RS} = \frac{(p^-)^2+ (q^-)^2}{\omega^2} \delta(\chi)+\frac{2 p^-q^-}{\omega^2}\delta \Big( \chi- \Big(\frac{\vec{q}_{\perp}}{q^-}+\frac{\vec{q}_{\perp}}{p^-}\Big)^2 \Big).
\eea
Note that the $L$ dependent term gets cancelled between Eqs.~\ref{eq:srwil2} and ~\ref{eq:srwil3}. Therefore, the overall contribution from these diagrams does not depend on the medium length. Adding all the above contributions, we get  
\begin{equation}
\text{B}32+\text{B}33+\text{B}34=\frac{(N_c^2-1)C_F}{N_c\pi^2}\int dq^{-}\int d^2q_{\perp}\frac{p^{-}}{q^{-}q_{\perp}^2\omega}\mathcal{M}_{RV}  
\end{equation}

Next we consider diagrams with collinear gluon emission from the Lagrangian insertion. In this case there are four diagrams out of which the dominant contribution come from first two diagrams shown below. The corresponding matrix elements after adding the mirror diagrams are
\begin{align}
\begin{minipage}{3cm}\includegraphics[width=\textwidth]{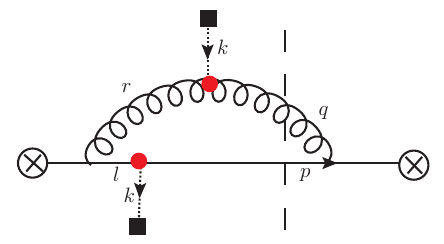} \end{minipage}+{\rm c.c.}=& -4g^2N_c\delta_{CD}\int \frac{dq^{-}}{(2\pi)^3}\int d^2q_{\perp} \frac{(p^-)^2}{q_{\perp}^2\kappa_{\perp}^2\omega^2}\bigg[\frac{\vec{q}_{\perp}\cdot\vec{\kappa}_{\perp}}{q^{-}}+\frac{\vec{q}_{\perp}\cdot\vec{\kappa}_{\perp}}{p^-}+\frac{(\vec{q}_{\perp}\cdot\vec{\kappa}_{\perp})q^-}{2(p^-)^2}  \bigg],\nonumber\\
&\bigg[F\bigg(\frac{L(\kappa_{\perp}^2-q_{\perp}^2)\omega}{2p^{-}q^{-}} \bigg)-F\bigg(\frac{Lq_{\perp}^2\omega}{2p^{-}q^{-}}\bigg)\bigg]\\
\begin{minipage}{3cm}\includegraphics[width=\textwidth]{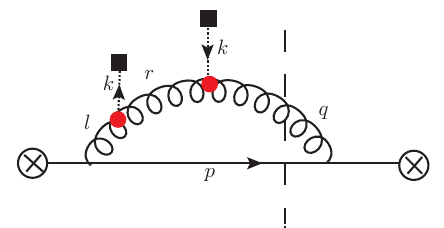} \end{minipage}+{\rm c.c.}=&4g^2N_c\delta_{CD}\int \frac{dq_{-}}{(2\pi)^3}\int d^2q_{\perp}\frac{(p^{-})^2}{q_{\perp}^2\omega^2}\bigg(1-F\bigg(\frac{Lq_{\perp}^2\omega}{2p^{-}q^{-}} \bigg) \bigg)\nonumber\\
&\bigg[\frac{1}{q^{-}}+\frac{1}{p^{-}}+\frac{q^-}{2(p^-)^2} \bigg].
\end{align}
Note that the contribution from the first diagram vanishes in the $L\to\infty$ limit.  Finally, the matrix elements for the diagram with glauber insertions on the collinear quark reads as 
\begin{align}
\begin{minipage}{3cm}\includegraphics[width=\textwidth]{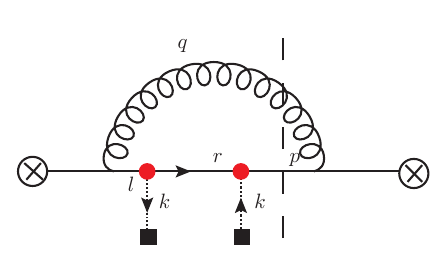} \end{minipage}+{\rm c.c.}=&\, -2g^2C_F\int \frac{dq^-}{(2\pi)^3}\int d^2q_{\perp}\frac{q^-}{q_{\perp}^2\omega^2}\bigg(1-F\bigg(\frac{Lq_{\perp}^2\omega}{2p^-q^-} \bigg) \bigg), \\
\begin{minipage}{3cm}\includegraphics[width=\textwidth]{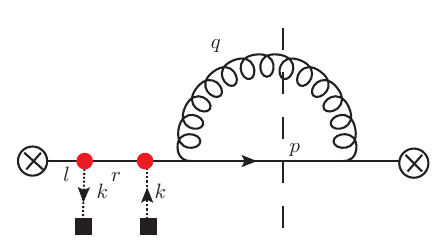} \end{minipage}+{\rm c.c.}=&\, -2g^2C_F \int\frac{dq^-}{(2\pi)^3}\int d^2q_{\perp}\frac{q^-}{q_{\perp}^2\omega^2}F\bigg(\frac{Lq_{\perp}^2\omega}{2p^-q^-} \bigg),\\
\begin{minipage}{3cm}\includegraphics[width=\textwidth]{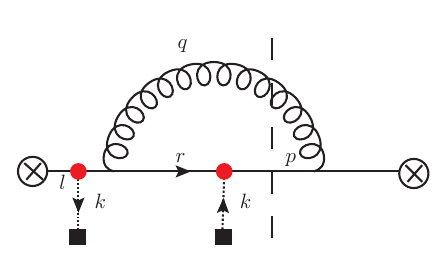} \end{minipage}+{\rm c.c.}=&\,0.
\end{align}

The contribution from the last diagram vanishes due to cancellation between the poles. The overall contribution to the observable  from the above diagrams is subleading even with finite $L$. Moreover, the contribution of these diagrams vanishes in the $L\to\infty$ limit. In the soft limit, i.e., $L \rightarrow \infty$ we get 
\bea
 J_{RV} = 4g^2\delta_{CD} \int \frac{dq^-}{q^-}\int d^2q_{\perp}\Bigg[ \frac{C_F}{q_{\perp}^2}+ \frac{N_c}{q_{\perp}^2}\Bigg]\mathcal{M}_{RV}
\eea
Next we consider virtual diagrams for glauber insertions on the same side of the cut.

\subsection{Virtual diagrams with insertions on the same side of the cut}\label{sec:virtualsame}

Here the measurement is only on the quark and so  
\bea
\mathcal{M}_{VV} = \delta(\chi).
\eea
Let us first consider the diagrams with collinear gluon emission from the Wilson line originating from the hard vertex. As mentioned earlier these kinds of diagrams originate from separating hard scale from the jet scale in the SCET factorization. In this case there are three independent diagrams that are shown below.  The corresponding matrix elements including the complex conjugate ones are given by
\begin{align}
\begin{minipage}{3cm}\includegraphics[width=\textwidth]{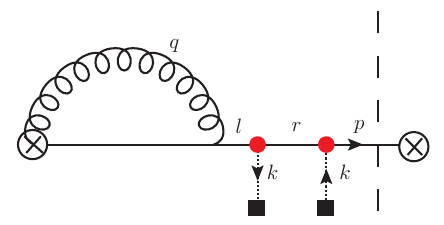} \end{minipage}+{\rm c.c.}=&2g^2C_F{\delta_{CD}}\int\frac{dq^-}{(2\pi)^3}\int d^2q_{\perp}\frac{p^-}{q_{\perp}^2q^-\omega}\left[-1+F\left(\frac{Lq_{\perp}^2\omega}{2q^- p^-} \right) \right]\mathcal{M}_{VV},\\
\begin{minipage}{3cm}\includegraphics[width=\textwidth]{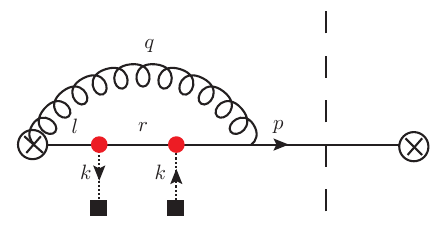} \end{minipage}+{\rm c.c.}=&-2g^2C_F{\delta_{CD}}\frac{dq^-}{(2\pi)^3}\int d^2q_{\perp}\frac{p^-}{q^-q_{\perp}^2\omega}F\left(\frac{Lq_{\perp}^2\omega}{2q^- p^-} \right) \mathcal{M}_{VV},\\
\begin{minipage}{3.5cm}\includegraphics[width=\textwidth]{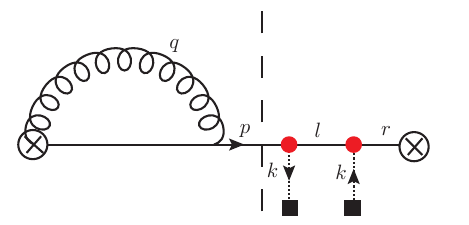} \end{minipage}+{\rm c.c.}=&-2g^2C_F\int \frac{dq^{-}}{(2\pi)^2}\int d^2q_{\perp}\frac{p^{-}}{q_{\perp}^2q^{-}\omega},
\end{align}
where again $p^-=\omega-q^-$. Note that once again the $L$ dependent terms gets cancelled. Therefore, the total contribution to the observable from these diagrams does not depend on the medium size. Finally we consider the last set of diagrams with collinear gluon emission from  Lagrangian insertions. Let us first look at the diagrams with one collinear gluon-glauber vertex. In this case there are two diagrams and the corresponding matrix elements including the mirror diagrams reads as
\begin{align}
\begin{minipage}{3cm}\includegraphics[width=\textwidth]{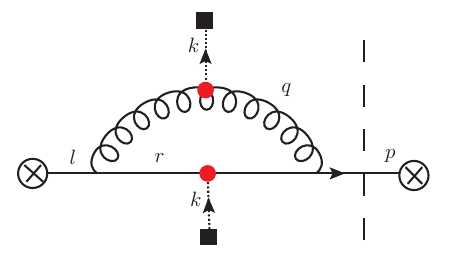} \end{minipage}+{\rm c.c.}=&4g^2N_c{\delta_{CD}}\int \frac{dq^{-}}{(2\pi)^3}\int d^2q_{\perp} \frac{(p^{-})^2}{\kappa_{\perp}^2q_{\perp}^2\omega^2}  \bigg[1-F\bigg(\frac{L\kappa_{\perp}^2\omega}{2p^{-}q^{-}} \bigg)\bigg]\nonumber\\
&\bigg[\frac{\vec{q}_{\perp}\cdot\vec{\kappa}_{\perp}}{q^{-}}+\frac{\vec{q}_{\perp}\cdot\vec{\kappa}_{\perp}}{p^{-}}+\frac{(\vec{q}_{\perp}\cdot\vec{\kappa}_{\perp})q^-}{2(p^-)^2} \bigg]\mathcal{M}_{VV},\label{eq:reals1} \\
\begin{minipage}{3cm}\includegraphics[width=\textwidth]{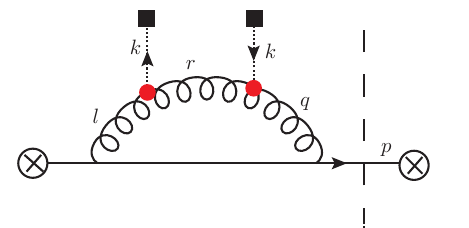} \end{minipage}+{\rm c.c.}=&8g^2N_c{\delta_{CD}}\int \frac{dq^{-}}{(2\pi)^3}\int d^2q_{\perp}\frac{(p^{-})^2}{q_{\perp}^2\omega^2}\bigg(1-F\bigg(\frac{Lq_{\perp}^2\omega}{2p^{-}q^{-}} \bigg) \bigg)\nonumber\\
&\bigg[\frac{1}{q_{-}}+\frac{1}{p_{-}}+\frac{q^-}{2(p^-)^2} \bigg]\mathcal{M}_{VV}.
\end{align}
Only the second diagram contributes in $L\to\infty$ limit. 

Finally we have diagrams with Lagrangian insertions and glauber insertion on the quark. In this case there are six diagrams. Let us first look at the following set of diagrams with matrix elements including the complex conjugate ones are
\begin{align}
\begin{minipage}{3cm}\includegraphics[width=\textwidth]{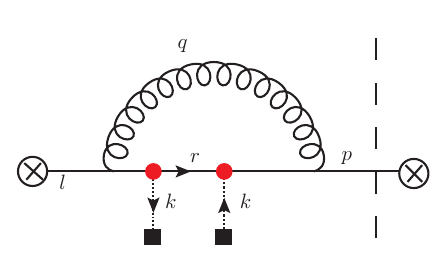} \end{minipage}+{\rm c.c.}=& \,2g^2C_F\int\frac{dq^-}{(2\pi)^3}\int d^2q_{\perp}\frac{q^-}{q_{\perp}^2\omega^2}\bigg(1-F\bigg(\frac{Lq_{\perp}^2\omega}{2q^-p^-} \bigg) \bigg)\mathcal{M}_{VV},\\
\begin{minipage}{3cm}\includegraphics[width=\textwidth]{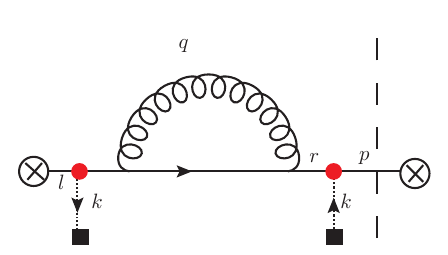} \end{minipage}+{\rm c.c.}=&-4g^2C_F\int\frac{dq^-}{(2\pi)^3}\int d^2q_{\perp}\bigg(\frac{\vec{k}\cdot\vec{\kappa}_{\perp}}{p^-\omega}+\frac{\kappa_{\perp}^2}{(p^-)^2} \bigg)\nonumber \\
&\times\frac{q^-(p^-)^2\omega^2}{[\omega(q_{\perp}^2p^-+\kappa_{\perp}^2q^-)-k_{\perp}^2p^-q^-]^2}\mathcal{M}_{VV},\\
\begin{minipage}{3cm}\includegraphics[width=\textwidth]{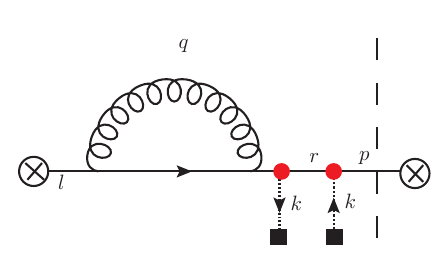} \end{minipage}+{\rm c.c.}=&\,2g^2C_F\int\frac{dq^-}{(2\pi)^3}\int d^2q_{\perp} \frac{q^-}{q_{\perp}^2\omega^2}\mathcal{M}_{VV},\\
\begin{minipage}{3cm}\includegraphics[width=\textwidth]{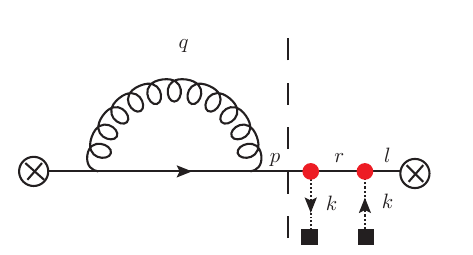} \end{minipage}+{\rm c.c.}=&-2C_F\int\frac{dq^-}{(2\pi)^3}\int d^2q_{\perp}\frac{q^-}{q_{\perp}^2\omega^2}\mathcal{M}_{VV},\\
\begin{minipage}{3cm}\includegraphics[width=\textwidth]{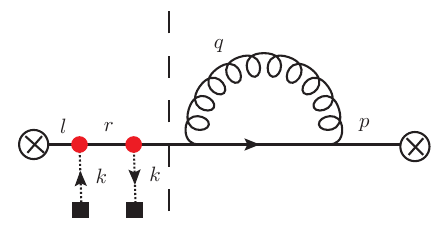} \end{minipage}+{\rm c.c.}=&-2C_F\int\frac{dq^-}{(2\pi)^3}\int d^2q_{\perp}\frac{q^-}{q_{\perp}^2\omega^2}\mathcal{M}_{VV}.
\end{align}
Further the contribution from the following two diagrams vanishes due to cancellations between the poles 
\begin{align}
\begin{minipage}{3cm}\includegraphics[width=\textwidth]{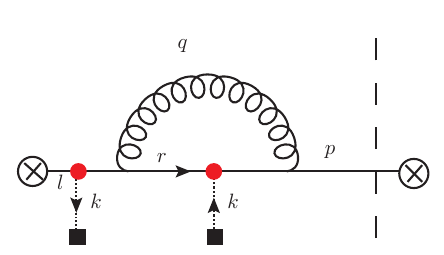} \end{minipage}+\begin{minipage}{3cm}\includegraphics[width=\textwidth]{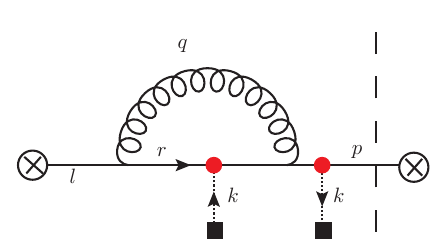} \end{minipage}=&0.
\end{align}
In the  soft limit the overall contribution from virtual diagrams with same side Glauber insertion takes the form
\bea
 J_{VV} &=& 4g^2\delta_{CD} \int \frac{dq^-}{q^-}\int d^2q_{\perp}\Bigg[ -\frac{C_F}{q_{\perp}^2}+\frac{N_c}{q_{\perp}^2}-\frac{N_c \vec{q}_{\perp} \cdot \vec{\kappa}_{\perp}}{q_{\perp}^2\kappa_{\perp}^2}\Bigg]\mathcal{M}_{VV}\nonumber \\
 &=& 4g^2\delta_{CD} \int \frac{dq^-}{q^-}\int d^2q_{\perp}\Bigg[ -\frac{C_F}{q_{\perp}^2}+\frac{N_c k_{\perp}^2}{2q_{\perp}^2\kappa_{\perp}^2}\Bigg]\mathcal{M}_{VV}.
\eea

\section{Plus distributions}
For a function $g(x)$ that is less singular than $1/x^2$, the plus distribution is defined as follows:
\begin{equation}
\label{appeq:PlusDefine}
\left[\theta(x)\, g(x)\right]_{+}^{x=x_0} = \lim_{\epsilon \rightarrow 0}\left[\theta(x-\epsilon)\, g(x) + \delta(x-\epsilon)\, \int_{x_0}^x\! x'\, g(x')\right]\, ,
\end{equation}
along with the boundary condition
\begin{equation}
\label{eq:BoundaryConditionDefine}
\int_{0}^{x_0}\! x\, \left[\theta(x)\, g(x)\right]_{+}^{x=x_0} = 0\, .
\end{equation}
Therefore, the distributions are given as
\begin{equation}
\label{appeq:PlusExpand}
\left[\frac{\theta(x)}{x^{1+\alpha}}\right]_{+}^{\infty} = -\frac{1}{\alpha}\delta(x)+\left[\frac{\theta(x)}{x}\right]_{+}\!\!\!-\alpha \left[\frac{\theta(x) \ln x}{x}\right]_{+}\!\!\! + {\cal O}(\alpha^2),
\end{equation}
and
\begin{equation}
\frac{1}{\zeta}\left[\frac{\theta(x)}{x/\zeta}\right]_{+}= \left[\frac{\theta(x)}{x}\right]_{+}\!\!\! - \ln \zeta\, \delta(x).    
\end{equation}

\bibliographystyle{utphys.bst}
\bibliography{main.bib}
\end{document}